\documentclass[11pt,oneside,a4paper,onecolumn]{elsarticle}
\usepackage{times}
\usepackage{fullpage}

\usepackage{gensymb}
\usepackage{tikz}
\usepackage{tikzscale}
\usepackage{alltt}
\usepackage{upquote}
\usepackage{float}
\usepackage{array}
\usepackage{pgfplots}
\usetikzlibrary{positioning}
\usepackage{graphicx}
\usepackage[toc,page]{appendix}
\usepackage[utf8]{inputenc}
\usepackage{amsmath,mathtools}
\usetikzlibrary{calc,decorations.markings,fit,shapes,chains,arrows}
\usepackage{amsthm}
\usepackage{amsfonts}
\usepackage{enumitem}
\usepackage{caption}
\usepackage{subcaption}
\usepackage{xcolor}
\usepackage{scalerel}
\usepackage{algorithm}
\usepackage{algorithmic}

\usepackage{multicol}
\usepackage{mathtools}
\usepackage{url}

\usepackage{amssymb}
\usepackage{soul}
\usepackage{slashed}
\usetikzlibrary{patterns}
\usetikzlibrary{shapes.arrows}
\usetikzlibrary{fillbetween,backgrounds}
\usepackage{placeins}
\pgfplotsset{compat=newest}
\usepackage{diagbox}
\usetikzlibrary{arrows.meta}

\pgfdeclarelayer{nodelayer}
\pgfdeclarelayer{edgelayer}

\tikzstyle{black_dot}=[fill=black, draw=black, shape=circle]
\tikzstyle{medium circle}=[fill=white, draw=black, shape=circle, radius=5cm]
\tikzstyle{medium box}=[fill=white, draw=black, shape=rectangle, minimum width=5.75cm, minimum height=2cm]
\tikzstyle{square_node}=[fill=white, draw=black, shape=rectangle]
\tikzstyle{grey_circle}=[fill={rgb,255: red,191; green,191; blue,191}, draw={rgb,255: red,191; green,191; blue,191}, shape=circle]
\tikzstyle{red_circle}=[fill=red, draw=red, shape=circle]
\tikzstyle{hollow_circle}=[fill=white, draw=black, shape=circle, radius=0.1cm]

\tikzstyle{latex}=[<->, fill=none, draw=black]
\tikzstyle{stealth}=[->]
\tikzstyle{stealth_rev}=[<-]
\tikzstyle{red_line}=[-, draw=red]
\tikzstyle{grey_line}=[-, fill=none, draw={rgb,255: red,128; green,128; blue,128}]
\tikzstyle{black_line}=[-, thickness=very thick]
\tikzstyle{black_dashed}=[-, dashed=dashed]

\input{tikz_figs/sample.tikzdefs}

\tikzstyle{none}=[inner sep=0pt]
\pgfsetlayers{background,nodelayer,edgelayer,main}

\usepackage[final]{hyperref} 
\hypersetup{
	colorlinks=true,       
	linkcolor=blue,        
	citecolor=blue,        
	filecolor=magenta,     
	urlcolor=blue         
}

\newcolumntype{M}[1]{>{\centering\arraybackslash}m{#1}}
\newcolumntype{N}{@{}m{0pt}@{}}

\tikzset{>=latex}
\def\@author#1{\g@addto@macro\elsauthors{\normalsize%
    \def\baselinestretch{1}%
    \upshape\authorsep#1\unskip\textsuperscript{%
      \ifx\@fnmark\@empty\else\unskip\sep\@fnmark\let\sep=,\fi
      \ifx\@corref\@empty\else\unskip\sep\@corref\let\sep=,\fi
      }%
    \def\authorsep{\unskip,\space}%
    \global\let\@fnmark\@empty
    \global\let\@corref\@empty
    \global\let\sep\@empty}%
    \@eadauthor={#1}
}

\graphicspath{{Images/}}

\begin{document}
\begin{frontmatter}
\title{An Interface-Driven Adaptive Variational Procedure for Fully Eulerian Fluid-Structure Interaction via Phase-field Modeling}
\author[ubc]{Biswajeet Rath}
\ead{biswajee@mail.ubc.ca}

\author[ubc]{Xiaoyu Mao}
\ead{dgcsmaoxiaoyu@gmail.com}

\author[ubc]{Rajeev K. Jaiman\corref{cor1}}
\ead{rjaiman@mail.ubc.ca}
\cortext[cor1]{Corresponding author}
\address[ubc]{Department of Mechanical Engineering, The University of British Columbia, Vancouver, BC V6T 1Z4}

\begin{abstract}
In this paper, we present a novel interface-driven adaptive variational procedure using a fully Eulerian description of fluid-structure interaction. The proposed fully-Eulerian procedure involves a fixed background unstructured mesh on which the fluid-structure interface is treated implicitly. We model the fluid-structure interaction by the phase-field finite element formulation relying on a partitioned staggered integration of the convective Allen-Cahn equation with the unified momentum equation for both solid and fluid dynamics. We employ the positivity preserving variational scheme for a bounded and stable solution of the Allen-Cahn phase-field equation. To evaluate the solid stresses, the left Cauchy-Green deformation tensor is convected at each time step to trace the evolution of the solid strain in the Eulerian reference frame. We utilize the residual based error indicators and the newest vertex bisection algorithm for the adaptive refinement/coarsening of the unstructured mesh. The proposed nonlinear adaptive partitioned procedure restricts the coarsening step to the last non-linear iteration while simultaneously ensuring convergence properties of the coupled governing equations. We perform a detailed convergence and accuracy analysis via two benchmark problems namely, the pure solid system and a coupled fluid-solid system with an interface in a rectangle domain. We next systematically assess the performance of the adaptive procedure in terms of conservation properties for the increasing complexity of problems. Finally, we demonstrate our fully-Eulerian interface-driven adaptive FSI model to simulate the contact and bouncing phenomenon between an elastic solid and a rigid wall.
\smallskip
\smallskip

\end{abstract}
\end{frontmatter}

\section{Introduction}
Fluid-structure interaction (FSI) is one of the ubiquitous multiphysics problems in nature and engineering applications. 
Two-way interactions between a flowing fluid and a deforming solid occur in applications ranging from aerospace engineering such as flapping and flexible wings for unmanned flying vehicles \cite{shyy1999flapping,li2018novel,joshi2020variational}, marine offshore platform and pipelines \cite{jaiman2016partitioned,joshi20183d} to biomedical fields such as flow of blood in arteries and bio-locomotion studies of fish, jellyfish, bat and among others \cite{griffith_annualreview_2020,jaimancomputational}.
The relative motion between a solid and the fluid in both internal and external flows leads to the application of loads on the solid resulting in deformations and/or displacements. These structural changes affect the fluid motion resulting in a bidirectional coupling between the solid and the fluid. This coupling is inherently nonlinear in nature, further complicating the dynamics of the problem for analytical methods \cite{jaimancomputational}. 
The motivation for this work primarily comes from bio-inspired locomotion. For example, octopus-based soft robotics \cite{trivedi2008soft, kier1985tongues} involve very flexible structures (e.g., muscular hydrostats) with large deformations and solid-to-solid contact which pose serious challenges to the mesh-based moving boundary and Eulerian-Lagrangian methods \cite{jaimancomputational}. There is a need for an alternative approach to deal with these specific types of FSI problems that involve fairly complicated structural motions and constitutive relations \cite{richer2017book}. In the present work, we explore the fully Eulerian approach to simulate such FSI problems with reasonable accuracy of the coupled dynamics.

For large-scale numerical FSI modeling, the continuum hypothesis for the physical domains is generally adopted. Using the continuum approach, the key challenges in modeling fluid-structure interactions pertain to the treatment of the boundary conditions at the fluid-solid interface; handling of dissimilar coordinate frames for the fluid and solid domains; and the construction of efficient and robust numerical techniques for the coupled nonlinear PDEs. While the mesh moves with the simulated continuum in the Lagrangian description, the physical field/continuum transports through a stationary or fixed grid in the Eulerian description. Based on the choice of coordinate frames, one can consider fully Lagrangian, fully Eulerian, or hybrid Lagrangian-Eulerian approaches for solving fluid-structure interaction problems. In a Lagrangian approach, the fluid-solid interface is defined explicitly on a boundary conforming mesh following the motion of the solid in the Lagrangian coordinates, such as fully Lagrangian \cite{belytschko1976fluid} and Arbitrary Lagrangian-Eulerian (ALE) \cite{hirt1974arbitrary, hughes1981lagrangian, hu2001direct} techniques. Body-fitted ALE techniques have long dominated the realm of numerical solutions for FSI problems. While these techniques have been very successful for accurate interface tracking of several complicated multiphase and multiphysics engineering applications, they face difficulties dealing with flexible multibody contact dynamics (e.g., self-contact between octopus arms) or those involving large topological changes (e.g., the rupture of an aortic aneurysm).

Hybrid Lagrangian-Eulerian and fully Eulerian approaches attempt to address these difficulties. Unlike a body-fitted approach, these techniques rely on a non-conforming discretization where the interface embeds or cuts across the fixed background grid. Such treatment can provide a better handling of large structural deformation while avoiding a frequent remeshing required in body-fitted methods. Hybrid techniques such as the immersed boundary method (IBM) by Peskin \cite{peskin2002immersed,mittal_annualreview_2005} make use of the regularized delta functions to describe the interface, hence smoothing out stress discontinuities that are usually associated with fluid-solid interfaces. In the context of  variational finite element framework, Boffi et al. \cite{boffi2015finite} and Zhang et al. \cite{zhang2004immersed} extended the immersed boundary method for fluid-solid interaction. In immersed interface methods (IIM) \cite{leveque1994immersed, li2016immersed}, the regularized kernels are avoided and the jump conditions are directly built into the finite difference approximations. IIMs can provide higher accuracy, although requiring coupling conditions that involve complex geometrical computations. 

Some hybrid approaches introduce additional interface variables and couple the fluid and solid system using Lagrange multipliers e.g., fictitious domain method. 
Glowinski et al. \cite{glowinski2001fictitious} proposed the fictitious domain approach to perform numerical simulations of fluid flow over moving rigid bodies. One can generate the solid mesh independently from the fluid mesh and the interface cuts the elements without the need for mesh alignment. There are variants of the fictitious domain approach such as the finite cell method which utilizes the high-order finite element and the framework of cut cells \cite{parvizian2007finite,burman2012fictitious}. The challenging aspects of the finite cell method are the robust treatment of complex cut elements and the numerical integration over the cut cells \cite{parvizian2007finite}. Similar to the fictitious domain method, the Eulerian-Lagrangian coupling methods based on the eXtended Finite Element Method (XFEM) do not require conformity between the meshes.
%
The XFEM was initially introduced for simulation of cracks in structures \cite{belytschko1999elastic} and later extended for solving two-phase flows \cite{chessa2003extended} and fluid-structure interactions \cite{wagner2003particulate, gerstenberger2008extended}. 
Similar to the finite cell method, the key problem associated with XFEM is the evaluation of integrals over discontinuous integrands on polyhedral domains. 
 Felippa et al. \cite{felippa2011classification} presented a classification of several interface coupling-based approaches for solving FSI problems. The majority of issues of the hybrid techniques arise from the coupling conditions between the Eulerian and Lagrangian variables. Most of the immersed methods may suffer from conservation issues for large topology changes and contact problems \cite{griffith2012volume, casquero2018non}. Some of these methods may also pose challenges in handling geometric complexities and numerical implementation. Techniques such as the finite cell method or XFEM can face the issue of extremely small cut cells which cause a problem during the linear solver step, thus requiring further rectifications \cite{badia2018robust}. A fully Eulerian approach is usually desirable for problems involving very large deformation of the solid.

As the name suggests, in a fully Eulerian approach both the solid and the fluid dynamics are represented in the Eulerian coordinates. 
This approach belongs to a broad category of fixed-mesh interface-capturing framework. However, solid mechanics is formulated in terms of Eulerian coordinates 
in contrast to immersed and fictitious domain methods. The coupled system can be computed on a single mesh and the coupling can be constructed in 
a consistent variational monolithic formulation.
Liu and Walkington \cite{liu2001eulerian} attempted to develop the evolution equation for the strain in an Eulerian framework for the flowing fluid containing visco-hyperelastic particles. The idea of a fully Eulerian approach for solving FSI problems was then demonstrated by Dunne \cite{dunne2006eulerian}, whereby the authors proposed an Eulerian framework for modeling FSI of an incompressible fluid and an elastic structure. In \cite{dunne2006eulerian}, the author used a technique called the initial position set to track the initial location of a point in the domain. While the fully Eulerian approach has been further extended by \cite{wick2013fully} for unsteady problems, Richter \cite{richter2013fully} demonstrated the method for the non-linear coupling of an incompressible fluid and a hyperelastic solid. In \cite{richter2013fully}, characteristic functions were introduced to identify the fluid and solid domains and the inverse map function was employed to trace the material coordinates at each time step. The inverse map function along with the characteristic function serves the purpose of the initial point set technique used in the previous studies of \cite{dunne2006eulerian} and \cite{wick2013fully}. In both works \cite{wick2013fully} and \cite{richter2013fully}, the authors suggested the need for adaptive mesh refinement to maintain convergence in this approach. 
During the solution procedure, the evolution of the solid strains can be performed by convecting the inverse map function \cite{valkov2015eulerian, dunne2006adaptive}, the deformation gradient tensor or the left Cauchy-Green deformation tensor \cite{liu2001eulerian,sun2014full}. 


Handling the fluid-solid interface is one of the primary challenges in modeling FSI problems. Over the past two decades, a substantial amount of effort has been devoted to addressing this challenge. There are two key issues involved, namely: (i) accurate interface description i.e., satisfying the boundary conditions along the interface, and (ii) evolution of the interface in space and time. One way to capture the interface is via the sharp-interface approach where the fluid-solid interface is treated as a sharp boundary with an infinitely thin interface between the fluid and solid domains. On the other hand, in a diffuse-interface representation, the interface capturing variable varies smoothly across the interface \cite{van1979thermodynamic}. The level set method, introduced by Sethian \cite{sethian1999level} and Fedkiw and Osher \cite{fedkiw2002level} for tracking physical interfaces in Eulerian coordinates, is an example of a diffuse interface approach. The level set function is usually a signed distance function, which is used to describe the physical domains.
\cite{fedkiw2002level} and \cite{pino2013finite} discussed the application of level set technique to fluid-structure interaction problems.
The re-initialization process \cite{sussman1994level} for the level set function may prove to be computationally expensive and lead to poor mass conservation properties \cite{zhao2014improved}. Higher-order discretization methods and improved re-initialization \cite{sussman1999efficient, peng1999pde, olsson2005conservative} introduced for correcting the mass conservation issue for the level set technique can further make the process complicated associated with a high computational cost.

\begin{figure}[h]
	\centering
	\begin{tikzpicture}[scale=0.95]
	
	\draw[thick] (0,0) rectangle (8,8);
	\draw[step=0.3cm,thin, gray] (0.05,0.05) grid (7.95,7.95);
	\draw[very thick] (4,4) circle (2cm);
	\draw[thick,gray,dashed] (4,4) circle (2.5cm);
	\draw[thick,gray,dashed] (4,4) circle (1.5cm);
	
	\node at (2,1) {$\Omega^f (\phi = -1)$};
	\node at (4,4) {$\Omega^s (\phi = 1)$};
	\node at (4,6.8) {Diffused Interface};
	
	\end{tikzpicture}
	\caption{Illustration of a fixed grid representation of an FSI problem using the diffuse interface description. The region between the two dashed lines represents the diffused interface 
where the value of $\phi$ varies gradually from 1 to -1.}
	\label{fig:eulerian_grid}
\end{figure}
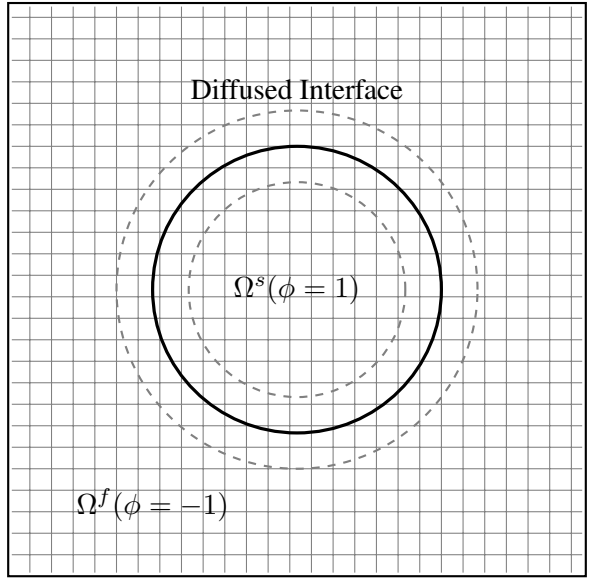

Originating from the thermodynamically consistent theories of phase transitions \cite{cahn1961spinodal, allen1979microscopic}, the phase-field models provide a natural way for the diffuse-interface description via minimization of the gradient energy. The phases are indicated by an order parameter $(\phi)$ which varies smoothly across the interface of a finite thickness as shown in Fig. \ref{fig:eulerian_grid}. Anderson et al. \cite{anderson1998diffuse} reviewed the development and application of diffuse interface models focusing on phase-field method for fluid-fluid interfaces. The phase-field method provides a regularized fluid-solid boundary, controlled by the interface thickness parameter $\varepsilon$. At the limit $\varepsilon \to 0$, the sharp interface limit can be recovered \cite{anderson1998diffuse}. The relative time scale between the coupled Navier-Stokes equations and the Allen-Cahn equation for a controlled diffusion in the interface was investigated in \cite{mao2021variational}. In the current work, we use the convective form of the Allen-Cahn equation to evolve the interface in space and time, in conjunction with the Lagrange multiplier term for mass conservation and positivity preserving terms \cite{joshi2018positivity} for a stable and bounded solution. The method is free from the re-initialization process from the level set method. Its free energy minimization process ensures energy stability and makes it more compatible with variational discretization \cite{joshi2018positivity}. As a result, we prefer the phase-field method for the interface capturing of fully Eulerian fluid-structure interaction. We introduce an adaptive refinement scheme to handle some of the numerical issues of the diffuse interface-based approach and enhance the conservation properties of the system.

The present study builds upon our previous work \cite{joshi2018adaptive}, where an adaptive refinement procedure was presented for two-phase flows, by integrating the Navier-Stokes and Allen-Cahn equations via a fully implicit partitioned approach. The recently proposed positivity preserving variational (PPV) technique \cite{joshi2018positivity} is employed to maintain the stable and bounded solution of the coupled nonlinear differential system. For the first time, the present work demonstrates the successful integration of the unified continuum equations for the entire domain and the Allen-Cahn equation for the phase-field function and accurate implementation of the adaptive procedure for capturing the fluid-solid interface. Since we do not explicitly solve for the solid displacements, we evolve the solid strains to evaluate the deformation and stress field. An adaptive algorithm based on the newest vertex bisection method \cite{chen2006afem} is employed on unstructured triangular meshes. This algorithm avoids complicated tree data structures to store the information about the refinement and coarsening nodes. The adaptive procedure restricts the coarsening step to the last non-linear iteration while maintaining the convergence properties of the coupled discretized equations for the fluid-structure interaction.
We examine the effect of adaptive mesh refinement and the role of interface thickness parameter $(\varepsilon)$ for different initial grid sizes $h_{init}$ through the use of performance indicators such as the mass conservation error $\left(e_{mass}\right)$, the volume conservation error $\left(e_{vol}\right)$, the elapsed time for the simulation and the residual error indicator $\left(\eta\right)$. The primary focus of the work is to study the use of the diffuse interface approach for solving fluid-structure interactions via the fully Eulerian method and evaluate the adaptive refinement procedure for improved accuracy and conservation properties.

The article is organized as follows. Section 2 presents the mathematical formulation of the fully Eulerian approach by describing the equations for the individual fields in the Eulerian frame of reference and further describes the associated equations. Section 3 discusses the semi-discrete variational form of the continuum field equations. Section 4 begins with a discussion on the need for adaptive refinement in a fully Eulerian approach for fluid-structure interaction that follows by the algorithmic details and numerical implementation. Section 5 demonstrates a systematic convergence and verification of the FSI framework. In Section 6, we first systematically examine the improved conservation properties obtained by the proposed adaptive refinement procedure and then demonstrate the efficacy and robustness of the procedure for contact and bouncing phenomena between an elastic solid and a rigid wall. Section 7 summarizes the key findings of the paper.

\section{Continuum Fully Eulerian Model}




This section deals with the development of the fluid and solid equations in the Eulerian frame for the one-field FSI formulation. We further elaborate on the evolution of the fluid-solid interface via the convective Allen-Cahn equation. Finally, we present the transport equation for the solid strain to evaluate the stress field.

\subsection{Continuity and Momentum Balance Equations}
From continuum mechanics conventions, let $\Omega$ denote the entire domain consisting of two parts: the fluid and the solid, such that $\Omega = \Omega^f \cup \Omega^s$, where $f$ and $s$ denote the fluid and solid domains respectively. Fluid is usually described in the Eulerian frame of reference, whereas solid is usually well suited in the Lagrangian frame of reference. However, in a fully Eulerian approach, we transform the solid equations into Eulerian coordinates to construct the unified continuum equations.

\subsubsection{Fluid Equations in Eulerian Frame}
For the incompressible and Newtonian fluid flow, the Navier-Stokes equations in the fluid domain can be written as:
\begin{equation}
	\begin{aligned}
		\rho^f \bigg( \frac{\partial \boldsymbol{v}^f}{\partial t} + (\boldsymbol{v}^f \cdot \nabla ) \boldsymbol{v}^f \bigg) &= \nabla \cdot \boldsymbol{\sigma}^f +  \boldsymbol{b}^f \ \ \ \  &\text{on   } \Omega^f, \\
		\nabla \cdot \boldsymbol{v}^f &= 0 \ \ \ \ \  &\text{on   } \Omega^f,
	\end{aligned}
\end{equation}
where  $\boldsymbol{v}^f = \boldsymbol{v}^f(\boldsymbol{x},t)$ and $p^f = p^f(\boldsymbol{x},t)$  denote the velocity and pressure at each spatial point $\boldsymbol{x} \in \Omega^f$, the Cauchy stress tensor is given by $\boldsymbol{\sigma}^f = -p^f\boldsymbol{I} + 2\mu^f\boldsymbol{D}(\boldsymbol{v}^f)$ with the rate of strain tensor $\boldsymbol{D}(\boldsymbol{v}^f) = \frac{1}{2}(\nabla \boldsymbol{v}^f + (\nabla \boldsymbol{v}^f)^T)$ and $\boldsymbol{b}^f$ represents the fluid body force. 
When a fluid flow interacts with a structure, the fluid loading changes the configuration of the structure by inducing deformations and/or displacements. 
Structural response can be treated in Lagrangian or Eulerian coordinates.

\subsubsection{Structure Equations in Lagrangian Frame}
We consider an incompressible, hypereleastic solid for the current work. The governing equations for the structure in Lagrangian coordinates are given by
\begin{equation}
	\begin{aligned}
		\hat{\rho^s} \frac{\partial \hat{\boldsymbol{v}}^s}{\partial t} \biggr\rvert_{\boldsymbol{X}} &= \hat{\nabla} \cdot (\hat{\boldsymbol{P}}) + \hat{\boldsymbol{b}^s} \ \ \ \  &\text{on   } \Omega^s, \\
		\text{det} (\boldsymbol{F}) &= 1 \ \ \ \  &\text{on   } \Omega^s,
	\end{aligned}
\end{equation}
where $(\hat{\cdot})$ represents quantities in the material configuration, $\hat{\boldsymbol{P}}$ is the first Piola-Kirchoff stress tensor, and $\boldsymbol{F}$ is the deformation gradient tensor given by $\boldsymbol{F} = (\boldsymbol{I} + \hat{\nabla} \boldsymbol{u}^s)$. Here $\boldsymbol{u}^s$ denotes the displacement vector and $\boldsymbol{I}$ is the identity matrix. $\hat{\boldsymbol{P}}$ is related to the Cauchy stress tensor $(\boldsymbol{\sigma}^s)$ by the relation $\hat{\boldsymbol{P}} = J \boldsymbol{\sigma}^s \boldsymbol{F}^{-T} $ and the material density is given by $\hat{\rho^s}=J \rho^s$, where $J = \text{det} \boldsymbol{F}$ is the volumetric stretch which is unity for incompressible materials, as mentioned above.



\subsubsection{Structure Equations in Eulerian Frame}
The above Lagrangian equations are converted to the Eulerian frame of reference using the deformation gradient tensor. The conversion of the time derivative from Lagrangian to Eulerian coordinates can be obtained as
\begin{equation}
		\hat{\rho^s} \frac{\partial \hat{\boldsymbol{v}}^s}{\partial t} \biggr\rvert_{\boldsymbol{X}} = J \rho^s \bigg( \frac{\partial \boldsymbol{v}^s}{\partial t} \biggr\rvert_{\boldsymbol{x}} +  (\boldsymbol{v}^s \cdot \nabla) \boldsymbol{v}^s \bigg), 
\end{equation}
where $J$ is the volumetric stretch defined in the previous section.  
Similarly, the stress term can be written as 
\begin{equation}
	\hat{\nabla} \cdot (\hat{\boldsymbol{P}}) = J \nabla \cdot (\boldsymbol{\sigma}^s).
\end{equation}
Incompressibility condition of $\text{det} \boldsymbol{F} = 1$ in the Lagrangian sense translates to $\nabla \cdot \boldsymbol{v}^s = 0$ in the Eulerian frame \cite{jain2019conservative}. 
Thus, the governing equations for the solid domain in the Eulerian frame are given by
\begin{equation}
	\begin{aligned}
		\rho^s \bigg( \frac{\partial \boldsymbol{v}^s}{\partial t} + (\boldsymbol{v}^s \cdot \nabla ) \boldsymbol{v}^s \bigg) &= \nabla \cdot \boldsymbol{\sigma}^s + \boldsymbol{b}^s \ \ \ \  &\text{on   } \Omega^s, \\
		\nabla \cdot \boldsymbol{v}^s &= 0 \ \ \ \  &\text{on   } \Omega^s,
	\end{aligned}
\end{equation}
where the Cauchy stress tensor can be expressed as
\begin{equation}
	\boldsymbol{\sigma}^s = -p^s \boldsymbol{I} + 2\mu^s \boldsymbol{D} (\boldsymbol{v}^s) + \boldsymbol{\sigma}^s_{sh}.
\end{equation}
In the above equation, $\boldsymbol{\sigma}^s_{sh} = \mu^s_{L} (\boldsymbol{B} - \boldsymbol{I})$ in accordance with the incompressible neo-Hookean model (refer to \ref{appendixB} for discretization of the solid shear stresses). In the previous relation, $\boldsymbol{B}$ is the left Cauchy-Green deformation tensor, given by $\boldsymbol{B}=\boldsymbol{F}\boldsymbol{F}^T$, $\mu^s_{L}$ is the shear modulus of the material (Lam\'e's second parameter), given by $\mu^s_{L} = \frac{E}{2(1+\nu)}$, where $E$ is the Young's modulus of the material and $\nu$ is the Poisson's ratio. 
The presence of the viscous term in the Cauchy stress tensor equation allows to model visco-hyperelastic solids.
It is well known that the conversion of these equations into the Eulerian frame introduces convective terms in the structural equation as well. This marks the primary point of difference compared to the Lagrangian approach. During the numerical implementation, these convective terms can cause accuracy and stability issues during the numerical solution if not discretized appropriately. 

\subsubsection{Unified Continuum Equations}
Next, we present the unified continuum equations for the fluid-solid system. Combining the developments from the previous sections, we arrive at the one-field continuum equations for the fluid and solid domains using the phase indicator $\phi(\boldsymbol{x},t)$ as
\begin{equation}
	\begin{aligned}
		\nabla \cdot \boldsymbol{v} &= 0 \ \ \ \  &\text{on   } \Omega, \\
		\rho(\phi) \left(\frac{\partial \boldsymbol{v}}{\partial t}\bigg\rvert_x + (\boldsymbol{v} \cdot \nabla)\boldsymbol{v}\right) &= \nabla \cdot (\boldsymbol{\sigma}(\phi)) + \boldsymbol{b}(\phi) \ \ \ \  &\text{on   } \Omega,
	\end{aligned}
\end{equation}
where the phase properties are defined as follows:
\begin{equation}
	\begin{aligned}
		\rho(\phi)&=\frac{1-\phi}{2}\rho^f + \frac{1+\phi}{2}\rho^s ,\\
		\mu(\phi)&=\frac{1-\phi}{2}\mu^f + \frac{1+\phi}{2}\mu^s ,\\
		\boldsymbol{\sigma}(\phi)&=\frac{1-\phi}{2}\boldsymbol{\sigma}^f + \frac{1+\phi}{2}\boldsymbol{\sigma}^s ,\\
		\boldsymbol{b}(\phi)&=\frac{1-\phi}{2}\boldsymbol{b}^f + \frac{1+\phi}{2}\boldsymbol{b}^s .
	\end{aligned}
\end{equation}
Hence the fully Eulerian formulation results into simplified equations for the combined fluid-solid domain.
We discuss more on the phase indicator function $\phi$ in the next section.

\subsection{Interface Representation via Phase-field Method}
In this section, we focus on an important aspect of any FSI problem i.e. dealing with the fluid-solid interface. As mentioned previously, we use the diffused interface based phase-field method for capturing the interface. In the phase-field representation, the continuum is considered as a mixture of the two phases and free energy is stored. Bulk phases are regions where there is a slow variation of the phase-field function and interface is the region where there is a high localized variation of the phase-field function. This avoids the requirement of the knowledge of exact interface location at every time step, further preventing setting boundary conditions at the interface for the bulk phases. This is one of the primary advantages of the phase-field method over sharp-interface methods. 
In both Cahn-Hilliard and Allen-Cahn equations for phase-field description, the phase separation and phase transition in the diffused interface result from the process of free energy minimization, where the Ginzburg-Landau free energy functional is given by:
\begin{equation}
	\mathcal{E}: \mathcal{H}^1 (\Omega) \cap \mathcal{L}^4 (\Omega) \to \mathbb{R}_{\geq 0}, \mathcal{E}(\phi(\boldsymbol{x},t)) = \int_\Omega \left( F(\phi(\boldsymbol{x},t)) + \frac{\varepsilon^2}{2} |\nabla \phi(\boldsymbol{x},t) |^2 \right) d\Omega , 
\end{equation}
where $\Omega$ is the bounded physical domain, $H^1(\Omega)$ denotes the space of square-integrable real-valued functions with square-integrable derivatives on $\Omega$, $L^4(\Omega)$ denotes the function space in which the fourth power of the function is integrable, $\mathbb{R}_{\geq 0}$ represents the set of non-negative real numbers and $\phi(\boldsymbol{x},t)$ is referred to as the order parameter or the phase-field function which indicates the two domains and has values of either +1 or -1. The first term in the RHS is the bulk or mixing energy and depends on the local composition of the mixture. The second term is the interfacial or gradient energy and depends on the composition of immediate environment. Hence there is a trade-off between preference of pure phases and a mixed uniform phase. The ratio of these two effects controlled by $\varepsilon$ decides the thickness of the diffused interface region. At equilibrium, the interface thickness is the distance over which $\phi$ varies from $-0.9$ to $0.9$ which can be estimated as $2\sqrt{2}\tanh^{-1} (0.9) \varepsilon \approx 4\varepsilon$.

The Allen-Cahn equation is computationally preferred for its lower order equations compared to the Cahn-Hilliard equation. The missing mass conservation property in the Allen-Cahn equation is made up for by adding a Lagrange multiplier \cite{rubinstein1992nonlocal,bretin2009modified} or an anti-curvature term \cite{sun2007sharp}. The final convective form of the Allen-Cahn equation after addition of the Lagrange multiplier can be derived as follows:
\begin{equation}
	\frac{\partial \phi}{\partial t} + \boldsymbol{v} \cdot \nabla \phi = -\gamma \left ( F'(\phi) - \varepsilon^2 \Delta \phi - \beta (t) \sqrt{F(\phi)} \right ) \text{  on  } \Omega \times [0,T] ,
\end{equation}
where $\gamma$ is the mobility parameter (of the order $\mathcal{O}(10^{-3})$ for all the test problems in this work) and $\beta(t)$ is the time-dependent part of the Lagrange multiplier, given by $\beta(t) = \frac{\int_\Omega F'(\phi) d\Omega}{\int_\Omega \sqrt{F(\phi)} d\Omega}$. The functional form for the bulk energy $F(\phi)$ has been chosen as the double-well potential function $F(\phi) = \frac{1}{4} (\phi^2 - 1)^2$ in the present study.
It is worth mentioning that this quartic polynomial function is an approximation to the more accurate logarithmic function that is derived from thermodynamic considerations. Analogous studies have been carried out for the more rigorous Cahn-Hilliard equation \cite{copetti1992numerical, barrett1995error} to illustrate its numerical properties under the assumption of logarithmic free energy. The polynomial approximation, though has been shown to perform reasonably well in literature \cite{kim2021unconditionally} and is more popular due to its numerical simplicity.

\noindent The boundary condition and initial condition for the Allen-Cahn equation are as follows:
\begin{align}	
		\frac{\partial \phi}{\partial n} \biggr\rvert_{\Gamma} = \boldsymbol{n} \cdot \nabla \phi = 0 \ \ \ \  &\text{on   } \Gamma \times [0,T] ,\\
		\phi|_{t=0} = \phi_0 \ \ \ \  &\text{on   } \Omega .	
\end{align}
Evaluation of the solid stresses accurately in an Eulerian frame of reference calls for an additional equation in the form of a deformation tensor transport equation, which is the point of discussion for the next section. 

\subsection{Evolution of the Solid Strain}
\begin{figure}[h]
     \centering
     \includegraphics[scale=0.9]{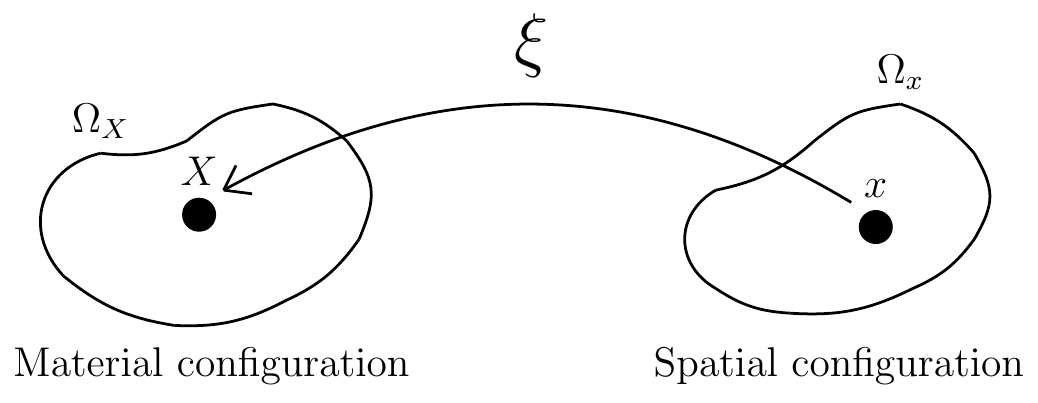}
     \caption{Representation of the inverse map function $(\boldsymbol{\xi})$ that maps from the current (spatial) configuration to the reference (material) configuration. $\Omega_X$ is the material configuration of the solid at $t=0$ with material coordinates $\boldsymbol{X}$ and $\Omega_x$ is the spatial configuration of the solid at the current time $t$ with spatial coordinates $\boldsymbol{x}$.}
     \label{xi_mapping}
 \end{figure}

Apart from solving the unified momentum equation, it is also essential to evolve the solid strain to evaluate the stress field accurately. This becomes crucial since we are not directly computing the solid displacements. This additional information closes the system of equations for the fully Eulerian model. 
The mapping from the spatial to the material coordinates is carried out through the inverse map function $\boldsymbol{\xi}$ such that $\boldsymbol{X} = \boldsymbol{\xi}(\boldsymbol{x},t)$. This mapping function helps to trace the material points on the solid during its motion, as shown in Fig.~\ref{xi_mapping}.

Consider initial undeformed configuration of the solid $\boldsymbol{\xi}(\boldsymbol{x},0) = \boldsymbol{x} = \boldsymbol{X}$.
By convecting the inverse map function \cite{valkov2015eulerian, dunne2006adaptive},
the evolution of $\boldsymbol{\xi}$ subject to this initial condition can be written as:
\begin{equation}
	\frac{\partial \boldsymbol{\xi}}{\partial t} + (\boldsymbol{v}^s \cdot \nabla) \boldsymbol{\xi} = 0 \ \ \ \  \text{on   } \Omega^s.
\end{equation}
In the continuum sense, one can show the equivalence between the convection of inverse map function and the deformation gradient tensor or left Cauchy-Green deformation tensor using simple manipulations and the fact that $\boldsymbol{F} = (\nabla \boldsymbol{\xi})^{-1}$ and $\boldsymbol{B} = \boldsymbol{FF}^T$ (refer to \ref{appendixA}). 
The transport equations for the  deformation gradient tensor and left Cauchy-Green tensor can be expressed as follows:
\begin{equation}
	\frac{\partial \boldsymbol{F}}{\partial t} + (\boldsymbol{v}^s \cdot \nabla ) \boldsymbol{F} = \nabla \boldsymbol{v}^s \boldsymbol{F} \ \ \ \  \text{on   } \Omega^s,	
\end{equation}
and
\begin{equation}
	\frac{\partial \boldsymbol{B}}{\partial t} + (\boldsymbol{v}^s \cdot \nabla) \boldsymbol{B} = \nabla \boldsymbol{v}^s \boldsymbol{B} + \boldsymbol{B}(\nabla \boldsymbol{v}^s)^T \ \ \ \  \text{on   } \Omega^s.
\end{equation}
The left Cauchy-Green tensor equation is chosen for the evolution of the solid strains for two important reasons: (i) if we solve for $\boldsymbol{\xi}$, we would still need to compute either $\boldsymbol{F}$ or $\boldsymbol{B}$ to evaluate the solid stresses. Since these quantities are not directly available, we will be required to calculate the gradient of $\boldsymbol{\xi}$ and use some interpolation/projection technique to obtain the nodal values at every time step; (ii) the preference of $\boldsymbol{B}$ over $\boldsymbol{F}$ arises from the simple fact that $\boldsymbol{B}$ is a symmetric tensor.
This completes the description of the mathematical formulation of the system. In the next section, we present the weak variational forms of the above equations.

\section{Semi-discrete Variational Eulerian FSI Formulation}


Next, we will turn our attention to the discretization of nonlinear coupled equations. This section presents the semi-discrete variational form using the stabilized finite element formulation. 

\subsection{Temporal Discretization}
We begin with the temporal discretization carried out using the generalized-$\alpha$ method \cite{chung1993time, jansen2000generalized, jaimancomputational} to evolve each of the partitoned systems implicitly in time. It provides a single user-controlled parameter called the spectral radius $\rho_{\infty}$ to dampen the undesirable high frequency oscillations in the solution. The generalized-$\alpha$ method for any generic variable $\varphi$ can be given as
\begin{align}
	\varphi^{n+1}&=\varphi^{n}+\Delta t \partial_t \varphi^n +\Delta t\varsigma (\partial_t \varphi^{n+1}-\partial_t \varphi^n),\\
	\partial_t \varphi^{n+\alpha_m} &= \partial_t \varphi^{n}+\alpha_m (\partial_t \varphi^{n+1}-\partial_t \varphi^n),\\
	\varphi^{n+\alpha}&=\varphi^n+\alpha(\varphi^{n+1}-\varphi^{n}),
\end{align}
where $\Delta t$ is the time step size, $\alpha_m$, $\alpha$ and $\varsigma$ are the generalized-$\alpha$ parameters defined as:
\begin{align}
	\alpha=\frac{1}{1+\rho_{\infty}},\ \alpha_m=\frac{1}{2}\left(\frac{3-\rho_{\infty}}{1+\rho_{\infty}}\right),\ \varsigma=\frac{1}{2}+\alpha_m-\alpha.
\end{align} 
The above method works in a predictor-multicorrector type technique between time steps $n+1$ and $n+\alpha$. For this study, we set $\rho_{\infty}=1$ in the simulations, which essentially recovers to the trapezoidal time integration.

\subsection{Unified Continuum Equations}
We first present the variational form of the unified continuum equations. Let $\mathcal{S}^h$ be the space of trial solutions, whose values satisfy the Dirichlet boundary conditions and $\mathcal{V}^h$ be the space of test functions whose values vanish on the Dirichlet boundary. The variational form for the unified continuum equation can be written as: find $[\boldsymbol{v}_h^{n+\alpha}, p_h^{n+1}] \in \mathcal{S}^h$ such that $\forall [\boldsymbol{\psi}_h, q_h] \in \mathcal{V}^h$,
\begin{equation}
	\begin{aligned}
		&\int_{\Omega} \rho(\phi)\left(\partial_{t} \boldsymbol{v}_{\mathrm{h}}^{\mathrm{n}+\alpha_{\mathrm{m}}}+\boldsymbol{v}_{\mathrm{h}}^{\mathrm{n}+\alpha} \cdot \nabla \boldsymbol{v}_{\mathrm{h}}^{\mathrm{n}+\alpha}\right) \cdot \boldsymbol{\psi}_{\mathrm{h}} \mathrm{d} \Omega+\int_{\Omega} \boldsymbol{\sigma}_{\mathrm{h}}^{\mathrm{n}+\alpha}: \nabla \boldsymbol{\psi}_{\mathrm{h}} \mathrm{d} \Omega +\int_{\Omega} q_{\mathrm{h}}\left(\nabla \cdot \boldsymbol{v}_{\mathrm{h}}^{\mathrm{n}+\alpha}\right) \mathrm{d} \Omega \\
		&+\sum_{\mathrm{e}=1}^{\mathrm{n}_{\mathrm{el}}} \int_{\Omega_{\mathrm{e}}} \frac{\tau_{\mathrm{m}}}{\rho(\phi)}\left(\rho(\phi) \boldsymbol{v}_{\mathrm{h}}^{\mathrm{n}+\alpha} \cdot \nabla \boldsymbol{\psi}_{\mathrm{h}}+\nabla q_{\mathrm{h}}\right) \cdot \boldsymbol{\mathcal{R}}_{\mathrm{m}}(\boldsymbol{v}, p) \mathrm{d} \Omega_{\mathrm{e}} +\sum_{\mathrm{e}=1}^{\mathrm{n}_{\mathrm{el}}} \int_{\Omega_{\mathrm{e}}} \nabla \cdot \boldsymbol{\psi}_{\mathrm{h}} \tau_{\mathrm{c}} \rho(\phi) \mathcal{R}_{\mathrm{c}}(\boldsymbol{v}) \mathrm{d} \Omega_{\mathrm{e}} \\
		&=\int_{\Omega} \boldsymbol{b}\left(t^{\mathrm{n}+\alpha}\right) \cdot \boldsymbol{\psi}_{\mathrm{h}} \mathrm{d} \Omega+\int_{\Gamma_{\mathrm{h}}} \boldsymbol{h} \cdot \boldsymbol{\psi}_{\mathrm{h}} \mathrm{d} \Gamma .
	\end{aligned}
\end{equation}
The first line consists of the Galerkin terms for the combined momentum and continuity equations. The second line contains the Petrov-Galerkin stabilization terms for the continuum equations. $\boldsymbol{\mathcal{R}}_m$ and $\mathcal{R}_c$ are the element-wise residuals for the momentum and continuity equations respectively. The stabilization parameters $\tau_m$ and $\tau_c$ \cite{shakib1991new, franca1992stabilized, johnson2012numerical} are given by
\begin{equation}
	\tau_m = \left[\left(\frac{2}{\Delta t}\right)^2 + \boldsymbol{v}_h \cdot \boldsymbol{G} \boldsymbol{v}_h + C_I \bigg(\frac{\mu(\phi)}{\rho(\phi)}\bigg)^2 \boldsymbol{G}:\boldsymbol{G}\right]^{-1/2},  \text{   } \tau_c = \frac{1}{\mathrm{tr}(\boldsymbol{G})\tau_m} ,
\end{equation}
where $C_I$ is a constant derived from the element-wise inverse estimate and $\boldsymbol{G}$ is the element contravariant metric tensor.

\subsection{Allen-Cahn Equation}
We next present the interface modeling aspect of the formulation.
The semi-discrete Allen-Cahn equation in the convective-diffusive-reactive form can be written as:
\begin{equation}
	G\left(\partial_{t} \phi^{\mathrm{n}+\alpha_{\mathrm{m}}}, \phi^{\mathrm{n}+\alpha}\right)=\partial_{t} \phi^{\mathrm{n}+\alpha_{\mathrm{m}}}+\boldsymbol{v} \cdot \nabla \phi^{\mathrm{n}+\alpha}-\gamma\left(k \nabla^{2} \phi^{\mathrm{n}+\alpha}-s \phi^{\mathrm{n}+\alpha}+f\right)=0 ,
\end{equation}
where 
$$
\begin{array}{l}
\text {Convection velocity }=\boldsymbol{v} ,\\
\text {Diffusion coefficient }=k=\varepsilon^{2} ,\\
\text {Reaction coefficient }=s=\frac{1}{4}\left[\frac{\left(\phi^{\mathrm{n}+\alpha}\right)^{2}}{\alpha^{3}}-\left(\frac{3}{\alpha^{3}}-\frac{4}{\alpha^{2}}\right) \phi^{\mathrm{n}+\alpha} \phi^{\mathrm{n}}+\left(\frac{3}{\alpha^{3}}-\frac{8}{\alpha^{2}}+\frac{6}{\alpha}\right)\left(\phi^{\mathrm{n}}\right)^{2}-\frac{2}{\alpha}\right] \\
-\frac{\beta(t)}{2}\left[\frac{\phi^{\mathrm{n}+\alpha}}{3 \alpha^{2}}+\frac{1}{3}\left(-\frac{2}{\alpha^{2}}+\frac{3}{\alpha}\right) \phi^{\mathrm{n}}\right] ,\\
\text {Source term }=f=-\frac{1}{4}\left[\left(-\frac{1}{\alpha^{3}}+\frac{4}{\alpha^{2}}-\frac{6}{\alpha}+4\right)\left(\phi^{\mathrm{n}}\right)^{3}+\left(\frac{2}{\alpha}-4\right) \phi^{\mathrm{n}}\right] \\
+\frac{\beta(t)}{2}\left[\frac{1}{3}\left(\frac{1}{\alpha^{2}}-\frac{3}{\alpha}+3\right)\left(\phi^{\mathrm{n}}\right)^{2}-1\right] .
\end{array}
$$
The above expressions have been obtained by simplifying terms using the generalized-$\alpha$ time integration \cite{joshi2018positivity}.
Let $\mathcal{S}^h$ be the space of trial solutions, whose values satisfy the Dirichlet boundary conditions and $\mathcal{V}^h$ be the space of test functions whose values vanish on the Dirichlet boundary. The final stabilized form of the Allen-Cahn equation can be stated as: find $\phi_h^{n+\alpha} \in \mathcal{S}^h$ such that $\forall w_h \in \mathcal{V}^h$,
\begin{equation} \label{AC_var}
	\begin{aligned}
		&\int_{\Omega}\left(w_{\mathrm{h}} \partial_{\mathrm{t}} \phi_{\mathrm{h}}^{\mathrm{n+\alpha_m}}+w_{\mathrm{h}}\left(\boldsymbol{v}^{\mathrm{n+\alpha}} \cdot \nabla \phi^{\mathrm{n+\alpha}}_{\mathrm{h}}\right)- \gamma \left(\nabla w_{\mathrm{h}} \cdot\left(k \nabla \phi_{\mathrm{h}}^{\mathrm{n+\alpha}}\right)-w_{\mathrm{h}} s \phi_{\mathrm{h}}^{\mathrm{n+\alpha}}+w_{\mathrm{h}} f\right)\right) \mathrm{d} \Omega + \\
		&\sum_{\mathrm{e}=1}^{\mathrm{n}_{\mathrm{el}}} \int_{\Omega_{e}}\left(\left(\boldsymbol{v}^{\mathrm{n+\alpha}} \cdot \nabla w_{\mathrm{h}}\right) \tau_{\phi}\left(\partial_{\mathrm{t}} \phi_{\mathrm{h}}^{\mathrm{n+\alpha}}+\boldsymbol{v}^{\mathrm{n+\alpha}} \cdot \nabla \phi_{\mathrm{h}}^{\mathrm{n+\alpha}}-\gamma \left(\nabla \cdot\left(k \nabla \phi_{\mathrm{h}}^{\mathrm{n+\alpha}}\right)-s \phi_{\mathrm{h}}^{\mathrm{n+\alpha}}+f\right)\right)\right) \mathrm{d} \Omega_{e}\\
		&+\sum_{e=1}^{n_{\mathrm{el}}} \int_{\Omega_{e}} \chi \frac{\left|\mathcal{R}\left(\phi_{\mathrm{h}}\right)\right|}{\left|\nabla \phi^{\mathrm{n+\alpha}}_{\mathrm{h}}\right|} k_{s}^{\mathrm{add}} \nabla w_{\mathrm{h}} \cdot\left(\frac{\boldsymbol{v}^{\mathrm{n+\alpha}} \otimes \boldsymbol{v}^{\mathrm{n+\alpha}}}{|\boldsymbol{v}^{\mathrm{n+\alpha}}|^{2}}\right) \cdot \nabla \phi_{\mathrm{h}}^{\mathrm{n+\alpha}} \mathrm{d} \Omega_{\mathrm{e}}\\
		&+\sum_{\mathrm{e}=1}^{n_{\mathrm{el}}} \int_{\Omega_{e}} \chi \frac{\left|\mathcal{R}\left(\phi_{\mathrm{h}}\right)\right|}{\left|\nabla \phi^{\mathrm{n+\alpha}}_{\mathrm{h}}\right|} k_{\mathrm{c}}^{\mathrm{add}} \nabla w_{\mathrm{h}} \cdot\left(\mathrm{\boldsymbol{I}}-\frac{\boldsymbol{v}^{\mathrm{n+\alpha}} \otimes \boldsymbol{v}^{\mathrm{n+\alpha}}}{|\boldsymbol{v}^{\mathrm{n+\alpha}}|^{2}}\right) \cdot \nabla \phi_{\mathrm{h}}^{\mathrm{n+\alpha}} \mathrm{d} \Omega_{\mathrm{e}}\\
		&=0 ,
	\end{aligned}
\end{equation}
where $(\otimes)$ represents the dyadic product and $\mathcal{R}\left(\phi_{\mathrm{h}}\right)$ is the element-wise residual of the Allen-Cahn equation.
The first line of Eq. (\ref{AC_var}) contains the Galerkin terms, the second line contains the linear stabilization terms with the stabilization parameter, $\tau_{\phi}$ \cite{shakib1991new, johnson2012numerical} given by
\begin{equation}
	\tau_{\phi} = \left[\left(\frac{2}{\Delta t}\right) + \boldsymbol{v} \cdot \boldsymbol{G}\boldsymbol{v} + 9k^2 \boldsymbol{G}: \boldsymbol{G} + s^2 \right]^{-1/2} .
\end{equation}
The third and fourth line consist of the positivity preserving terms. The expressions for the added diffusions $k_s^{add}$, $k_c^{add}$ and $\chi$ are as follows: 
\begin{align}	
	k_s^\mathrm{add} &= \mathrm{max} \bigg\{ \frac{||\boldsymbol{v}| - \tau|\boldsymbol{v}|s|h}{2} - (k + \tau|\boldsymbol{v}|^2) + \frac{sh^2}{6}, 0 \bigg\},\\
	k_c^\mathrm{add} &= \mathrm{max} \bigg\{ \frac{|\boldsymbol{v}|h}{2} - k + \frac{sh^2}{6}, 0 \bigg\},\\
	\chi &= \frac{2}{|s|h + 2|\boldsymbol{v}|},
\end{align}
where $|\boldsymbol{v}|$ is the magnitude of the convection velocity and $h$ is the characteristic element length. The details of the derivation of these quantities can be found in \cite{joshi2018positivity}.
The positivity condition is enforced to prevent oscillations in $\phi$ at regions of high gradient which might have otherwise led to negative values of density or viscosity. Since upwinding makes the solution first-order accurate, hence the addition of diffusion is restricted to regions of oscillations only. To achieve this criterion, the non-linear positivity preserving terms are designed to depend on the residual of the equation \cite{joshi2018positivity}. The implementation of the PPV method makes the scheme at least second-order accurate in space and the method allows to capture the high gradient internal and boundary layers in multi-dimensions.
The $\chi \frac{\left|\mathcal{R}\left(\phi_{\mathrm{h}}\right)\right|}{\left|\nabla \phi_{\mathrm{h}}\right|}$ factor adds a nonlinear property and provides the functionality of a limiter for the upwinding near oscillating regions.

\subsection{Left Cauchy-Green Tensor Transport Equation}
Finally, we proceed to the variational form of the $\boldsymbol{B}$ tensor equation to evaluate the solid stresses accurately. Let $\mathcal{S}^h$ be the space of trial solutions, whose values satisfy the Dirichlet boundary conditions and $\mathcal{V}^h$ be the space of test functions whose values vanish on the Dirichlet boundary. The variational form of the left Cauchy-Green tensor equation can be written as: find $\boldsymbol{B}_h^{n+\alpha} \in \mathcal{S}^h$ such that $\forall \boldsymbol{m}_h \in \mathcal{V}^h$,
\begin{equation}
	\begin{aligned}
		&\int_{\Omega} (\boldsymbol{m}_h):\bigg( \partial_t \boldsymbol{B}_h^{\mathrm{n+\alpha_m}} + \bigg( \frac{1+\phi}{2} \bigg) (\boldsymbol{v}^{s,{\mathrm{n+\alpha}}} \cdot \nabla) \boldsymbol{B}_h^{\mathrm{n+\alpha}} \bigg) \mathrm{d} \Omega - \\
		&\int_{\Omega} \bigg( \bigg( \frac{1+\phi}{2} \bigg) \nabla \boldsymbol{v}^{s,{\mathrm{n+\alpha}}} \boldsymbol{B}_h^{\mathrm{n+\alpha}} + \bigg( \frac{1+\phi}{2} \bigg) \boldsymbol{B}_h^{\mathrm{n+\alpha}} (\nabla \boldsymbol{v}^{s,{\mathrm{n+\alpha}}} )^T \bigg) : (\boldsymbol{m}_h) \mathrm{d}\Omega = 0 ,
	\end{aligned}
\end{equation}
where $(:)$ represents double dot product. We obtain updated values of $\boldsymbol{B}$ by solving the above variational equation in the solid domain which can then be used to calculate the new solid stress values before substituting in the unified continuum equations in the next time step. 
Next, we turn our attention to the linearized form of the equations.

\subsection{Linarized System of Equations}
This section presents the linearized system of equations (i.e., matrix form) for the numerical implementation of the proposed fully Eulerian FSI framework. The velocity, pressure and order parameter increments are evaluated in each time step using Newton-Raphson type iterations. The linearized system for the unified continuum equations can be expressed by:
\begin{align} \label{LS_UCeq}
	\begin{bmatrix}
		\boldsymbol{K}_\Omega &  & \boldsymbol{G}_\Omega \\
		& \\
		-\boldsymbol{G}^T_\Omega &  &\boldsymbol{C}_\Omega
	\end{bmatrix} 
	\begin{Bmatrix}
		\Delta \boldsymbol{v}^\mathrm{n+\alpha} \\
		\\
		\Delta p^\mathrm{n+1}
	\end{Bmatrix}
	= \begin{Bmatrix} 
		-\widetilde{\boldsymbol{\mathcal{R}}}_\mathrm{m}(\boldsymbol{v},p) \\
		\\
		-\widetilde{\mathcal{R}}_\mathrm{c}(\boldsymbol{v})
	\end{Bmatrix} ,
\end{align}
where $\boldsymbol{K}_\Omega$ is the stiffness matrix of the momentum equation consisting of inertia, convection, diffusion and stabilization terms, $\boldsymbol{G}_\Omega$ is the discrete gradient operator, $\boldsymbol{G}^T_\Omega$ is the divergence operator and $\boldsymbol{C}_\Omega$ is the pressure-pressure stabilization term. Here, $\Delta \boldsymbol{v}$ and $\Delta p$ are the increments in velocity and pressure respectively and $\widetilde{\boldsymbol{\mathcal{R}}}_\mathrm{m}(\boldsymbol{v},p)$ and $\widetilde{\mathcal{R}}_\mathrm{c}(\boldsymbol{v})$ represent the weighted residuals of the stabilized momentum and continuity equations respectively. Let the updated quantities at $t^\mathrm{n+1}_\mathrm{(k)}$ be represented as $\boldsymbol{X}^\mathrm{n+1}_\mathrm{(k)}$, $\mathrm{k}$ being the non-linear iteration index and the increments in these quantities be represented as $\Delta\boldsymbol{X}$. The error in solving the unified continuum equations can be defined as
\begin{align} \label{e_UC}
	e_{UC} = \frac{||\Delta \boldsymbol{X}||}{||\boldsymbol{X}^\mathrm{n+1}_{\mathrm{(k)}}||}.
\end{align}  
Similarly, the linearized form for the Allen-Cahn equation can be expressed as:
\begin{align} \label{LS_AC}
	\begin{bmatrix}
		\boldsymbol{K}_{AC}
	\end{bmatrix} 
	\begin{Bmatrix}
		\Delta \phi^\mathrm{n+\alpha}
	\end{Bmatrix}
	= \begin{Bmatrix} 
		-\widetilde{\mathcal{R}}(\phi)
	\end{Bmatrix} ,
\end{align}
where $\boldsymbol{K}_{AC}$ consists of the inertia, convection, diffusion, reaction and stabilization terms and $\widetilde{\mathcal{R}}(\phi)$ represents the weighted residual for the stabilized conservative Allen-Cahn equation.  The numerical error in solving the Allen-Cahn equation can be written as
\begin{align} \label{e_AC}
	e_{AC} = \frac{||\Delta \phi^\mathrm{n+\alpha}||}{||\phi^\mathrm{n+1}_{\mathrm{(k)}}||}.
\end{align} 
Finally, the linearized form for the transport of the left Cauchy-Green tensor can be written as:
\begin{align} \label{LS_CGT}
	\begin{bmatrix}
		\boldsymbol{K}_{CGT}
	\end{bmatrix} 
	\begin{Bmatrix}
		\Delta \boldsymbol{B}^\mathrm{n+\alpha}
	\end{Bmatrix}
	= \begin{Bmatrix} 
		-\widetilde{\mathcal{R}}(\boldsymbol{B})
	\end{Bmatrix} ,
\end{align}
where $\boldsymbol{K}_{CGT}$ is the stiffness matrix and $\widetilde{\mathcal{R}}(\boldsymbol{B})$ represents the weighted residual for the left Cauchy Green tensor equation. Similar to the error evaluation for the order parameter, the numerical error in solving the above equation can be written as
\begin{align} \label{e_CGT}
	e_{CGT} = \frac{||\Delta \boldsymbol{B}^\mathrm{n+\alpha}||}{||\boldsymbol{B}^\mathrm{n+1}_{\mathrm{(k)}}||}.
\end{align} 
Using the aforementioned framework of the fully Eulerian FSI, we proceed to the central topic of this paper i.e.,  the adaptive mesh refinement procedure.

\section{Adaptive Variational Formulation}




In this section, we discuss the adaptivity procedure for the coupled solver based on the unified momentum and the Allen-Cahn equations.
Using the mesh adaptivity procedure, we aim to reduce the nonlinear errors and improve the conservation properties.
Since the fluid-solid interface is central to the coupled dynamic, it is natural to consider an interface-driven adaptivity so that the non-linear convergence can be improved while reducing the conservation errors and the computational effort.  The Galerkin variational form of the momentum and the Allen–Cahn equations  are essentially the differential operator multiplied by the weighting function and integrated by parts as appropriate, which tends to minimize the residual of the equations in a chosen set of weighting functions. 
While the variational error tends to be small convergence when the solution is smooth, the residual error can be large when the solution exhibits oscillations near sharp gradients owing to dominant convection and reaction effects on an under-resolved mesh. To capture the sharp gradients, one needs to consider the refinement of the under-resolved mesh and the residual of the differential equation can serve as a useful indicator for the refinement strategy.
In this work, we therefore employ a residual-based error indicator for the phase-field equation to minimize oscillations that occur 
due to convection and reaction effects on an under-resolved mesh. 

Based on discussions in Joshi and Jaiman \cite{joshi2018adaptive}, the adaptive error indicator can be derived from the discretized Allen-Cahn equation. Consider the domain $\Omega$ which consists of elements $\Omega_\mathrm{e}$, chosen such that $\Omega = \cup_\mathrm{e=1}^\mathrm{n_{el}} \Omega_\mathrm{e}$ and $\emptyset = \cap_\mathrm{e=1}^\mathrm{n_{el}} \Omega_\mathrm{e}$ where $\mathrm{n}_\mathrm{el}$ is the number of elements. Let $\Gamma$ be the Lipschitz continuous boundary of the domain $\Omega$, $\Gamma_D$ and $\Gamma_N$ be the Dirichlet and Neumann boundaries of $\Omega$ respectively such that $\Gamma = \Gamma_D \cup \Gamma_N$. Furthermore, let $\mathcal{E}$ denote the set of edges for all the elements in the domain, $\mathcal{E}_\Gamma$, $\mathcal{E}_{\Gamma_D}$ and $\mathcal{E}_{\Gamma_N}$ be the set of edges on the boundary $\Gamma$, $\Gamma_D$ and $\Gamma_N$ respectively. Consider $\mathcal{E}_{\Omega_\mathrm{e}}$ to be the set of edges of an element $\Omega_\mathrm{e}$ and $\mathcal{E}_\Omega$ be the set of all the interior edges of $\Omega$. The Galerkin terms of the variational Allen-Cahn equation can be written after integration by parts as
\begin{align}
	&\int_\Omega w_\mathrm{h}\partial_t\phi_\mathrm{h} \mathrm{d}\Omega + \int_\Omega w_\mathrm{h}(\boldsymbol{v}\cdot\nabla\phi_\mathrm{h}) \mathrm{d}\Omega -\int_{\Omega} \nabla w_\mathrm{h}\cdot(k \nabla\phi_\mathrm{h}) \mathrm{d}\Omega + \int_\Omega w_\mathrm{h}s\phi_\mathrm{h} \mathrm{d}\Omega - \int_\Omega w_\mathrm{h}f\mathrm{d}\Omega \nonumber \\
	=& \displaystyle\sum_\mathrm{e=1}^\mathrm{n_{el}} \int_{\Omega_\mathrm{e}} w_\mathrm{h} \mathcal{R}_{\Omega_\mathrm{e}}(\phi_\mathrm{h}) \mathrm{d}\Omega_\mathrm{e} + \displaystyle\sum_{E\in \mathcal{E}} \int_{E} w_\mathrm{h} \mathcal{R}_{E}(\phi_\mathrm{h}) \mathrm{d}E ,
\end{align}
where $\mathcal{R}_{\Omega_\mathrm{e}}$ and $\mathcal{R}_{E}$ are the element and edge based residuals given as, 
\begin{align}
	\mathcal{R}_{\Omega_\mathrm{e}} &= \partial_t\phi_\mathrm{h} + \boldsymbol{v}\cdot\nabla\phi_\mathrm{h} - k\nabla^2\phi_\mathrm{h} + s\phi_\mathrm{h} - f,\\
	\mathcal{R}_{E} &= \begin{cases}
		-\mathbb{J}_{E}(\mathbf{n}_{E}\cdot k\nabla\phi_\mathrm{h})\ \ \ \ \ &\mathrm{if}\ E\in \mathcal{E}_\Omega ,\\
		-\mathbf{n}_{E}\cdot k\nabla\phi_\mathrm{h}\ \ \ \ \ &\mathrm{if}\ E\in \mathcal{E}_{\Gamma_N} ,\\
		0\ \ &\mathrm{if}\ E\in \mathcal{E}_{\Gamma_D} ,
	\end{cases}
\end{align}
where $\mathbb{J}_E(\varphi)$ is the jump of the quantity $\varphi$ across the element edge $E$ and $\mathbf{n}_E$ is the normal to the edge $E$.
After a few algebraic manipulations and using the Cauchy-Schwarz inequality \cite{joshi2018adaptive}, we can define the error estimate between the true $(\phi)$ and approximate $(\phi_h)$ solutions using the following relation:
\begin{equation}\label{eta}
	||\phi - \phi_h|| \leq \eta = \left ( \sum_{e=1}^{n_{el}} \eta_{\Omega_e}^2 \right )^{1/2} ,
\end{equation}
where the error indicator $\eta_{\Omega_e}$ for a single triangular element $\Omega_e$ is given by
\begin{equation}\label{eta_omega_e}
	\eta_{\Omega_e}^2 = h_{\Omega_e}^2 ||R_{\Omega_e}||_{\Omega_e}^2 + \sum_{E \in \varepsilon_{\Omega_e} \cap \varepsilon} h_E^2 ||R_E||_E^2 .
\end{equation}
In the above equations, $\eta$ is the adaptive error indicator for the entire domain, $h_{\Omega_\mathrm{e}}$ is the characteristic length of the element $\Omega_\mathrm{e}$ and $h_E$ is the length of the edge $E$. In actual numerical implementation, we evaluate the volume residuals by computing the elementwise integral mean of the residual $(f_{\Omega_e})$. This is done by evaluating the residual at the quadrature points and summing it over all the points for each element. Similarly, the characteristic length of the element is evaluated as: $h_{\Omega_e}=|\Omega_e|^{1/d}$, where $d$ is the dimension of the problem (2-D or 3-D). The edge-based residuals are computed by summing the jump conditions across the edges and the Neumann boundary conditions. However, since the Neumann boundary condition for the order parameter is zero, hence their contribution can be neglected. Also, since the Dirichlet conditions are exact values, they don't contribute towards the error estimation.  So we finally arrive at
\begin{equation}
	\tilde{\eta}_{\Omega_e}^2 = h_{\Omega_e}^2 (f_{\Omega_e})^2 + \sum_{E \in \varepsilon_{\Omega_e} \cap \varepsilon} h_E^2 (\mathbb{J}_{E}(\mathbf{n}_{E}\cdot k\nabla\phi_\mathrm{h}))^2 .
\end{equation}
The equivalence between $\tilde{\eta}_{\Omega_e}$ and $\eta_{\Omega_e}$ has been illustrated in \cite{funken2011efficient}.


After obtaining the error estimates from the Allen-Cahn equation, the newest vertex bisection algorithm \cite{chen2006afem} is employed to carry out the adaptivity procedure. The algorithm is based on appropriately marking the elements and edges that need to be coarsened or refined and then carrying out the required procedure. The D{\"o}rfler criterion \cite{dorfler1996convergent} is used to mark the elements for refinement which finds the minimum set of elements $\Omega^M$ such that
\begin{equation}\label{dorfler}
	\theta \sum_{\Omega_e \in \Omega} \eta_{\Omega_e}^2 \leq \sum_{\Omega_e \in \Omega^M} \eta_{\Omega_e}^2 ,
\end{equation}
where $\theta$ is a user defined parameter in the range $(0,1)$. This criterion ensures that the elements which have a larger contribution to the error estimate are chosen for refinement. It is worthy to note that $\theta \to 0$ refers to the case of highly localized/adaptive refinement, whereas $\theta \to 1$ indicates the case of uniform refinement in the entire domain. During the numerical implementation of this criterion, we first sort the elements in descending order of their corresponding error indicators and then pick the set of elements $\Omega^M$ that satisfy the criterion (Eq. \ref{dorfler}).

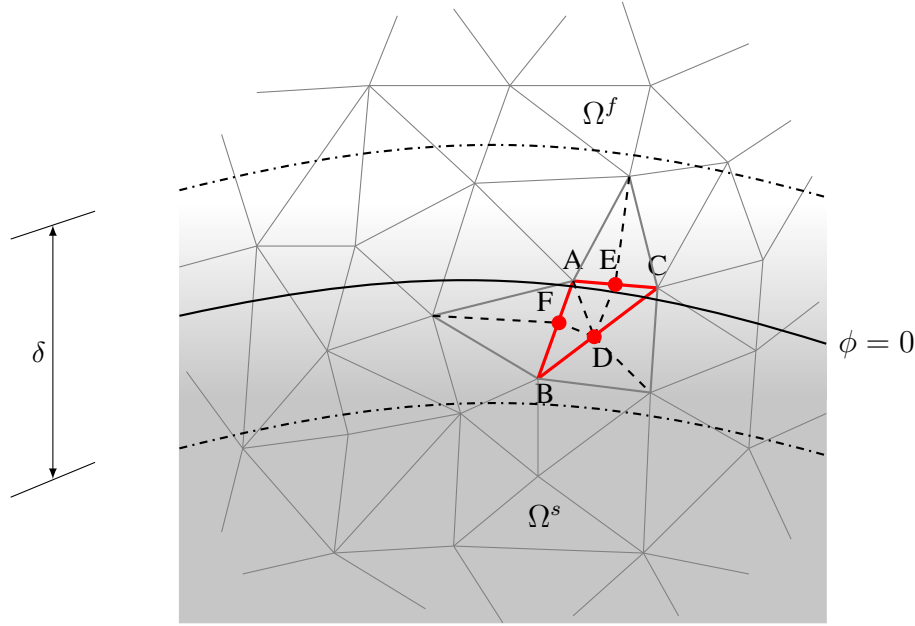
\begin{figure}[h]
	\centering
	\begin{tikzpicture}[scale=0.37]
	\begin{scope}[on background layer]
	\shade[top color=white,bottom color=gray!45] (-12,-3) rectangle (11,4);
	\draw[fill=gray!45,draw=gray!45] (-12,-11) rectangle (11,-3);
	\end{scope}	
	\begin{pgfonlayer}{nodelayer}
		\node [style=none] (0) at (-12, 0) {};
		\node [style=none, label={right:\large $\phi=0$}] (1) at (11, -1) {};
		\node [style=none] (3) at (-12, 4.5) {};
		\node [style=none] (4) at (-12, -4.75) {};
		\node [style=none] (5) at (11, -5) {};
		\node [style=none] (6) at (-2, -3.5) {};
		\node [style=none] (7) at (-3, 0) {};
		\node [style=none] (8) at (0.75, -2.25) {};
		\node [style=none, label={above:A}] (9) at (2, 1.25) {};
		\node [style=none] (10) at (5, 1) {};
		\node [style=none] (11) at (4, 5) {};
		\node [style=none] (12) at (-1.5, 4.75) {};
		\node [style=none] (13) at (4.75, -2.75) {};
		\node [style=none] (14) at (0.75, -5.75) {};
		\node [style=none] (15) at (8.5, -1.25) {};
		\node [style=none] (18) at (-5.75, 2.5) {};
		\node [style=none] (20) at (4.5, -8.5) {};
		\node [style=none] (22) at (-2.25, -8.25) {};
		\node [style=none] (23) at (-0.25, 8.25) {};
		\node [style=none] (24) at (8.75, 2) {};
		\node [style=none] (25) at (8.25, -4.75) {};
		\node [style=none] (26) at (-6.75, -1.25) {};
		\node [style=none] (27) at (4.75, 8.25) {};
		\node [style=none] (28) at (2.75, -0.75) {};
		\node [style=none] (29) at (1.5, -0.25) {};
		\node [style=none] (30) at (3.5, 1.125) {};
		\node [style=none, label={below:$\delta$}] (31) at (-17, -0.5) {};
		\node [style=none, label={above:C}] (32) at (5, 1) {};
		\node [style=none, label={right:D}] (33) at (2.25, -1.5) {};
		\node [style=none, label={below:B}] (34) at (1, -2) {};
		\node [style=none, label={above:F}] (35) at (1, -0.25) {};
		\node [style=none, label={above:E}] (36) at (3.25, 1.25) {};
		\node [style=none] (38) at (7.5, 5.5) {};
		\node [style=none] (39) at (-5.25, 8.25) {};
		\node [style=none] (40) at (-6, -4.25) {};
		\node [style=none] (41) at (-9.75, -4.75) {};
		\node [style=none] (42) at (-6.5, -8.5) {};
		\node [style=none] (43) at (-9.25, 2.5) {};
		\node [style=none, label={above:\large $\Omega^s$}] (44) at (1, -8) {};
		\node [style=none, label={above:\large $\Omega^f$}] (45) at (3, 6.45) {};
		\node [style=none] (46) at (11, 4.25) {};
		\node [style=none] (47) at (-16.5, 3.25) {};
		\node [style=none] (49) at (10.75, 0.5) {};
		\node [style=none] (50) at (10.5, 5) {};
		\node [style=none] (51) at (9.75, 7) {};
		\node [style=none] (52) at (8.25, 9.75) {};
		\node [style=none] (53) at (3.75, 11.25) {};
		\node [style=none] (54) at (1.75, 10.75) {};
		\node [style=none] (55) at (-2.25, 11) {};
		\node [style=none] (56) at (-4.25, 10.75) {};
		\node [style=none] (57) at (-9.25, 8) {};
		\node [style=none] (58) at (-10.5, 6.5) {};
		\node [style=none] (59) at (-12, 1.75) {};
		\node [style=none] (60) at (-11.75, -2) {};
		\node [style=none] (61) at (-10.5, -7.75) {};
		\node [style=none] (62) at (-3.25, -10.5) {};
		\node [style=none] (63) at (-0.5, -11) {};
		\node [style=none] (64) at (2, -10.75) {};
		\node [style=none] (65) at (5.75, -10.75) {};
		\node [style=none] (66) at (9.75, -8) {};
		\node [style=none] (67) at (11, -3.5) {};
		\node [style=none] (69) at (-16.5, -5.85) {};
		\node [style=none] (70) at (-18, 2.75) {};
		\node [style=none] (71) at (-18, -6.5) {};
		\node [style=none] (72) at (-15, 3.75) {};
		\node [style=none] (73) at (-15, -5.25) {};
		\node [style=none] (74) at (11, -11) {};
		\node [style=none] (75) at (11, 11) {};
		\node [style=none] (76) at (-12, -11) {};
		\node [style=none] (77) at (-12, 11) {};
		\node [style=none] (78) at (-9.75, -9.25) {};
		\node [style=none] (79) at (8.25, -9.25) {};
	\end{pgfonlayer}
	\begin{pgfonlayer}{edgelayer}
		\draw [style={grey_line}] (6.center) to (7.center);
		\draw [thick, style={grey_line}] (7.center) to (8.center);
		\draw [style={grey_line}] (6.center) to (8.center);
		\draw [very thick, style={red_line}] (8.center) to (10.center);
		\draw [very thick, style={red_line}] (8.center) to (9.center);
		\draw [very thick, style={red_line}] (9.center) to (10.center);
		\draw [thick, style={grey_line}] (10.center) to (11.center);
		\draw [thick, style={grey_line}] (9.center) to (11.center);
		\draw [thick, style={grey_line}] (7.center) to (9.center);
		\draw [style={grey_line}] (7.center) to (12.center);
		\draw [style={grey_line}] (12.center) to (9.center);
		\draw [style={grey_line}] (12.center) to (11.center);
		\draw [style={grey_line}] (10.center) to (15.center);
		\draw [thick, style={grey_line}] (10.center) to (13.center);
		\draw [thick, style={grey_line}] (8.center) to (13.center);
		\draw [style={grey_line}] (8.center) to (14.center);
		\draw [style={grey_line}] (6.center) to (14.center);
		\draw [style={grey_line}] (14.center) to (13.center);
		\draw [style={grey_line}] (13.center) to (15.center);
		\draw [style={grey_line}] (12.center) to (18.center);
		\draw [style={grey_line}] (12.center) to (23.center);
		\draw [style={grey_line}] (13.center) to (20.center);
		\draw [style={grey_line}] (14.center) to (22.center);
		\draw [style={grey_line}] (7.center) to (26.center);
		\draw [style={grey_line}] (13.center) to (25.center);
		\draw [style={grey_line}] (15.center) to (24.center);
		\draw [style={grey_line}] (11.center) to (27.center);
		\draw [thick, dashed] (28.center) to (30.center);
		\draw [thick, dashed] (28.center) to (29.center);
		\draw [thick, dashed] (11.center) to (30.center);
		\draw [thick, dashed] (7.center) to (29.center);
		\draw [thick, dashed] (28.center) to (13.center);
		\draw [style={grey_line}] (14.center) to (20.center);
		\draw [style={grey_line}] (15.center) to (25.center);
		\draw [style={grey_line}] (10.center) to (24.center);
		\draw [style={grey_line}] (11.center) to (38.center);
		\draw [style={grey_line}] (38.center) to (10.center);
		\draw [style={grey_line}] (38.center) to (24.center);
		\draw [style={grey_line}] (7.center) to (18.center);
		\draw [style={grey_line}] (18.center) to (26.center);
		\draw [style={grey_line}] (26.center) to (6.center);
		\draw [style={grey_line}] (6.center) to (22.center);
		\draw [style={grey_line}] (39.center) to (23.center);
		\draw [style={grey_line}] (39.center) to (18.center);
		\draw [style={grey_line}] (39.center) to (12.center);
		\draw [style={grey_line}] (23.center) to (11.center);
		\draw [style={grey_line}] (23.center) to (27.center);
		\draw [style={grey_line}] (27.center) to (38.center);
		\draw [style={grey_line}] (26.center) to (40.center);
		\draw [style={grey_line}] (40.center) to (6.center);
		\draw [style={grey_line}] (40.center) to (41.center);
		\draw [style={grey_line}] (41.center) to (26.center);
		\draw [style={grey_line}] (41.center) to (42.center);
		\draw [style={grey_line}] (40.center) to (42.center);
		\draw [style={grey_line}] (42.center) to (6.center);
		\draw [style={grey_line}] (42.center) to (22.center);
		\draw [style={grey_line}] (43.center) to (18.center);
		\draw [style={grey_line}] (43.center) to (26.center);
		\draw [style={grey_line}] (25.center) to (66.center);
		\draw [style={grey_line}] (15.center) to (49.center);
		\draw [style={grey_line}] (24.center) to (50.center);
		\draw [style={grey_line}] (38.center) to (51.center);
		\draw [style={grey_line}] (27.center) to (52.center);
		\draw [style={grey_line}] (27.center) to (53.center);
		\draw [style={grey_line}] (23.center) to (54.center);
		\draw [style={grey_line}] (55.center) to (23.center);
		\draw [style={grey_line}] (39.center) to (56.center);
		\draw [style={grey_line}] (39.center) to (57.center);
		\draw [style={grey_line}] (43.center) to (58.center);
		\draw [style={grey_line}] (43.center) to (59.center);
		\draw [style={grey_line}] (41.center) to (60.center);
		\draw [style={grey_line}] (41.center) to (61.center);
		\draw [style={grey_line}] (42.center) to (62.center);
		\draw [style={grey_line}] (22.center) to (63.center);
		\draw [style={grey_line}] (20.center) to (64.center);
		\draw [style={grey_line}] (20.center) to (65.center);
		\draw [style={grey_line}] (25.center) to (67.center);
		\draw [style={grey_line}] (43.center) to (39.center);
		\draw [style={grey_line}] (43.center) to (41.center);
		\draw [style={grey_line}] (22.center) to (20.center);
		\draw [style={grey_line}] (20.center) to (25.center);
		\draw [style=latex] (47.center) to (69.center);
		\draw (72.center) to (70.center);
		\draw (73.center) to (71.center);
		\draw [thick, style={black_dashed}] (9.center) to (28.center);
		\draw [thick, bend left=15] (0.center) to (1.center);
		\draw [thick, dash dot, bend left=15] (3.center) to (46.center);
		\draw [thick, dash dot, bend left=15] (4.center) to (5.center);
		\draw [style={grey_line}] (42.center) to (78.center);
		\draw [style={grey_line}] (20.center) to (79.center);
	\end{pgfonlayer}
	\filldraw[red] (28) circle (7pt);
	\filldraw[red] (29) circle (7pt);
	\filldraw[red] (30) circle (7pt);
\end{tikzpicture}

	\caption{Application of the newest vertex bisection method for refinement of a portion of the representative unstructured grid along the diffused interface: $\Delta$ ABC has been marked for refinement by the Dorfler criterion. Nodes D, E and F are mid-points of the edges to which bisectors are drawn to create additional elements, starting with node D since BC is the longest edge. In the above figure, $\delta \approx 4\varepsilon$ at equilibrium (where $\varepsilon$ is the interface thickness parameter).}	
	\label{fig:bisection}
\end{figure}

The newest vertex bisection algorithm is used for refining the triangular elements of the unstructured mesh overlaying the diffused interface as shown in Fig. \ref{fig:bisection}. The longest edge (BC) of the element ABC is chosen first and bisected by placing a node (D) at the mid-point. Subsequently, the remaining edges are bisected by placing nodes E and F at their mid-points respectively, following the corresponding edge being opposite to the newest vertex created in the respective elements. Once the particular element is refined, the hanging nodes are avoided by introducing additional elements adjoining the triangle. The same procedure is then repeated for all the marked elements. To evaluate the physical quantities on the new nodes, the value is simply linearly interpolated from the adjoining nodes on the corresponding edge. An additional array is created in the process which maintains the identities of the nodes added by bisection and those nodes that belong to the initial grid. This step considerably simplifies the numerical implementation and helps in the coarsening step as well. Once all the marked elements are refined, the governing equations are solved on the refined mesh until convergence.

Using the coarsening algorithm, elements with very small residual errors can be coarsened to reduce the computational effort. The nodes added later during the bisection process are eligible for removal during the coarsening procedure, while initial nodes cannot be coarsened \cite{chen2010coarsening}. Similar to the refinement criterion, a D{\"o}rfler criterion is employed for the coarsening with $\theta_c$ as the user-defined coarsening parameter. We arrange the error indicators in their increasing order and identify the minimum set of elements that satisfy the criterion for coarsening.  We cannot coarsen below the initial number of elements that we start with and the coarsening step is only carried out once before moving to the next time step, unlike the refinement step which is carried out multiple times to achieve convergence of the mesh. To ensure that the non-linearities are captured properly,   the coarsening/refinement steps are not carried out in the intermediate iterations.


\subsection{Algorithm}
The proposed algorithm for carrying out adaptive procedure for a phase-field based fully Eulerian FSI solver has been presented in Algorithm \ref{algorithm_1}. 
At a given time step, we start with all the primitive variables: velocities $\boldsymbol{v}^\mathrm{n}$, pressure $p^\mathrm{n}$, order parameter $\phi^\mathrm{n}$ and left Cauchy-Green tensor $\boldsymbol{B}^{\mathrm{n}}$ at the nodal points on grid $\mathcal{T}^\mathrm{n}$. Before entering the non-linear iteration loop, we predict the solutions for the next time step: $\boldsymbol{v}^{\mathrm{n}+1}_{(0)}$, $p^{\mathrm{n}+1}_{(0)}$, $\phi^{\mathrm{n}+1}_\mathrm{(0)}$ and $\boldsymbol{B}^{\mathrm{n}+1}_{(0)}$. Inside the non-linear iteration loop, we first solve the unified continuum equations to obtain the velocities and pressures in the entire domain as well as evaluate the corresponding error $e_{UC}$. The updated velocities are then used for evolving the solid strain by solving the transport equation for $\boldsymbol{B}$. The corresponding error $e_{CGT}$ is evaluated at this step. Finally, we solve the Allen-Cahn equation to evolve the fluid-solid interface in space and time and estimate the error $e_{AC}$. 

Next, the error estimator for the adaptive procedure is computed using Eq. (\ref{eta}) and Eq. (\ref{eta_omega_e}). The succeeding steps are used to decide if the present triangulation (mesh) needs refinement/coarsening based on the user-defined tolerances for the convergence criteria. $nIterMax$ represents the maximum number of nonlinear iterations and $nElemMax$ represents the maximum number of elements allowed in the domain. The boundary conditions are satisfied at the end of each loop (line 27) and the updated solution values are copied to the previous time step variables upon exiting the non-linear loop (line 31).

\begin{algorithm}
	\caption{Adaptive procedure for a fully Eulerian FSI solver}
	\label{algorithm_1}
	\begin{algorithmic}[1]
		\STATE Given $\boldsymbol{v}^0$, $p^0$, $\phi^0$ on $\mathcal{T}^0$ \\
		
		\STATE Initial refinement step  
		\STATE Loop over time steps $\mathrm{n}=0,1,\cdots$ \\
		\STATE \quad Start from known variables $\boldsymbol{v}^\mathrm{n}$, $p^\mathrm{n}$, $\phi^\mathrm{n}$, $\boldsymbol{B}^\mathrm{n}$  on $\mathcal{T}^\mathrm{n}$\\
		\STATE \quad Predict the solution on $\mathcal{T}^\mathrm{n+1}_\mathrm{(0)} = \mathcal{T}^\mathrm{n}$: $\boldsymbol{v}^\mathrm{n+1}_{(0)}=\boldsymbol{v}^\mathrm{n}$, $p^\mathrm{n+1}_{(0)} = p^\mathrm{n}$, $\phi^\mathrm{n+1}_{(0)} = \phi^\mathrm{n}$ and $\boldsymbol{B}^\mathrm{n+1}_{(0)}=\boldsymbol{B}^\mathrm{n}$ \\
		\STATE \quad Loop over nonlinear iterations $\mathrm{k}=0,1,\cdots ,nIterMax$ until convergence \\
		\STATE \qquad\quad Solve Eq. (\ref{LS_UCeq}) for $\boldsymbol{v}^\mathrm{n+1}_\mathrm{(k)}$ and $p^\mathrm{n+1}_\mathrm{(k)}$ on $\mathcal{T}^\mathrm{n+1}_\mathrm{(k)}$ and evaluate $e_{UC}$ via Eq. (\ref{e_UC}) \\
		\STATE \qquad\quad Solve Eq. (\ref{LS_CGT}) for $\boldsymbol{B}^\mathrm{n+1}_\mathrm{(k)}$ on $\mathcal{T}^\mathrm{n+1}_\mathrm{(k)}$ and evaluate $e_{CGT}$ via Eq. (\ref{e_CGT})\\
		\STATE \qquad\quad Solve Eq. (\ref{LS_AC}) for $\phi^\mathrm{n+1}_\mathrm{(k)}$ on $\mathcal{T}^\mathrm{n+1}_\mathrm{(k)}$ and evaluate $e_{AC}$ via Eq. (\ref{e_AC})\\
		\STATE \qquad\quad Evaluate error estimator $\eta$\\
		\STATE \qquad\quad \textbf{if}\ ($e_{UC} \leq tol_{UC}$) \textbf{\&} ($e_{CGT} \leq tol_{CGT}$) \textbf{\&} ($e_{AC} \leq tol_{AC}$) \textbf{\&} ($\eta \leq tol_{R}$) \textbf{then}\\
		\STATE \qquad\qquad refine = 0; coarsen = 1; break = 1\\
		\STATE \qquad\quad \textbf{elseif}\ ($e_{UC} > tol_{UC}$) \textbf{\&} ($e_{CGT} > tol_{CGT}$) \textbf{\&} ($e_{AC} > tol_{AC}$) \textbf{\&} ($\eta \leq tol_{R}$) \textbf{then}\\
		\STATE \qquad\qquad refine = 0; coarsen = 0\\
		\STATE \qquad\quad \textbf{elseif}\ ($\eta > tol_{R}$) \textbf{then}\\
		\STATE \qquad\qquad refine = 1; coarsen = 0\\
		\STATE \qquad\quad \textbf{elseif}\ ($\mathrm{k} = nIterMax$) \textbf{then}\\
		\STATE \qquad\qquad refine = 0; coarsen = 1; break = 1\\
		\STATE \qquad\quad \textbf{endif}\\
		\STATE \qquad\quad \textbf{if}\ ($\mathrm{n}_\mathrm{el} > nElemMax$) \textbf{\&} ( $\mathrm{k}=1$ \textbf{or} $\mathrm{k}=nIterMax$ ) \textbf{then}\\
		\STATE \qquad\qquad refine = 0; coarsen = 1\\
		\STATE \qquad\quad \textbf{elseif}\ ($\mathrm{n}_\mathrm{el} > nElemMax$) \textbf{\&} ($\mathrm{k}\neq 1$) \textbf{\&} (break $\neq$ 1) \textbf{then}\\
		\STATE \qquad\qquad refine = 0; coarsen = 0\\
		\STATE \qquad\quad \textbf{endif}\\
		\STATE \qquad\quad \textbf{if}\ (coarsen = 1) \textbf{then}, coarsen the mesh\\
		\STATE \qquad\quad \textbf{if}\ (refine = 1) \textbf{then}, refine the mesh\\
		\STATE \qquad\quad Satisfy boundary conditions on $\mathcal{T}^\mathrm{n+1}_\mathrm{(k+1)}$\\
		\STATE \qquad\quad \textbf{if}\ (break = 1) \textbf{then}\\
		\STATE \qquad\qquad exit nonlinear interation loop\\
		\STATE \qquad\quad \textbf{endif}\\
		\STATE \quad Update the solution of current time step to previous time step on $\mathcal{T}^\mathrm{n+1}_\mathrm{(k+1)}$ 	
	\end{algorithmic}
\end{algorithm}

\subsection{Implementation Details}
The above algorithm is implemented on an unstructured mesh imported from Gmsh \cite{geuzaine2009gmsh}. 
The mesh file is first pre-processed to separate the nodal data, connectivity data, and boundary data. This step is followed by the initialization of velocities, pressure, order parameter and deformation tensors in the domain and the application of boundary conditions. The present triangulation is then passed through an iterative initial refinement step before entering the time loop. To avoid high gradients for velocity and stress field, we start with a sufficiently refined mesh along the interface. 
The system then enters the time loop where the nonlinear set of equations are solved in a partitioned manner \cite{joshi2018positivity}. Implicit discretizations are carried out for the underlying equations for improved efficiency and robustness of the solver. The incremental quantities are computed by employing Newton Raphson type iterations and the generalized-$\alpha$ time integration is used to update the solver. The Jacobians are calculated and the matrices are assembled for each of the equations separately and then solved sequentially as outlined in the previous sub-section. The updated velocities from the solution of the unified continuum equations are used for evolving the solid strains via the $\boldsymbol{B}$ transport equation and updating the order parameter values through the Allen-Cahn equation. Solid stresses are computed from the new deformation tensor $(\boldsymbol{B})$ values. The updated order parameter values are further utilized in the interpolation of density $\rho(\phi)$, viscosity $\mu(\phi)$ and stresses $\boldsymbol{\sigma}(\phi)$. These updated quantities are fed into the unified continuum equation for the next iteration. Gaussian quadrature technique is used for the integration of the variational equations. The present code is vectorized to reduce the excess time spent in loops.

The solver is designed such that refinement of the mesh is carried out till the criteria for $tol_R$ is satisfied and coarsening is only carried out in the last non-linear iteration. The governing equations are then solved on the refined mesh to obtain updated solution values.
This is done to reduce the individual errors of the corresponding equations below their tolerance values on the same background triangulation to capture the non-linearities accurately and ensure proper convergence properties of the governing equations. We also coarsen the grid if the number of elements $n_{el}$ exceeds $nElemMax$. Based on the proposed procedure, we present several numerical tests of increasing complexity to demonstrate the effectiveness of the solver.

\section{Convergence and Verification Study}

%

This section elaborates upon the convergence and error analysis of the fully Eulerian formulation. Two different test cases have been solved using the present solver with increasing order of complexity, namely a pure solid system and a fluid-solid interaction system. In each case, the convergence of the solver with respect to the Eulerian grid spacing is analyzed. 

\subsection{Pure Solid Wall}
%
The first attempted benchmark problem is that of a pure solid system which was carried out to verify the structural components of the solver. The simulation is performed on a square domain consisting entirely of a hyper-elastic Neo-Hookean solid. The physical parameters of the solid are considered as: solid density $\rho^s=1$, shear modulus $\mu^s_L=1$ and solid viscosity $\mu^s=0.1$. The method of manufactured solutions is used to validate the solver against the reference velocity field whose streamfunction representation is as follows:
\begin{equation}
	\psi (\boldsymbol{x},t) = \psi_0 \mathrm{sin}(\omega t) \mathrm{sin}(k_x x) \mathrm{sin}(k_y y) ,
\end{equation}
where $\psi_0=0.25, \omega=2\pi$ and $k_x = k_y =2\pi$.

The external body force that is applied to reproduce the above motion is:
\begin{equation}
	\boldsymbol{b}^s_{ex} = \frac{\partial \boldsymbol{v}^s}{\partial t} + \boldsymbol{v}^s\cdot \nabla \boldsymbol{v}^s - \mu^s \nabla^2 \boldsymbol{v}^s - \nabla \cdot \boldsymbol{\sigma}_{sh}^s ,
\end{equation}
where the stress term is calculated using the left Cauchy-Green deformation tensor and forwarded in time via the transport equation for the same with the manufactured velocity.
\begin{figure}[htbp]
	\centering
	\begin{subfigure}{0.46\textwidth}
	\centering
	\includegraphics[width=\textwidth]{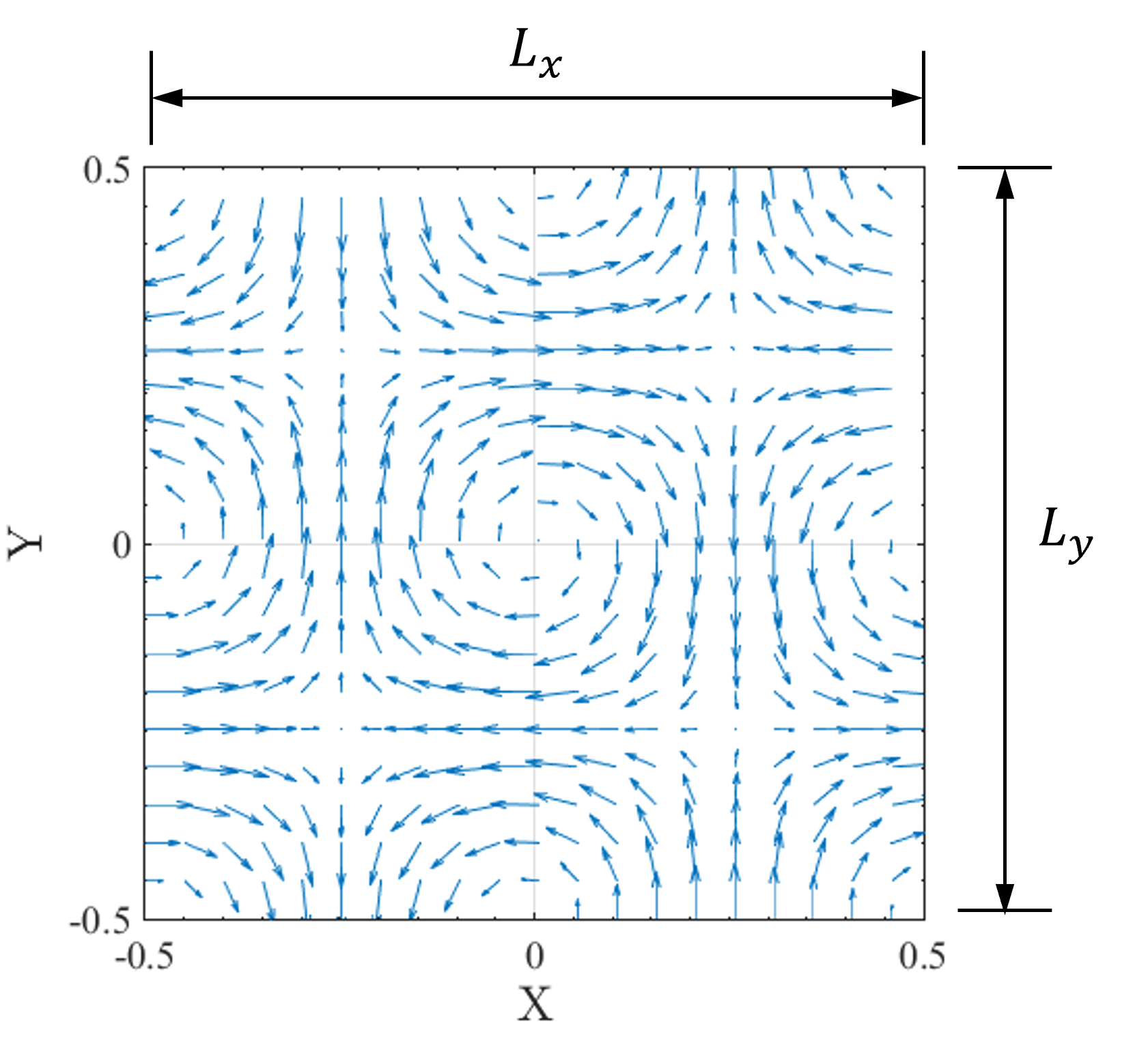}
	\caption{}
		\label{solid_wall_domain}
	\end{subfigure}
	\hfill
	\begin{subfigure}{0.46\textwidth}
	\centering
	\includegraphics[width=\textwidth]{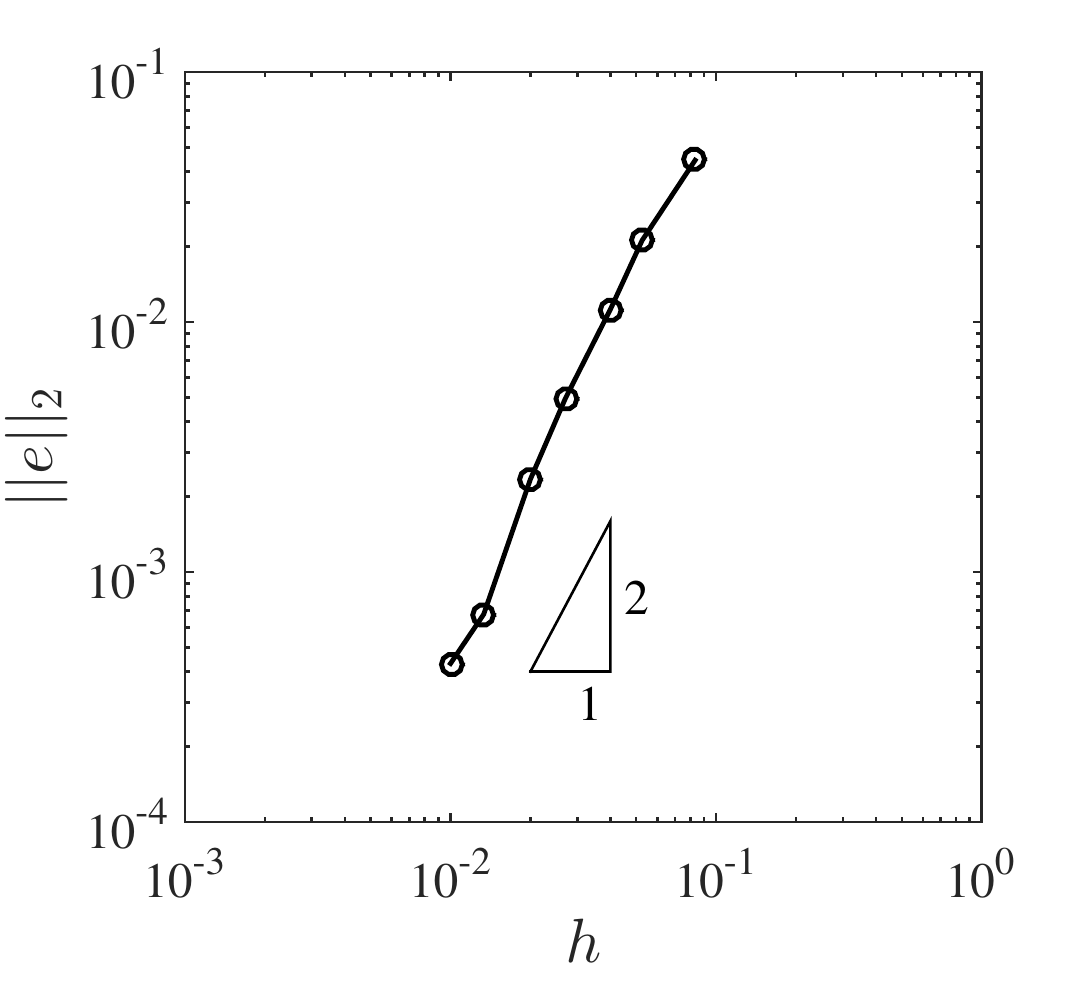}%
	\caption{}
		\label{solid_wall_conv}
	\end{subfigure}
	\caption{Pure solid wall: (a) schematic of the computational domain with the imposed velocity field (length scales are non-dimensionalised) and (b) plot of L2 norm of the error in x-velocity field}
	\label{solid_wall}
\end{figure}
The problem is solved on the above domain for various grid sizes ranging from $h=0.083$ to $h=0.01$. The $L_2$ norm of the difference between the expected (reference) and obtained velocities for different grid sizes is plotted in Fig. \ref{solid_wall_conv}, which can be defined as:
\begin{equation}
	||e||_2 = \frac{||v_x - v_{x,ref}||_2}{||v_{x,ref}||_2}
\end{equation}
We observe second order convergence for this problem with spatial order of accuracy $\mathcal{O}(h^{2.2})$.

\subsection{Flow Induced Deformation of an Elastic Wall}


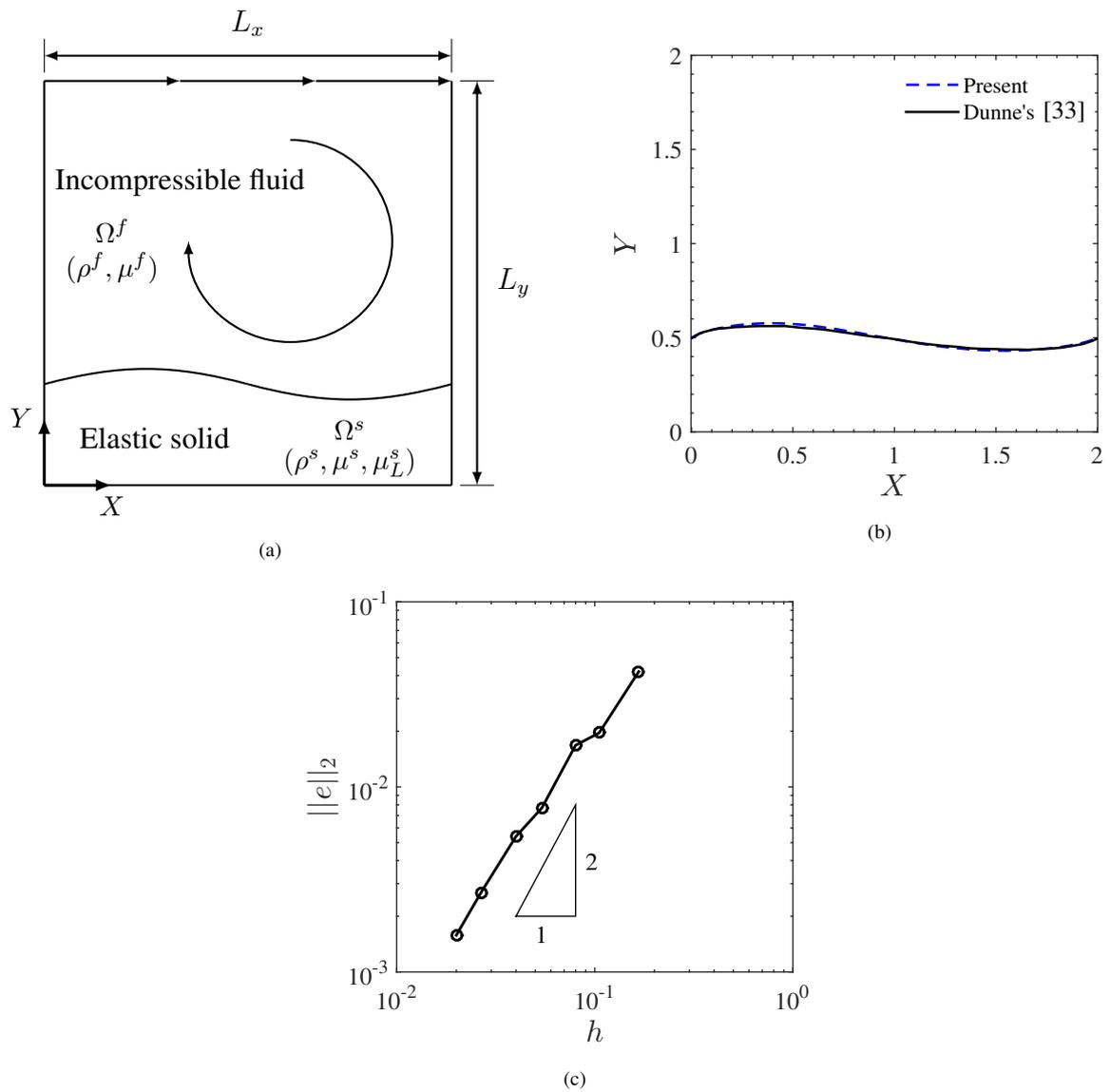
\begin{figure}[p]
	\centering
	\begin{subfigure}{0.47\textwidth}
		\centering
		\begin{tikzpicture}[scale=0.47]
	\begin{pgfonlayer}{nodelayer}
		\node [style=none] (0) at (0, 0) {};
		\node [style=none] (1) at (12, 0) {};
		\node [style=none] (2) at (12, 12) {};
		\node [style=none] (3) at (0, 12) {};
		\node [style=none] (4) at (4, 12) {};
		\node [style=none] (5) at (8, 12) {};
		\node [style=none] (6) at (12.25, 0) {};
		\node [style=none] (7) at (13.25, 0) {};
		\node [style=none] (8) at (12.25, 12) {};
		\node [style=none] (9) at (13.25, 12) {};
		\node [style=none] (10) at (12.75, 12) {};
		\node [style=none] (11) at (12.75, 0) {};
		\node [style=none] (12) at (0, 12.25) {};
		\node [style=none] (13) at (0, 13.25) {};
		\node [style=none] (14) at (12, 12.25) {};
		\node [style=none] (15) at (12, 13.25) {};
		\node [style=none] (16) at (12, 12.75) {};
		\node [style=none] (17) at (0, 12.75) {};
		\node [style=none, label={above:\large $L_x$}] (18) at (6, 13) {};
		\node [style=none, label={right:\large $L_y$}] (19) at (13, 6) {};
		\node [style=none] (20) at (0, 3) {};
		\node [style=none] (21) at (12, 3) {};
		\node [style=none] (22) at (6, 3) {};
		\node [style=none] (23) at (7.25, 10.25) {};
		\node [style=none] (24) at (9, 9.25) {};
		\node [style=none] (25) at (9.25, 7.5) {};
		\node [style=none] (26) at (8, 6) {};
		\node [style=none] (27) at (6, 5.75) {};
		\node [style=none] (28) at (4.5, 7.25) {};
		\node [style=none, label={above:\large Incompressible fluid}] (29) at (4, 8.25) {};
		\node [style=none, label={above:\large Elastic solid}] (30) at (3.25, 0.75) {};
		\node [style=none, label={below: $X$}] (31) at (2, 0) {};
		\node [style=none, label={left: $Y$}] (32) at (0, 2) {};
		\node [style=none, label={below: $\Omega^f$}] (33) at (2, 8.25) {};
		\node [style=none, label={below: $(\rho^f,\mu^f)$}] (34) at (2, 7.25) {};
		\node [style=none, label={below: $\Omega^s$}] (35) at (9, 2.25) {};
		\node [style=none, label={below: $(\rho^s,\mu^s,\mu^s_{L})$}] (36) at (9, 1.45) {};
	\end{pgfonlayer}
	\begin{pgfonlayer}{edgelayer}
		\draw [thick] (0.center) to (1.center);
		\draw [thick] (1.center) to (2.center);
		\draw [thick] (0.center) to (3.center);
		\draw [thick, style=stealth] (3.center) to (4.center);
		\draw [thick, style=stealth] (4.center) to (5.center);
		\draw [thick, style=stealth] (5.center) to (2.center);
		\draw (8.center) to (9.center);
		\draw (6.center) to (7.center);
		\draw (13.center) to (12.center);
		\draw (15.center) to (14.center);
		\draw [thick, style=latex] (10.center) to (11.center);
		\draw [thick, style=latex] (17.center) to (16.center);
		\draw [thick, bend left=15] (20.center) to (22.center);
		\draw [thick, bend right=15] (22.center) to (21.center);
		\draw [very thick, style=stealth] (0.center) to (31.center);
		\draw [very thick, style=stealth] (0.center) to (32.center);
		\draw [-latex, thick] (23) arc (90:-180:3);
	\end{pgfonlayer}
\end{tikzpicture}
		\caption{}
		\label{fig_first_case}
	\end{subfigure}
	\hfill
	\begin{subfigure}{0.47\textwidth}
		\centering
		\includegraphics[width=\textwidth]{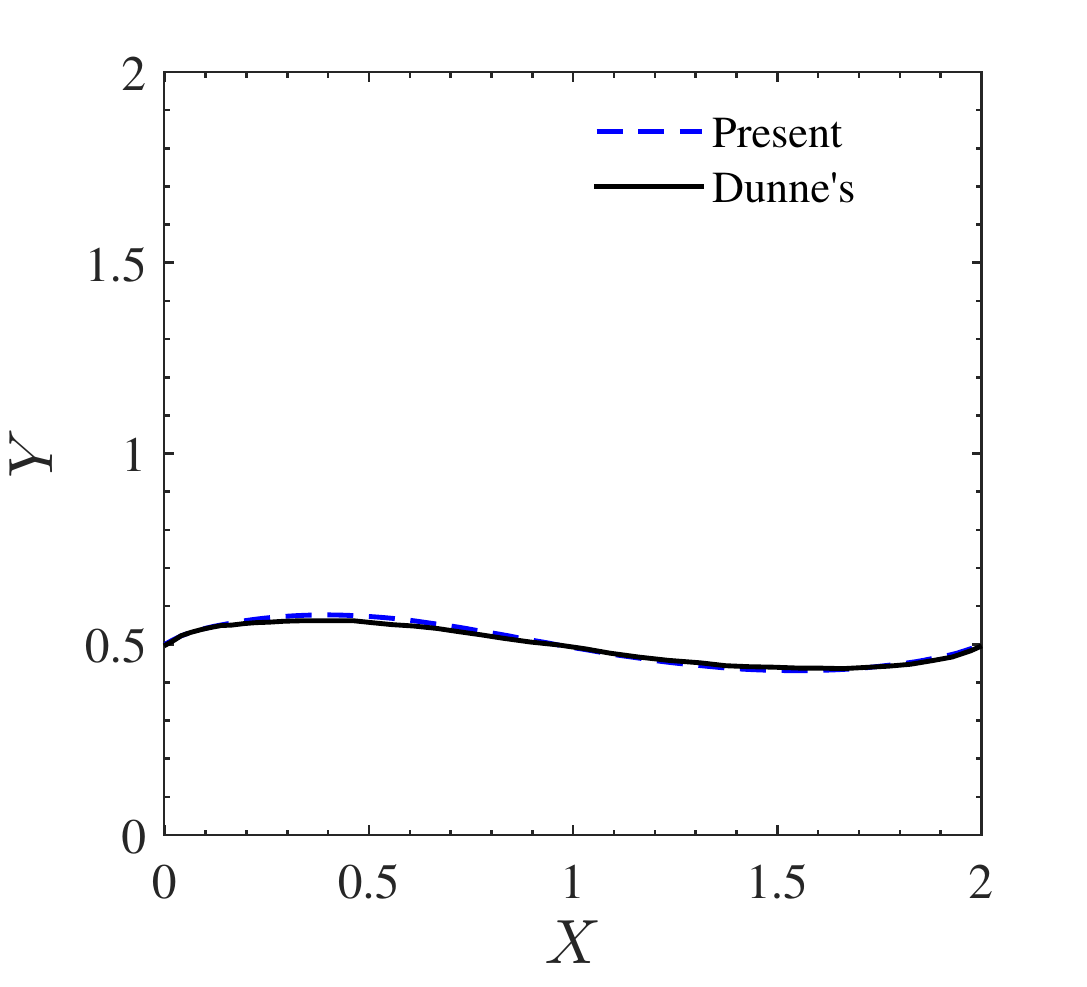}
		\put (-42,154) {\small{[33]}}
		\caption{}
		\label{fig_first_case}
	\end{subfigure}
	\begin{subfigure}{0.47\textwidth}
		\centering
		\includegraphics[width=\textwidth]{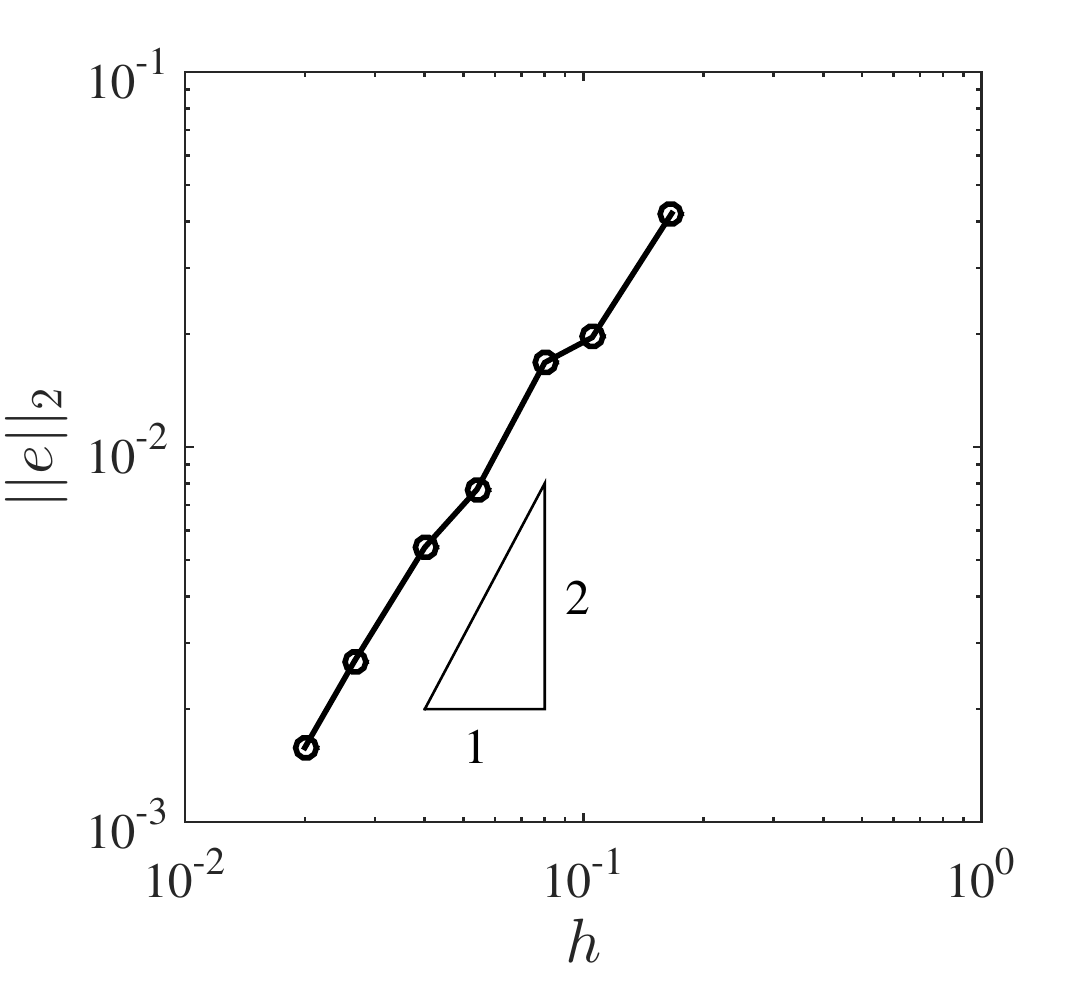}
		\caption{}
		\label{slosh_wall_conv}
	\end{subfigure}
	\caption{Flow induced deformation of an elastic wall: (a) schematic of the computational domain; (b) comparison of final interface position with \cite{dunne2006eulerian} for the non-adaptive case (length scales are non-dimensionalised) and (c) plot of L2 norm of the error in x-velocity field }
	\label{slosh_wall}
\end{figure}

The second test problem involves the flow induced deformation of an elastic wall. This benchmark is carried out to verify the full FSI solver using the proposed fully Eulerian approach. This is essentially a modified lid-driven cavity problem where the bottom part is occupied by a soft hyper-elastic solid to capture the deformations in the domain. The stiffness of the solid is reduced greatly and convection is removed from the system to visualize the deformations accurately. The size of the domain is $[0,2] \times [0,2]$. The physical parameters for the problem are: fluid viscosity $\mu^f=0.2$, solid viscosity $\mu^s=0$, shear modulus $\mu^s_L=0.2$, fluid density $\rho^f=1$ and solid density $\rho^s=1$. The velocity of the top lid is given by:
\begin{equation}
	v = 0.5  
	\begin{cases}
		\mathrm{sin}^2(\pi x/0.6)  & x \in [0,0.3], \\
		1  & x \in (0.3, 1.7), \\
		\mathrm{sin}^2(\pi (x - 2)/0.6)  & x \in [1.7,2].
	\end{cases}
\end{equation}
The no-slip condition is imposed on the rest of the boundaries and Neumann condition is applied for the order-parameter on all the boundaries. The initial condition for the phase-field function $\phi$ is given by,
\begin{equation}
	\phi(x,y,0) = -\tanh\bigg(\frac{y-0.5}{\sqrt{2}\varepsilon}\bigg) ,
\end{equation}
where $\varepsilon$ is the interface thickness parameter. 
Figure \ref{slosh_wall} illustrates the order of accuracy plot and a comparison of the present results with those of \cite{dunne2006eulerian} for the non-adaptive case. The accurate interface capturing for this benchmark problem ensures the applicability of the present method. A systematic convergence study is carried out for $\varepsilon=0.01$ for different mesh sizes ranging from $h=0.16$ to $h=0.005$. The x-velocity solution on the finest mesh ($h=0.005$) is taken as the reference solution. The L2 norm of the error between the velocity field for the given mesh with respect to the reference solution is plotted in Fig. \ref{slosh_wall_conv}. The spatial order of accuracy comes out to be around $\mathcal{O}(h^{1.6})$. The decrease in the order of accuracy from the previous case can be attributed to the presence of the fluid-solid interface.

\section{Assesment of Adaptive Fully Eulerian FSI Framework}
In this section, we present numerical test problems which have been carried out via the interface-driven adaptivity procedure. We begin with the simple problem of flow-induced deformation of an elastic wall, for which we presented convergence studies in the previous section, and evaluate the adaptive procedure in terms of computational efficiency and conservation properties. We gradually move to more complicated problems, finally concluding with the contact problem between an elastic solid and a rigid wall.

\subsection{Flow Induced Deformation of an Elastic Wall}
\begin{figure}[htbp]
	\centering
	\includegraphics[scale=0.35]{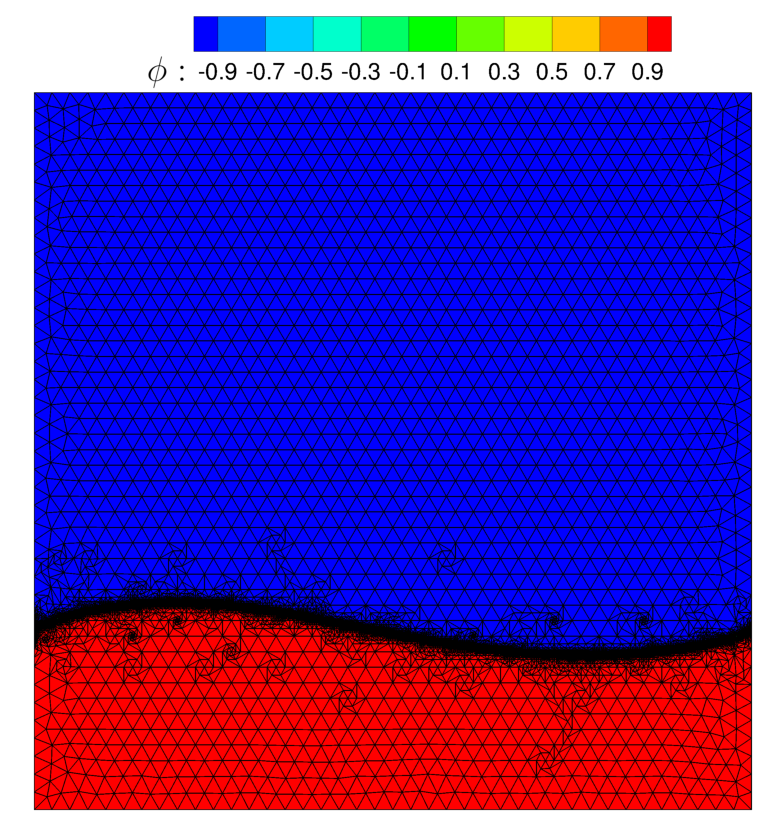}
	\caption{Adaptive case for flow induced deformation of an elastic wall problem: contour plot of the order parameter at the steady state solution superimposed on the adaptive grid}
	\label{adaptive_mesh}
\end{figure}

\begin{figure}[htbp]
	\centering
	\begin{subfigure}[b]{0.47\textwidth}	
		\centering	
		\includegraphics[width=\textwidth]{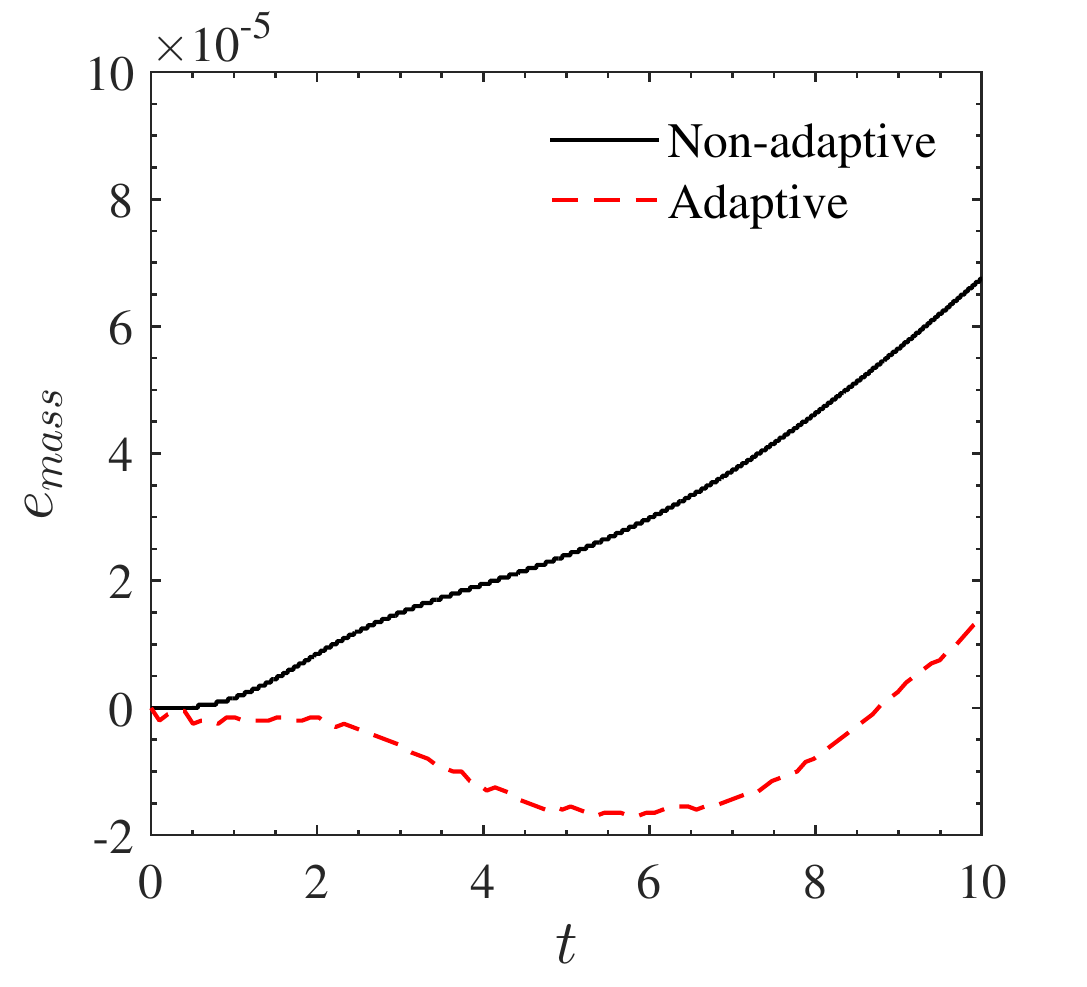}
		\caption{}		
		\label{fig_first_case2}
	\end{subfigure}
	\hfill
	\begin{subfigure}[b]{0.47\textwidth}	
		\centering	
		\includegraphics[width=\textwidth]{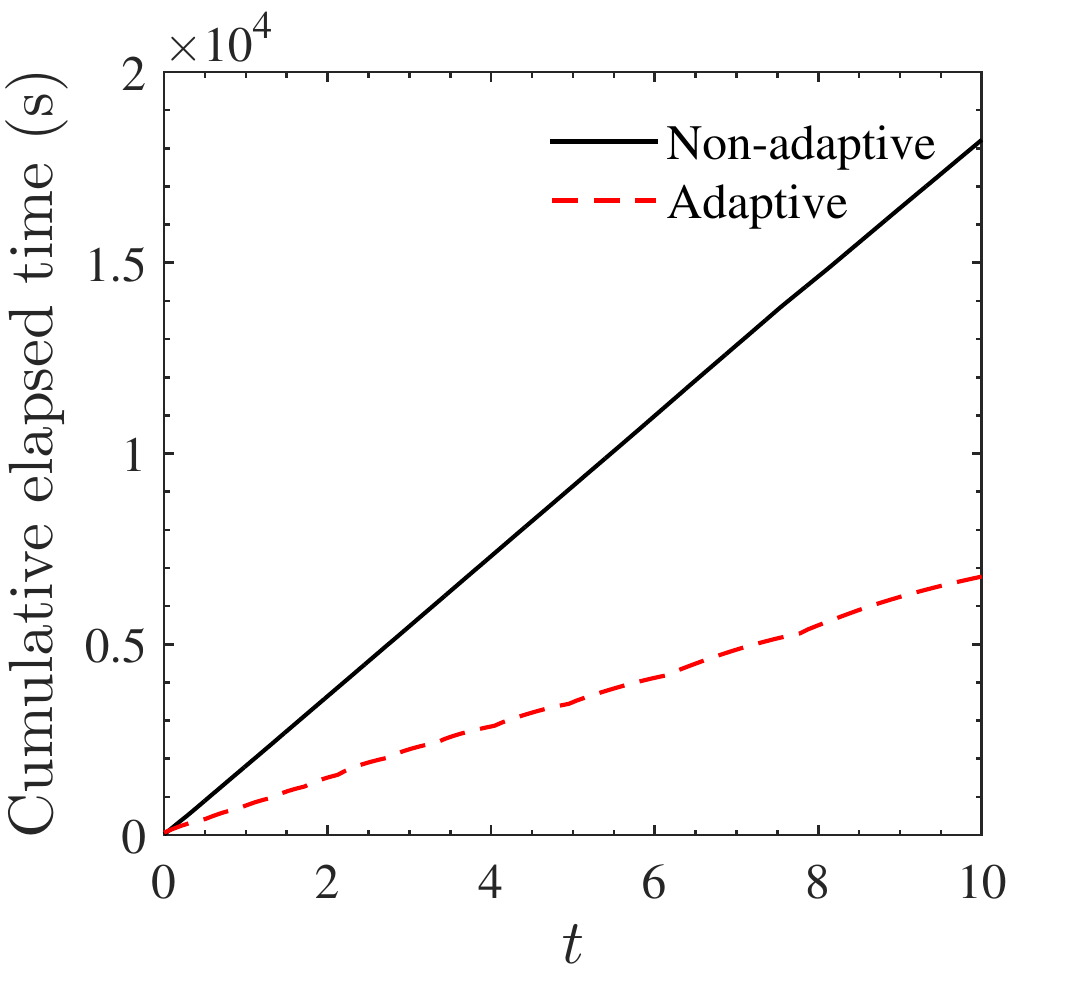}
		\caption{}		
		\label{fig_first_case2}
	\end{subfigure}	
	\caption{Comparison plots for flow induced deformation of an elastic wall: (a) total mass loss and (b) cumulative elapsed time (in seconds) for the adaptive and non-adaptive cases}
	\label{adapt_compare}
\end{figure}

We carry out this problem with adaptive mesh refinement as well with the same physical parameters as mentioned previously. The user-defined parameters for the D{\"o}rfler criteria are chosen as: $\theta=0.5$ and $\theta_c=0.05$. Figure \ref{adaptive_mesh} illustrates the refinement of the mesh along the fluid-solid interface almost exclusively. A comparison between the adaptive and non-adaptive cases has been carried out in Fig. \ref{adapt_compare}. It can be observed that the total mass conservation error reduces by nearly 4.5 times for the adaptive case compared to the non-adaptive case. The mass conservation error of the order parameter is evaluated as:
\begin{equation}
	e_{mass} = \frac{m - m_{t=0}}{m_{t=0}} ,
	\label{eq:e_mass}
\end{equation}
where $m$ is defined as $m = \int_{\Omega} \phi^n \mathrm{d}\Omega$ and $m_{t=0}$ is the mass of the order parameter at $t=0$. The increased density of the elements helps in accurate capturing of the fluid-solid interface, hence resulting in reduced error for mass conservation. Similarly the cumulative elapsed time decreases by a factor of nearly three for the adaptive case, hence indicating the reduction in computational costs.


\subsubsection{Effect of Interfacial Resolution}

\begin{figure}[htbp]
	\centering
	\includegraphics[scale=0.8]{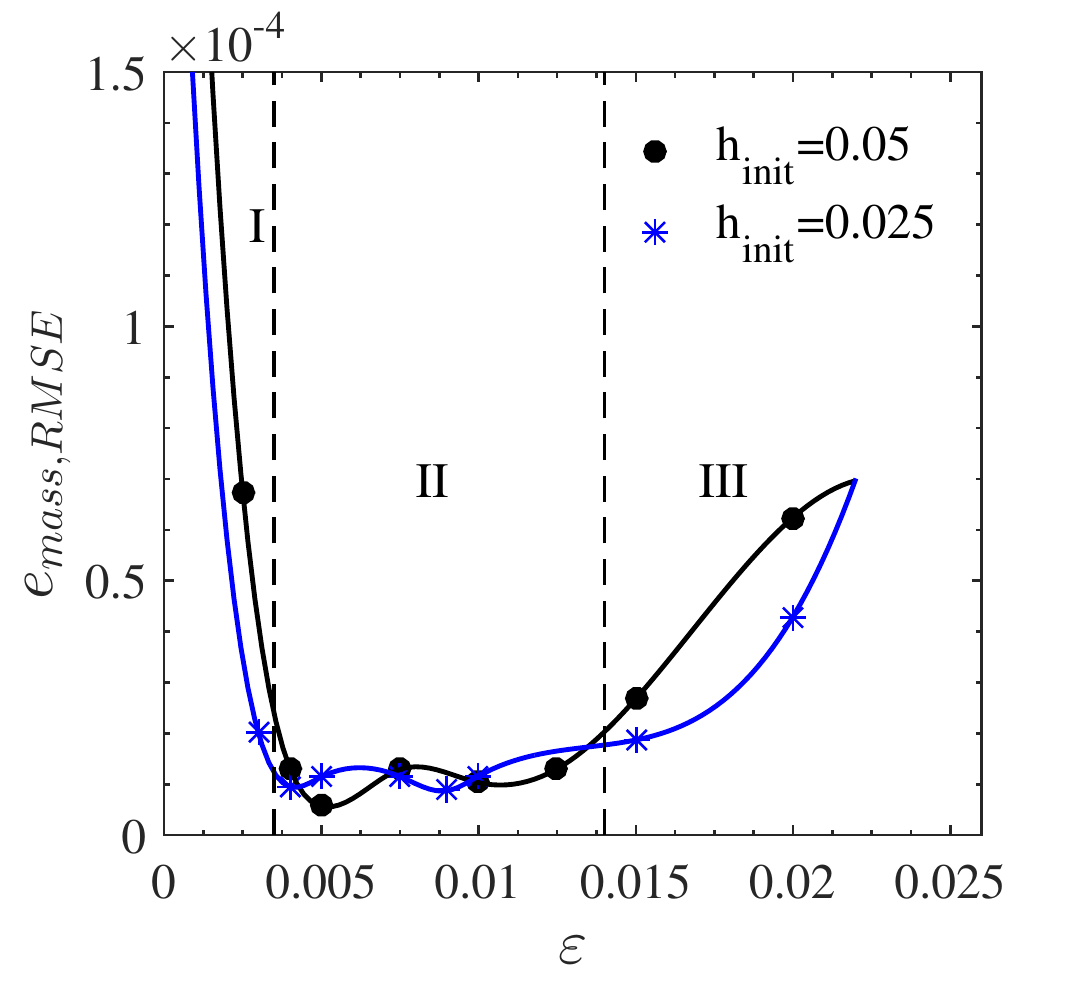}
	\caption{Flow induced deformation of an elastic wall: variation of the RMS mass loss error for different values of the interface thickness parameter, $\varepsilon$ and initial grid size $h_{init}$}
	\label{e_mass_RMSE}
\end{figure}
\begin{figure}[htbp]
	\centering
	\includegraphics[scale=0.8]{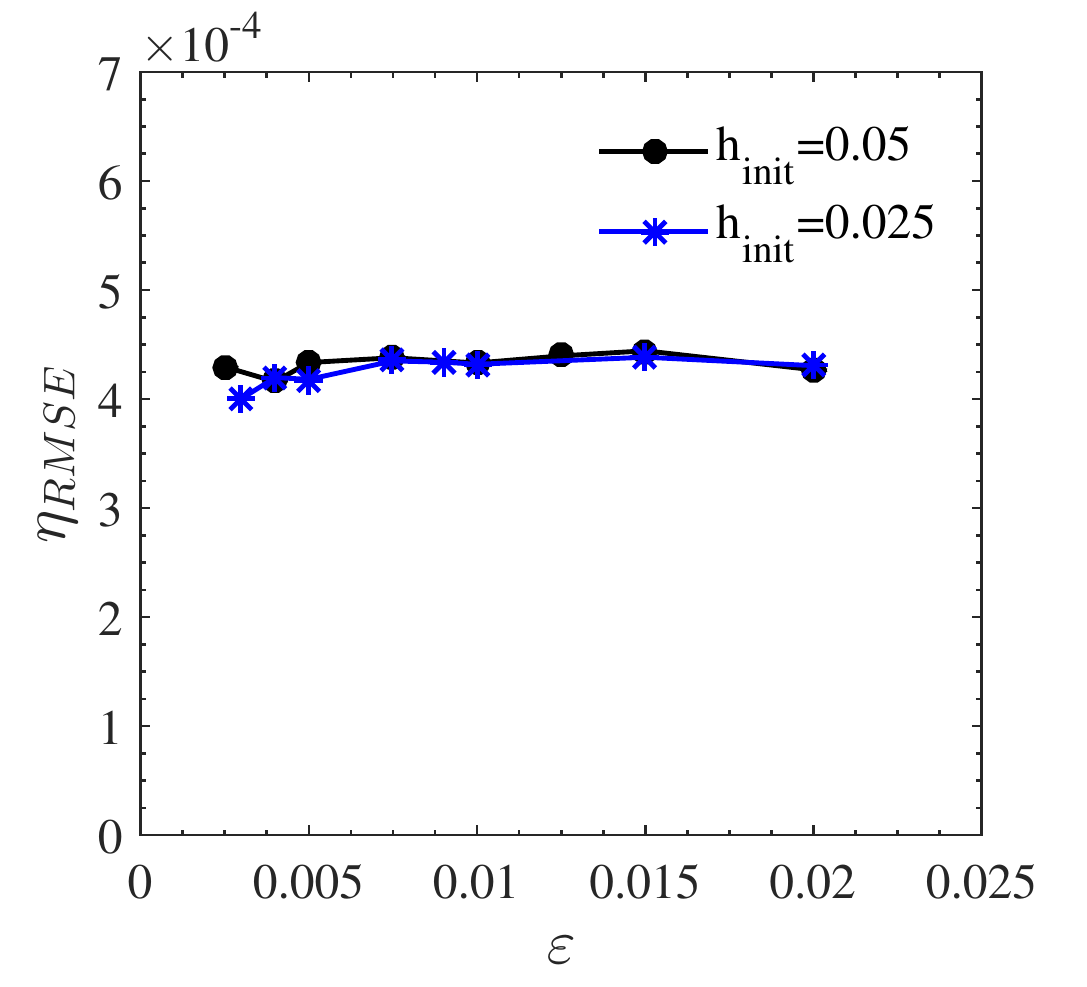}
	\caption{Flow induced deformation of an elastic wall: variation of the adaptive error indicator for different values of the interface thickness parameter, $\varepsilon$ and initial grid size $h_{init}$}
	\label{eta_RMSE}
\end{figure}

A parametric study was carried out to investigate the impact of interfacial resolution using the performance indicators of RMS errors for $e_{mass}$ and $\eta$ (adaptive error indicator). The RMS value of any generic quantity $\lambda$ varying in time can be evaluated as:
\begin{equation}
	\lambda_{RMS} = \sqrt{\frac{\int_0^T \lambda(t)^2 \mathrm{d}t}{\int_0^T \mathrm{d}t}} ,
\end{equation}
Figure \ref{e_mass_RMSE} illustrates the variation of the RMS mass conservation error with the interface thickness parameter $\varepsilon$. At very low values of the interface thickness parameter (Region I), we do not have enough grid cells to capture the interface. Even after adaptive refinement, we are not able to accurately describe the interface with such low $\varepsilon$ values leading to high mass conservation errors. On the other hand, when the interfacial thickness parameter is large (Region III), the interface is way too diffused. The grid refinement does not have a significant effect on the final resolution of the interface as we are constrained by the high $\varepsilon$ values, again leading to high mass errors. Hence, from these plots,  we are able to come up with an optimum range for $\varepsilon$ (Region II) in which simulations can be carried out. Based on Fig. \ref{e_mass_RMSE}, we can conclude that $\varepsilon$ should lie between $0.004-0.014$ to minimize the total error in mass conservation. The small value of this parameter can be attributed to the initial coarse mesh which is possible due to the adaptive refinement in the later time steps. Hence despite starting with a fairly coarse mesh, we recover similar levels of accuracy by specifically refining the grid along the fluid-solid interface. The RMS adaptive error indicator $\eta_{RMSE}$ does not vary significantly with the given parameter (Fig. \ref{eta_RMSE}). Hence, we can conclude that the adaptive error indicator is almost independent of the interface thickness parameter and the initial grid size.


\subsection{Deformable Solid in a Driven Cavity}

\begin{figure}[htbp]
	\centering
	\begin{tikzpicture}[scale=0.65]
	\begin{pgfonlayer}{nodelayer}
		\node [style=none] (0) at (0, 0) {};
		\node [style=none] (1) at (12, 0) {};
		\node [style=none] (2) at (12, 12) {};
		\node [style=none] (3) at (0, 12) {};
		\node [style=none] (6) at (12.25, 0) {};
		\node [style=none] (7) at (13.25, 0) {};
		\node [style=none] (8) at (12.25, 12) {};
		\node [style=none] (9) at (13.25, 12) {};
		\node [style=none] (10) at (12.75, 12) {};
		\node [style=none] (11) at (12.75, 0) {};
		\node [style=none] (12) at (0, 12.25) {};
		\node [style=none] (13) at (0, 13.25) {};
		\node [style=none] (14) at (12, 12.25) {};
		\node [style=none] (15) at (12, 13.25) {};
		\node [style=none] (16) at (12, 12.75) {};
		\node [style=none] (17) at (0, 12.75) {};
		\node [style=none, label={above:\large $L_x$}] (18) at (6, 13) {};
		\node [style=none, label={right:\large $L_y$}] (19) at (13, 6) {};
		\node [style=none, label={below: $X$}] (31) at (2, 0) {};
		\node [style=none, label={left: $Y$}] (32) at (0, 2) {};
		\node [style=none] (33) at (6, 6) {};
		\node [style=none, label={above: Incompressible fluid}] (34) at (4.5, 9.25) {};
		\node [style=none, label={below: Elastic solid}] (35) at (9.25, 2) {};
		\node [style=none] (36) at (8.5, 3.5) {};
		\node [style=none, label={below: $\Omega^f$}] (37) at (2, 8.25) {};
		\node [style=none, label={below: $(\rho^f,\mu^f)$}] (38) at (2, 7.25) {};
		\node [style=none, label={below: $\Omega^s$}] (39) at (7.2, 7.55) {};
		\node [style=none, label={below: $(\rho^s,\mu^s,\mu^s_{L})$}] (40) at (7.2, 6.5) {};
		\node [style=none] (41) at (4, 12) {};
		\node [style=none] (42) at (8, 12) {};
	\end{pgfonlayer}
	\begin{pgfonlayer}{edgelayer}
		\draw [thick] (0.center) to (1.center);
		\draw [thick] (1.center) to (2.center);
		\draw [thick] (0.center) to (3.center);
		\draw (8.center) to (9.center);
		\draw (6.center) to (7.center);
		\draw (13.center) to (12.center);
		\draw (15.center) to (14.center);
		\draw [thick, style=latex] (10.center) to (11.center);
		\draw [thick, style=latex] (17.center) to (16.center);
		\draw [very thick, style=stealth] (0.center) to (31.center);
		\draw [very thick, style=stealth] (0.center) to (32.center);
		\draw [thick] (3.center) to (2.center);
		\draw [style=stealth] (35.center) to (36.center);
		\draw [thick] (7.2,6) circle (2.8cm);
		\draw [thick, style=stealth] (3.center) to (41.center);
		\draw [thick, style=stealth] (41.center) to (42.center);
		\draw [thick, style=stealth] (42.center) to (2.center);
	\end{pgfonlayer}	
\end{tikzpicture}
\caption{Computational domain for deformable solid in a driven cavity.}
\label{fig:soft_solid_domain}
\end{figure}
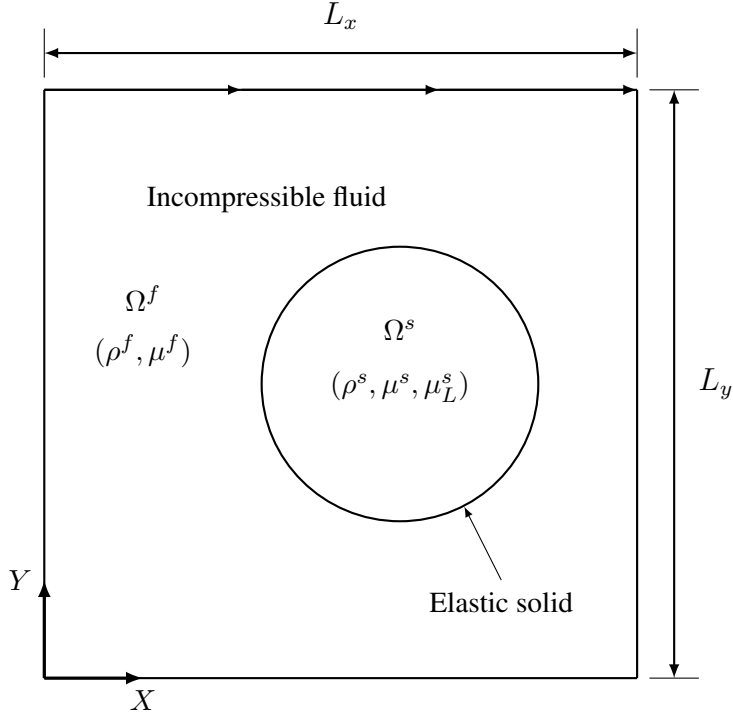

We simulate a soft solid undergoing deformation and motion inside a lid-driven cavity to further validate the solver and evaluate the conservation properties. This problem has been solved previously using the finite-difference \cite{sugiyama2011full} and finite volume \cite{jain2019conservative} methods via the fully Eulerian approach. Several other authors have also attempted this problem via immersed methods \cite{wang2010interpolation, roy2015benchmarking, griffith2017hybrid}.
The size of the domain is $[0,1] \times [0,1]$ and the radius of the circular solid is 0.2 units, placed at $(0.6,0.5)$ initially. The physical parameters of the problem are: fluid viscosity $\mu^f=10^{-2}$, solid viscosity $\mu^s=10^{-2}$, fluid density $\rho^f=1$ and solid density $\rho^s=1$. The problem is simulated for several values of the shear modulus of the solid ranging from $\mu^s_L=0.05$ (soft) to $\mu^s_L=5$ (hard). Volume conservation issues have been reported for extremely soft solids in the literature \cite{wang2010interpolation, roy2015benchmarking} for this canonical problem. We demonstrate the robustness of our adaptive fully Eulerian solver by highlighting the minimization of volume conservation error for this problem. The adaptive refinement for the domain is carried out for $\varepsilon=0.01$ with an initial grid size $h_{init}=0.025$. The parameters for the D{\"o}rfler criteria are again chosen as: $\theta=0.5$ and $\theta_c=0.05$. The initial condition for the order parameter is considered as follows:
\begin{equation}
\phi(x,y,0) = \tanh\bigg( \frac{0.2-\sqrt{(x-0.6)^2 + (y-0.5)^2}}{\sqrt{2}\varepsilon}\bigg) .
\end{equation}
The solid is initially at rest inside the square domain and the top lid is given a horizontal velocity of 1 unit, $V_{top}=1$. No-slip boundary condition is imposed on the other three walls for velocity and Neumann boundary condition is imposed on all the boundaries for the order parameter. The solid moves due to the motion of the surrounding fluid and deforms accordingly. The hydrodynamic repulsion due to the presence of the liquid layer can prevent any unwanted overlap/penetration between the soft solid and the rigid wall. It is to be noted that no special terms have been implemented to avoid particle-wall overlap or to enhance volume conservation in the present approach.

\begin{figure}[htbp]
	\centering
	\begin{subfigure}[b]{0.24\textwidth}
		\centering
		\includegraphics[width=\textwidth]{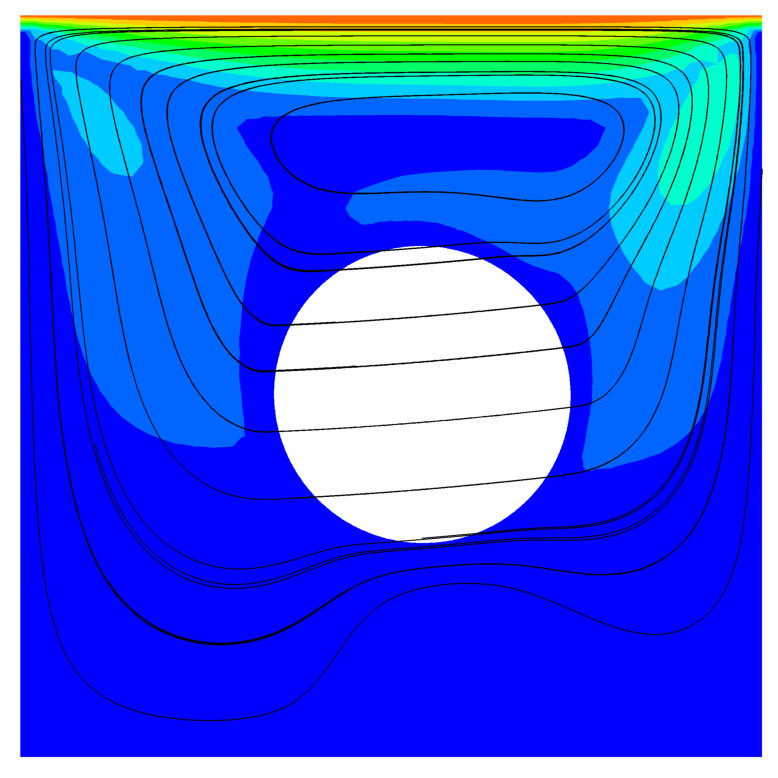}
		\caption{$t=1$}
	\end{subfigure}
	\hfill
	\begin{subfigure}[b]{0.24\textwidth}
		\centering
		\includegraphics[width=\textwidth]{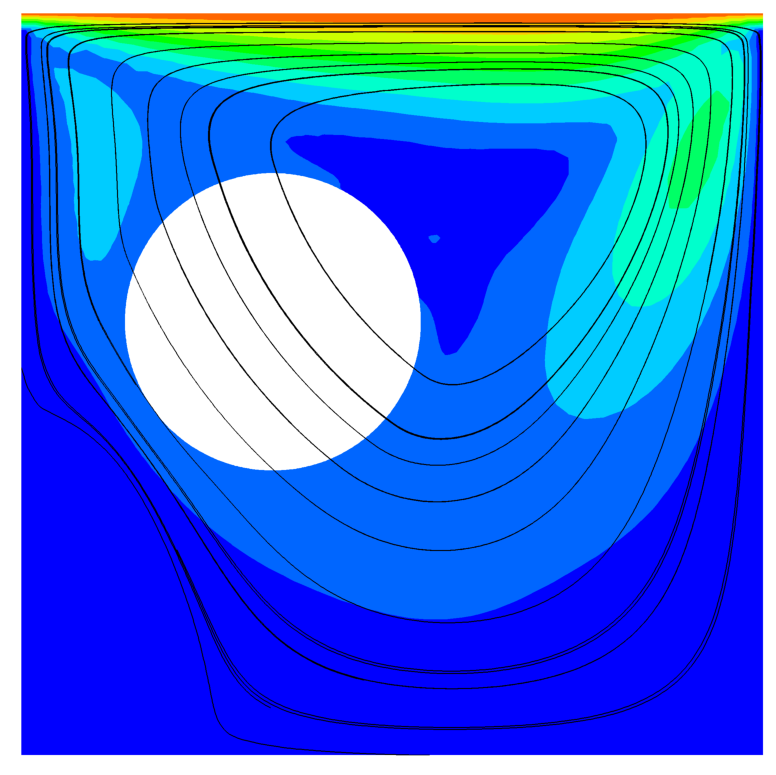}
		\caption{$t=3$}
	\end{subfigure}
	\hfill
	\begin{subfigure}[b]{0.24\textwidth}
		\centering
		\includegraphics[width=\textwidth]{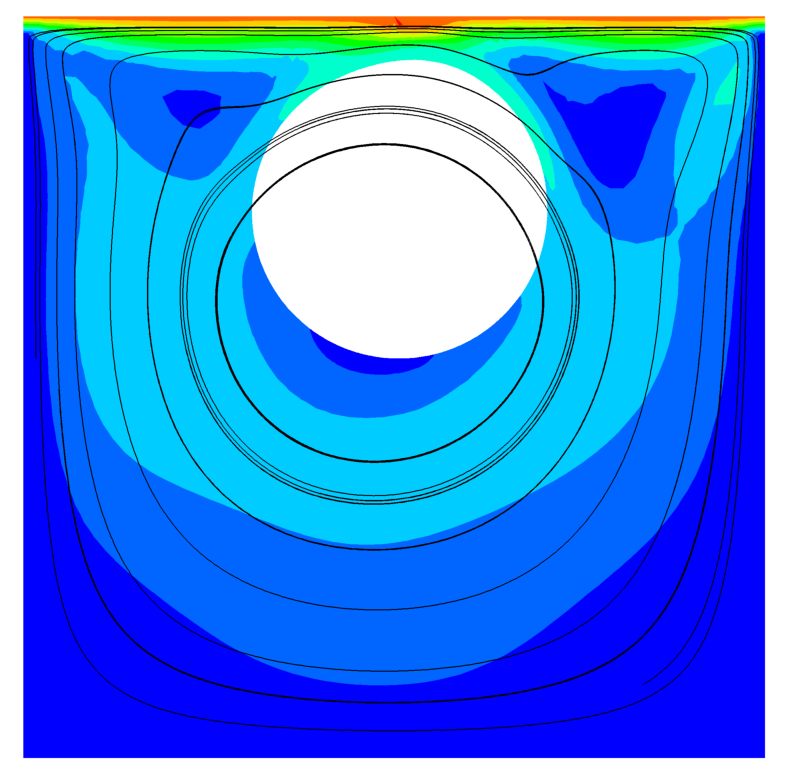}
		\caption{$t=6$}
	\end{subfigure}
	\hfill
	\begin{subfigure}[b]{0.24\textwidth}
		\centering
		\includegraphics[width=\textwidth]{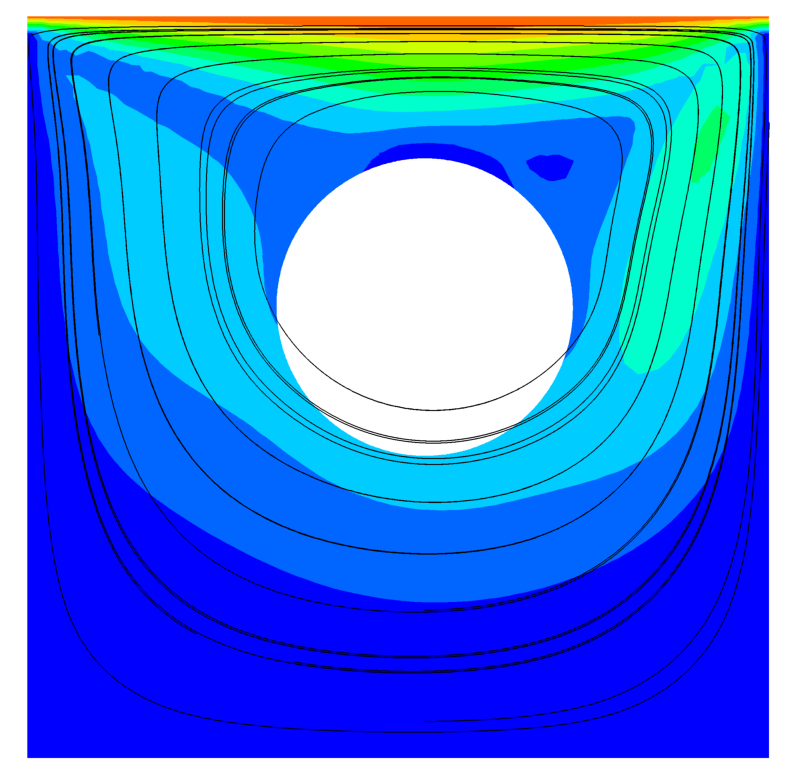}
		\caption{$t=9$}
	\end{subfigure}
	\hfill
	\includegraphics[width=0.6\textwidth]{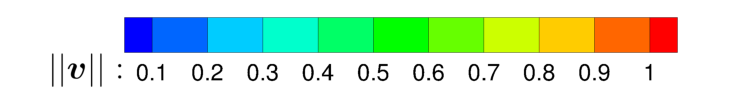}
	\caption{Snapshots for deformable solid in a driven cavity problem: velocity magnitude contours at different time instances for the hard solid case with $\mu^s_L=5$. Solid lines represent the streamlines inside the domain.}
	\label{hard_solid_snapshot}
\end{figure}

\begin{figure}[htbp]
	\centering	
	\begin{subfigure}[b]{0.24\textwidth}
		\centering
		\includegraphics[width=\textwidth]{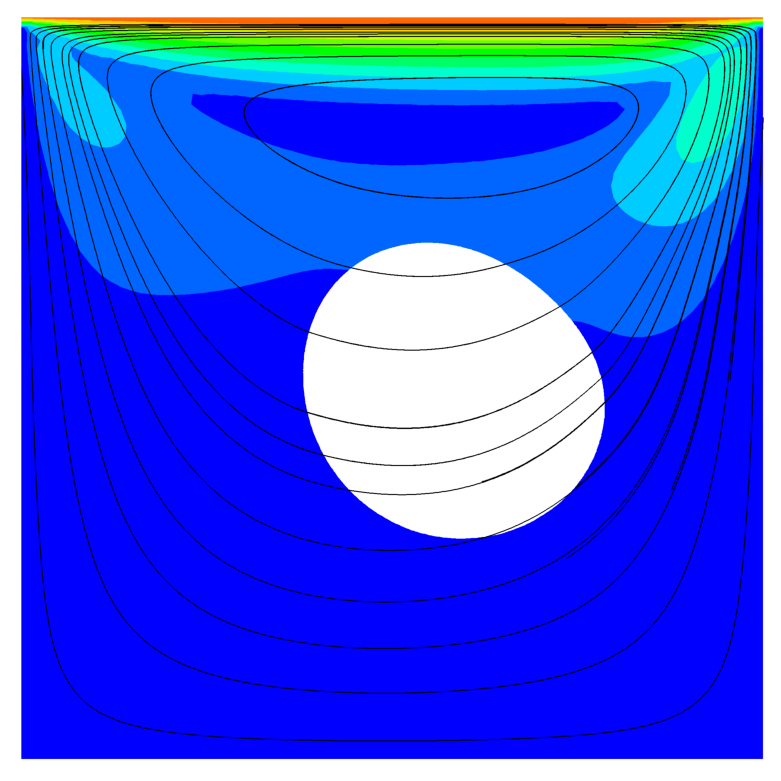}
		\caption{$t=0.5$}
	\end{subfigure}
	\hfill
	\begin{subfigure}[b]{0.24\textwidth}
		\centering
		\includegraphics[width=\textwidth]{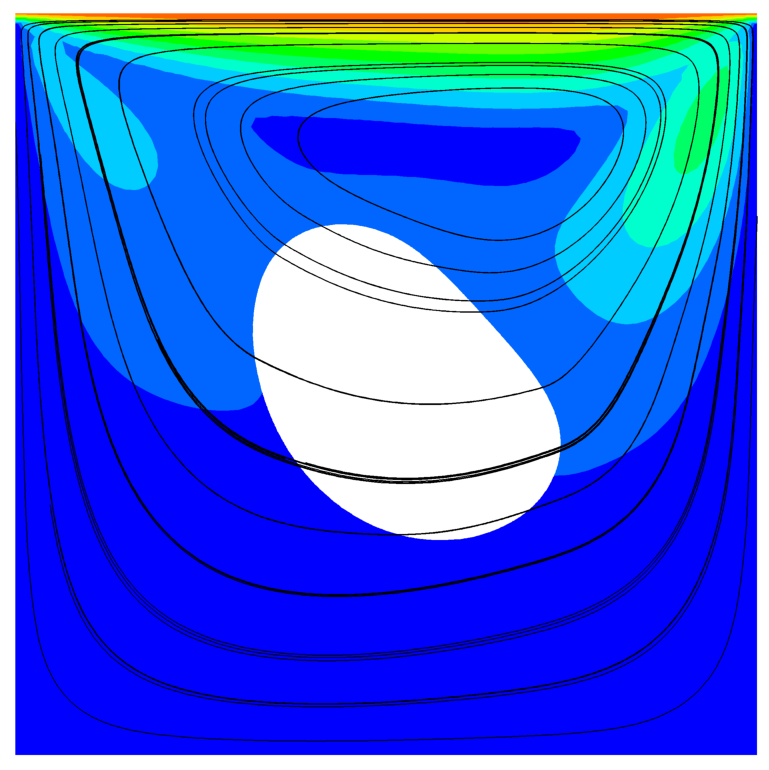}
		\caption{$t=1.2$}
	\end{subfigure}
	\hfill
	\begin{subfigure}[b]{0.24\textwidth}
		\centering
		\includegraphics[width=\textwidth]{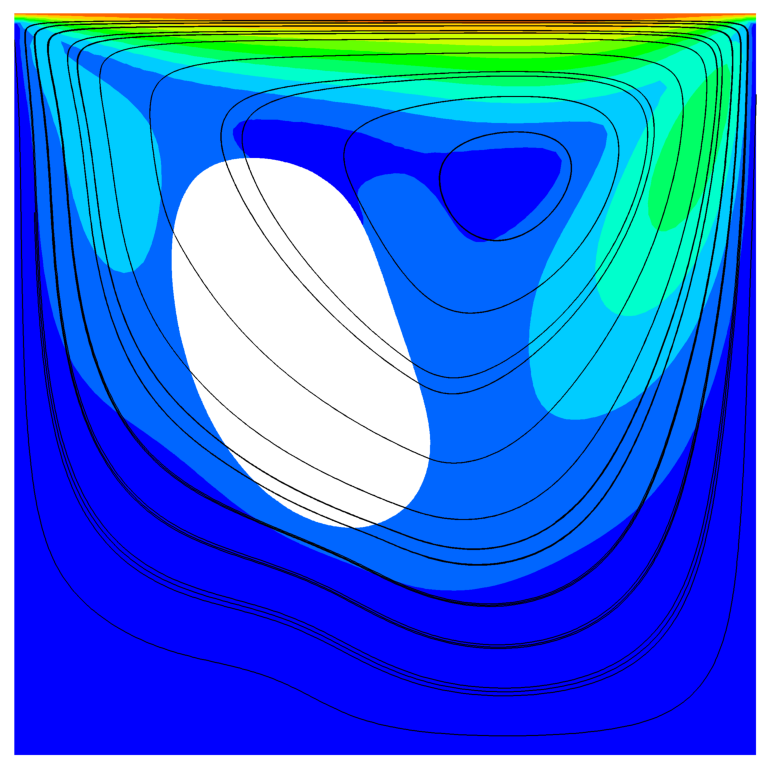}
		\caption{$t=2.4$}
	\end{subfigure}
	\hfill
	\begin{subfigure}[b]{0.24\textwidth}
		\centering
		\includegraphics[width=\textwidth]{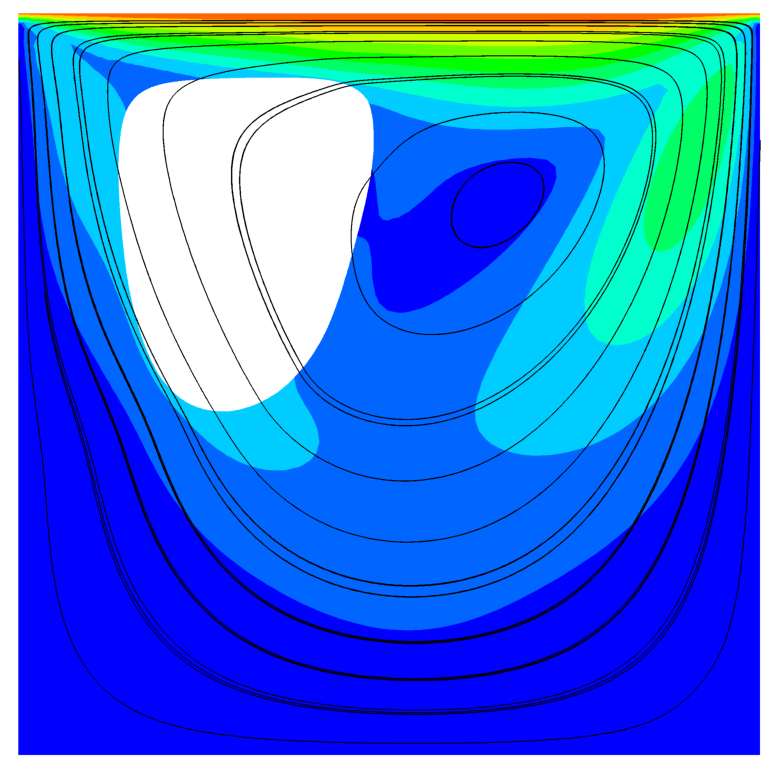}
		\caption{$t=3.6$}
	\end{subfigure}
	
	\begin{subfigure}[b]{0.24\textwidth}
		\centering
		\includegraphics[width=\textwidth]{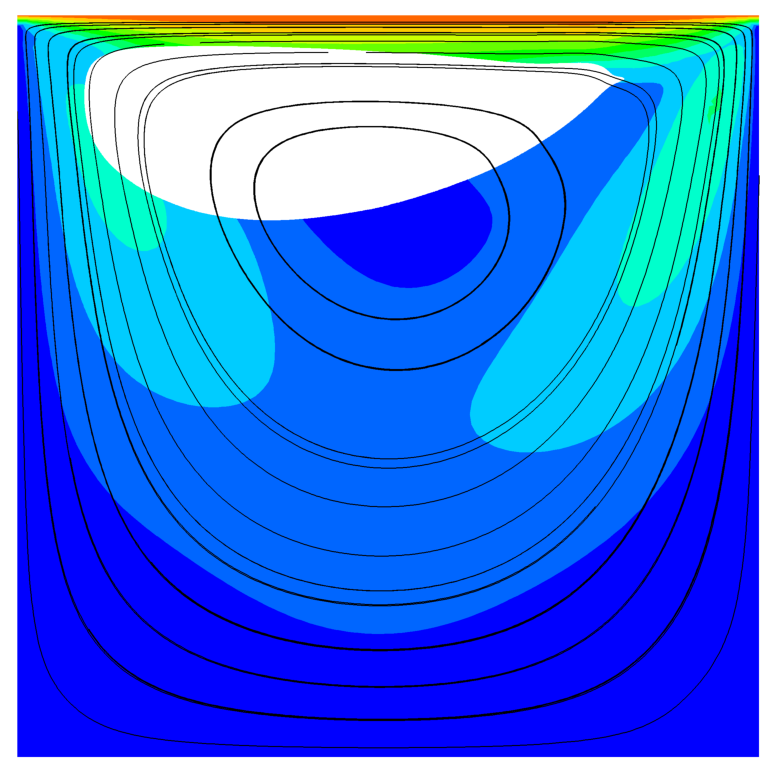}
		\caption{$t=4.8$}
	\end{subfigure}
	\hfill
	\begin{subfigure}[b]{0.24\textwidth}
		\centering
		\includegraphics[width=\textwidth]{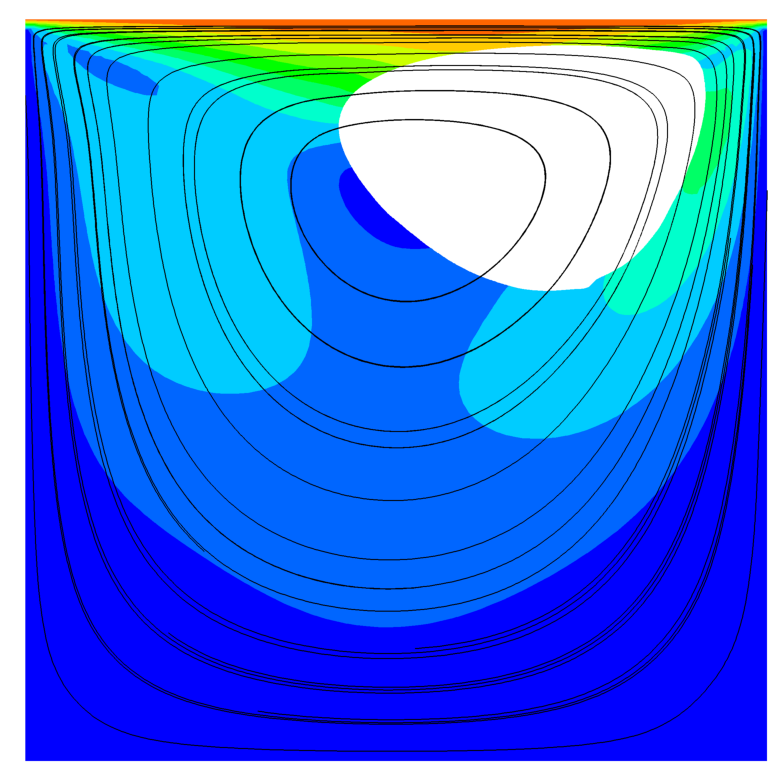}
		\caption{$t=6$}
	\end{subfigure}
	\hfill
	\begin{subfigure}[b]{0.24\textwidth}
		\centering
		\includegraphics[width=\textwidth]{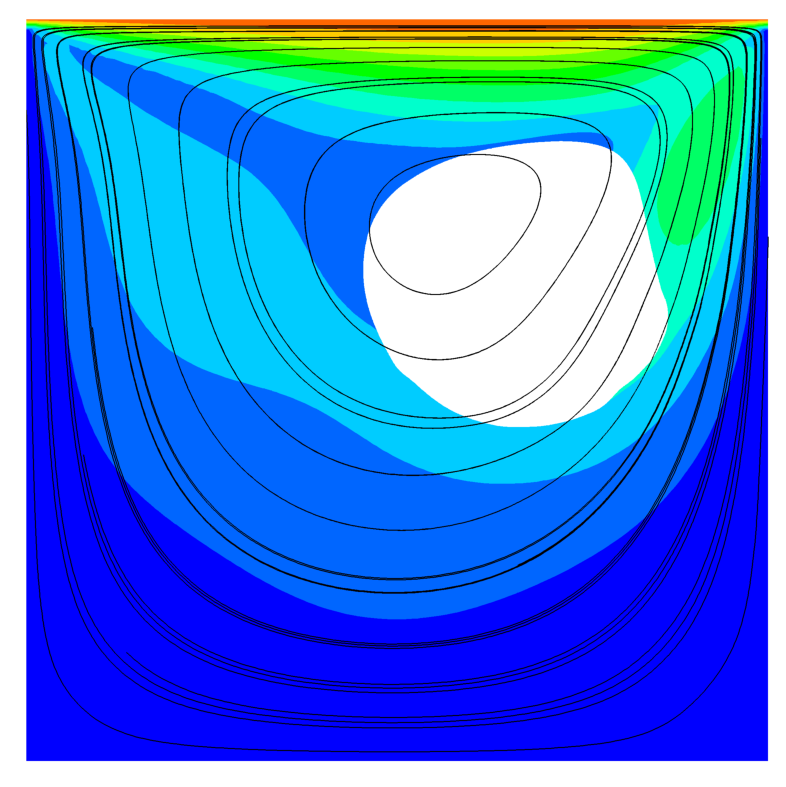}
		\caption{$t=7.2$}
	\end{subfigure}
	\hfill
	\begin{subfigure}[b]{0.24\textwidth}
		\centering
		\includegraphics[width=\textwidth]{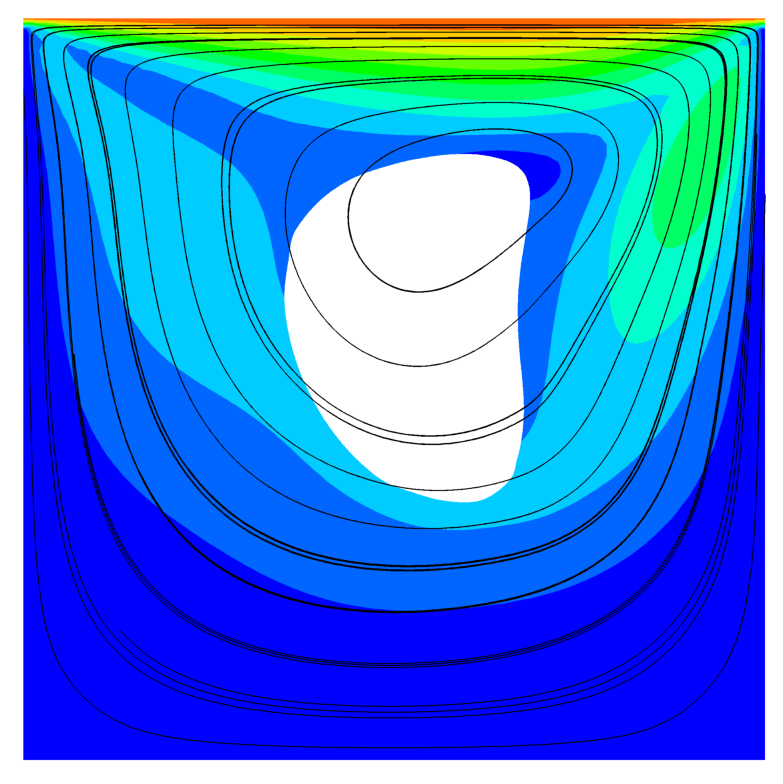}
		\caption{$t=8.4$}
	\end{subfigure}
	\hfill
	\includegraphics[width=0.6\textwidth]{snapshots_new/legend2.png}
	\caption{Snapshots for deformable solid in a driven cavity problem: velocity magnitude contours at different time instances for the soft solid case with $\mu^s_L=0.05$. Solid lines represent the streamlines inside the domain.}
	\label{soft_solid_snapshot}
\end{figure}
%

\begin{figure}[htbp]
	\centering	
	\begin{subfigure}[b]{0.49\textwidth}
		\centering	
		\includegraphics[width=\textwidth]{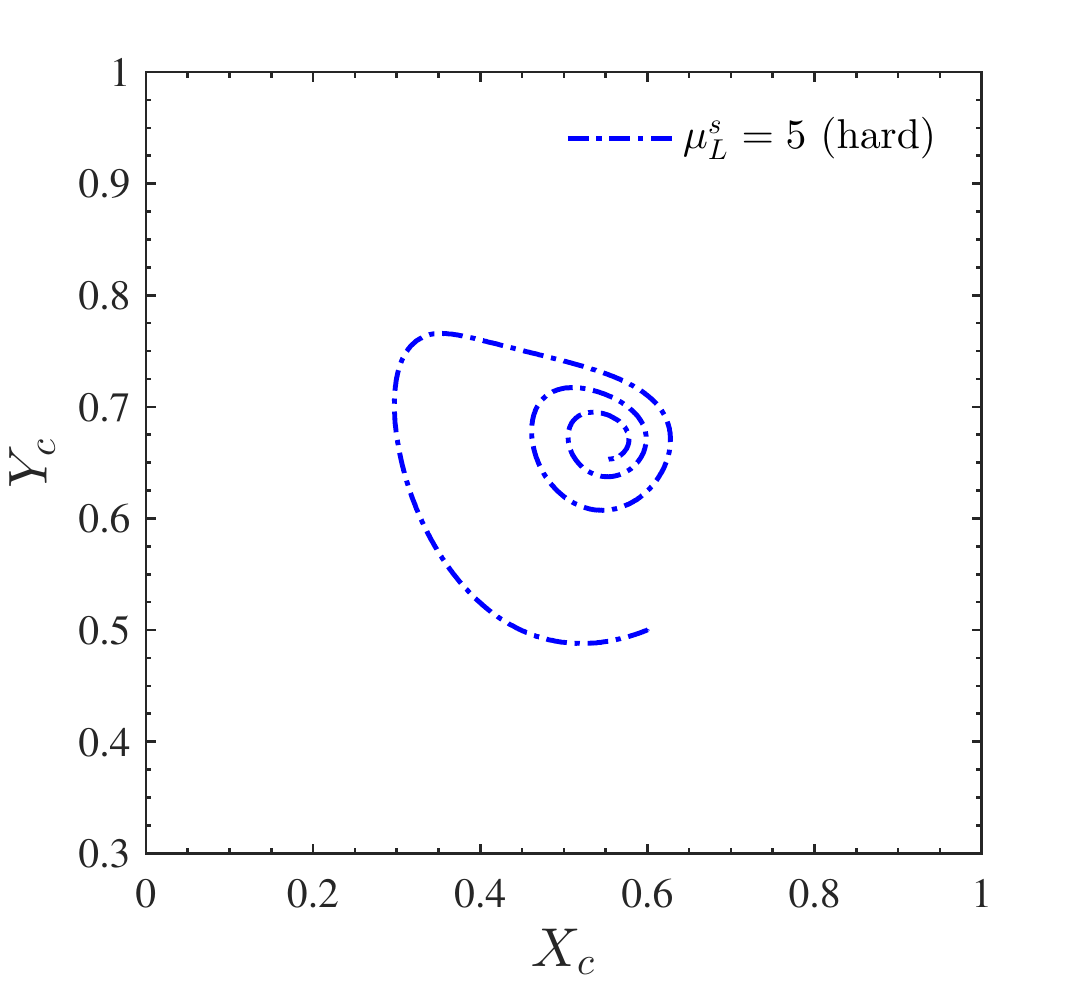}
		\caption{}
		\label{trajectory_hard_solid}
	\end{subfigure}
	\hfill
	\begin{subfigure}[b]{0.49\textwidth}
		\centering	
		\includegraphics[width=\textwidth]{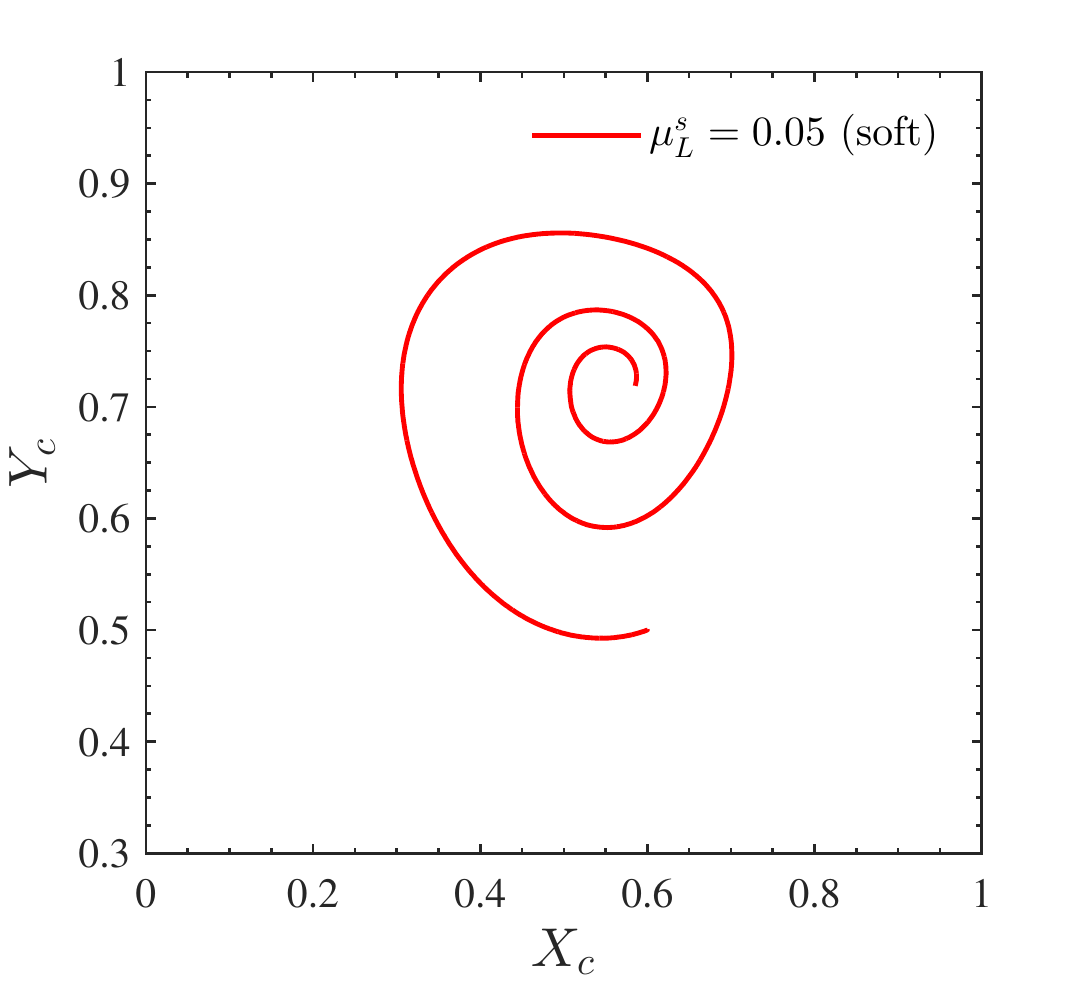}
		\caption{}
		\label{trajectory_soft_solid}
	\end{subfigure}
		\caption{Trajectory of COM for deformable solid in a driven cavity problem: trajectory of the centre of mass of (a) the hard solid $(\mu^s_L=5)$ and (b) the soft solid $(\mu^s_L=0.05)$ in the time interval $t \in [0,20]$ (length scales are non-dimensionalised) }
		\label{trajectory_compare}
\end{figure}

\begin{figure}[htbp]
	\centering
	\begin{subfigure}[b]{0.49\textwidth}
		\centering		
		\includegraphics[width=\textwidth]{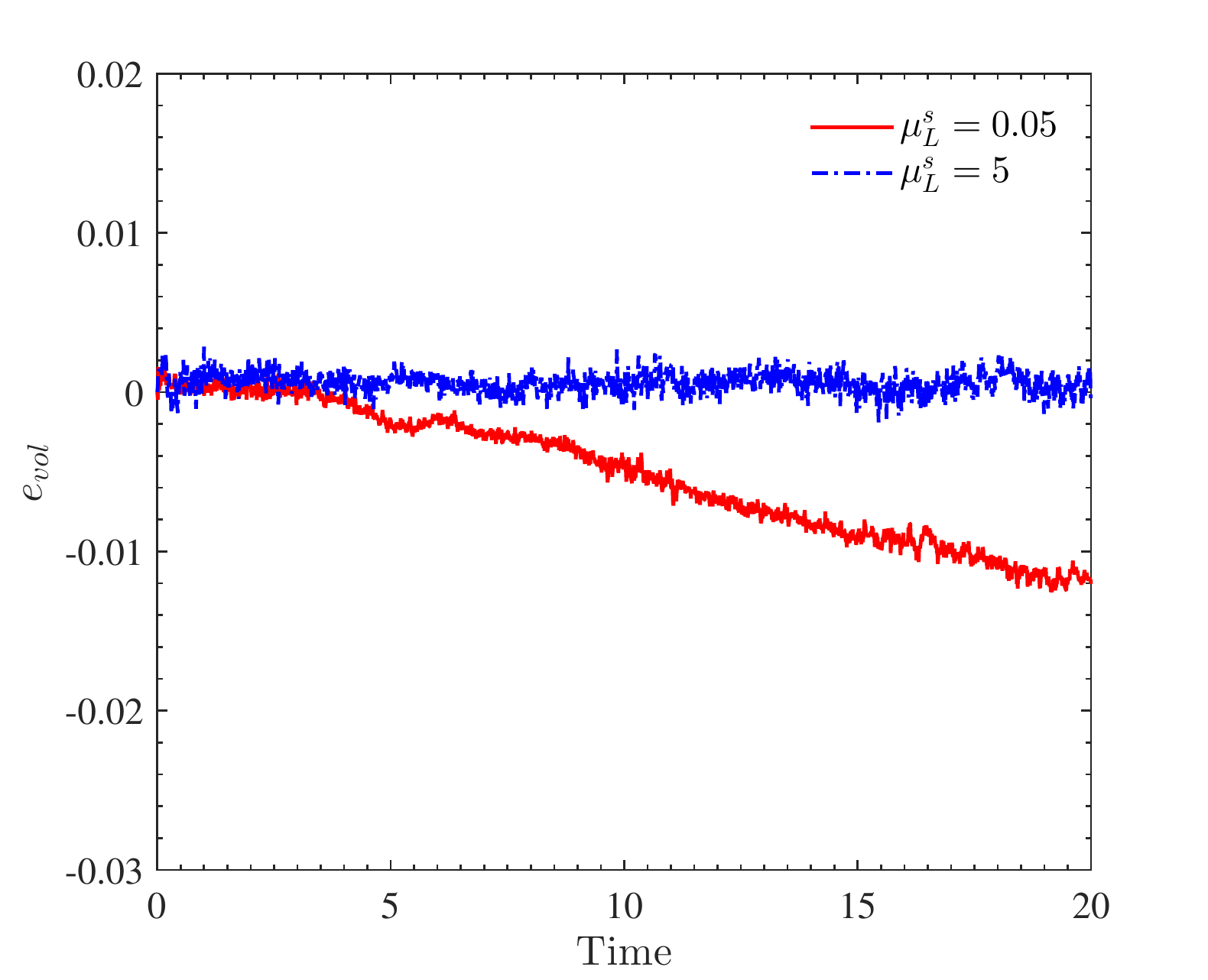}
		\caption{}
		\label{solid_cavity_e_vol_mu_s}
	\end{subfigure}
	\hfill
	\begin{subfigure}[b]{0.49\textwidth}
		\centering		
		\includegraphics[width=\textwidth]{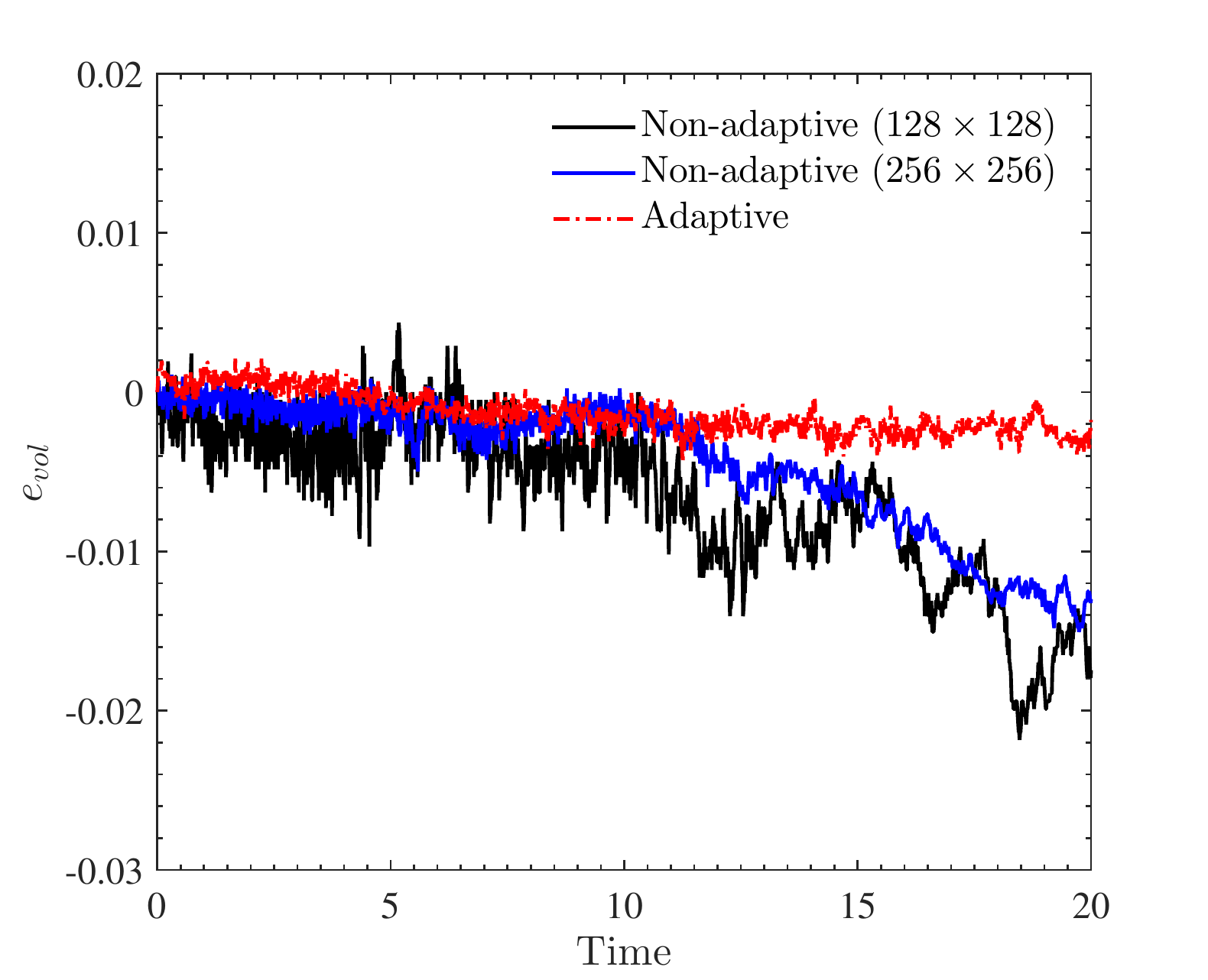}
		\caption{}
		\label{solid_cavity_e_vol_algo}
	\end{subfigure}
	\caption{Volume conservation errors for deformable solid in a driven cavity: (a) comparison of $e_{vol}$ for different values of the shear modulus $(\mu^s_L)$ and (b) comparison of $e_{vol}$ among non-adaptive and adaptive cases for $\mu^s_L=0.3$}
	\label{solid_cavity_e_vol}
\end{figure}

\begin{figure}[htbp]
	\centering
	\begin{subfigure}[b]{0.49\textwidth}
		\centering		
		\includegraphics[width=\textwidth]{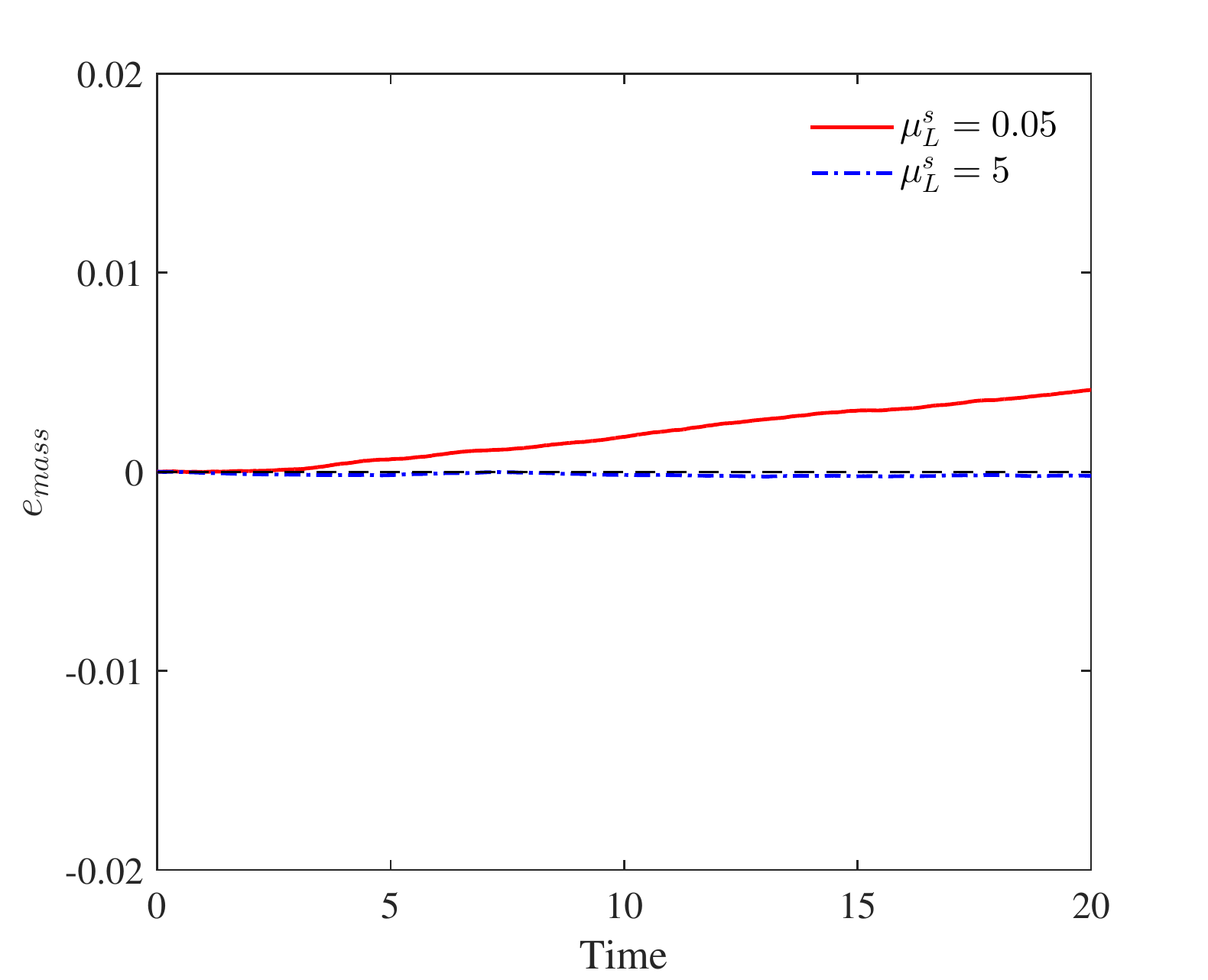}
		\caption{}
		\label{solid_cavity_e_mass_mu_s}
	\end{subfigure}
	\hfill
	\begin{subfigure}[b]{0.49\textwidth}
		\centering		
		\includegraphics[width=\textwidth]{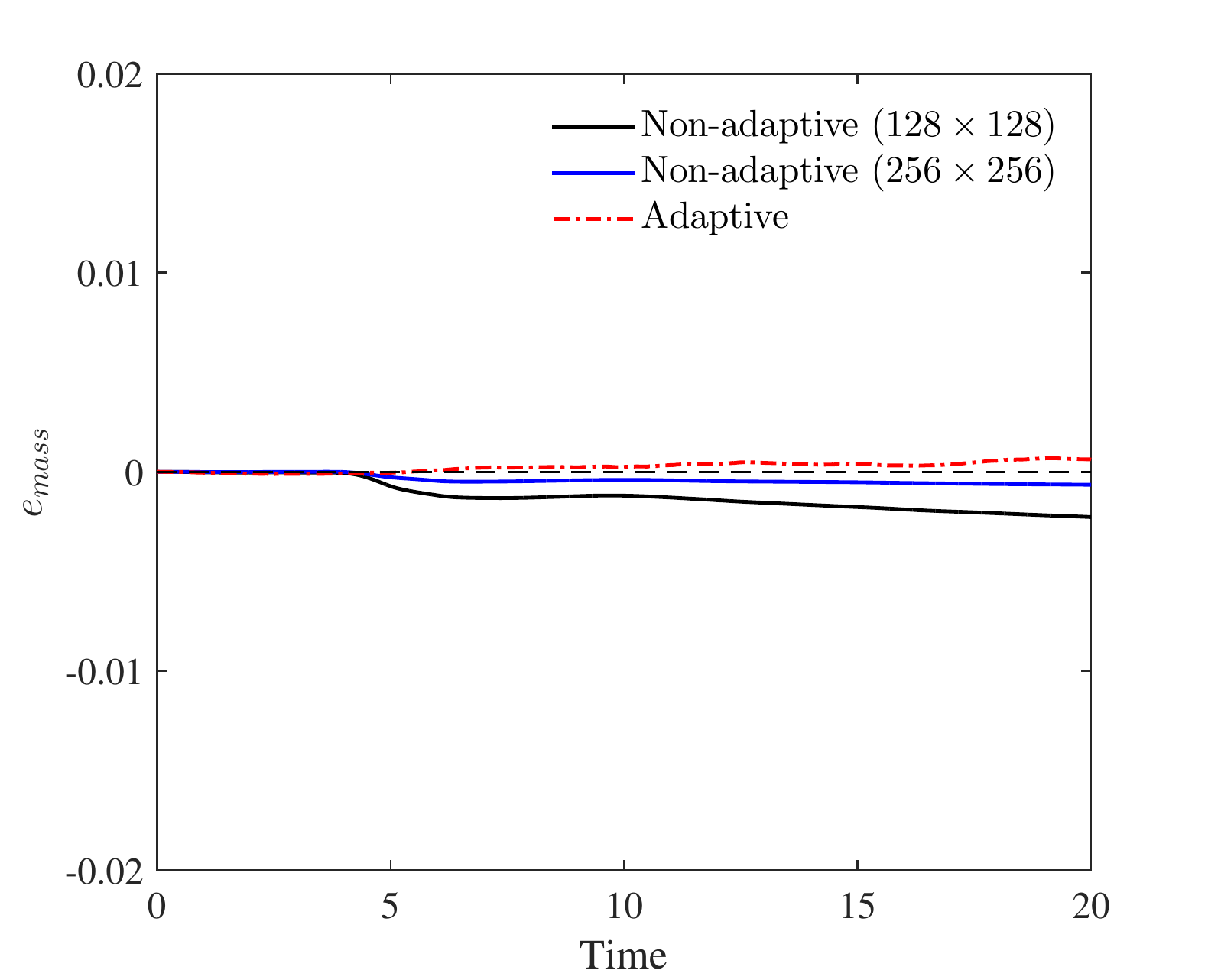}
		\caption{}
		\label{solid_cavity_e_mass_algo}
	\end{subfigure}
	\caption{Mass error via order parameter for deformable solid in a driven cavity: (a) comparison of $e_{mass}$ for different values of the shear modulus $(\mu^s_L)$ and (b) comparison of $e_{mass}$ among non-adaptive and adaptive cases for $\mu^s_L=0.3$}
	\label{solid_cavity_e_mass}
\end{figure}

Figure \ref{hard_solid_snapshot} illustrates the location and deformation of the hard solid inside the lid-driven cavity at various time instants. The black solid lines indicate the streamlines inside the domain. Similarly, Fig. \ref{soft_solid_snapshot} shows the motion of the softer solid with $\mu^s_L=0.05$ inside the lid-driven cavity. The deformation of the soft solid is much more prominent and it moves more quickly inside the domain compared to the hard solid. Figure \ref{trajectory_compare} presents the trajectories of the center of mass of the solid in each case. We can observe a clear difference in the motion of the solid for the two cases.

\subsubsection{Conservation Properties}
We investigate two important conservation properties: error for conservation of mass of the order parameter $e_{mass}$ as defined earlier (Eq. \ref{eq:e_mass}) and the volume conservation error of the solid $e_{vol}$ which can be defined as:
\begin{equation}
	e_{vol} = \frac{\int_{\Omega^s_t}d\Omega - \int_{\Omega^s_0}d\Omega}{\int_{\Omega^s_0}d\Omega}.
\end{equation}
The above error is evaluated by computing the volume integral for the cells whose phase indicator $\phi$ is greater than $0$ at each time step and comparing it with the initial volume of the solid. Issues in volume conservation are well known for this problem with a decrease in stiffness of the solid. This fact is clearly illustrated in Fig. \ref{solid_cavity_e_vol_mu_s}, where an increase in stiffness of the solid causes a reduction in the volume conservation error. The maximum volume error for the extremely soft solid case $(\mu^s_L=0.05)$ is around $1.2\%$. Whereas for the stiffer ball case $(\mu^s_L=5)$, the error in volume conservation is only around $0.25\%$ via the present approach. Similarly, introduction of interface-driven adaptivity significantly improves the volume conservation properties (Fig. \ref{solid_cavity_e_vol_algo}). The minimum stiffness of the solid for which we can obtain a converging solution on a $128 \times 128$ non-adaptive Eulerian mesh was $\mu^s_L=0.3$. However, with the introduction of adaptivity, we can also simulate the extremely flexible ball case of $\mu^s_L=0.05$ with fairly good conservation properties.

For the non-adaptive case of $\mu^s_L=0.3$ with $128 \times 128$ elements, the max absolute error in the volume conservation comes around $2.2\%$, which appears much later in the time domain. This value improves to around $1.5\%$ with the uniform refinement of the domain. The adaptive procedure on a coarse initial grid of $h_{init}=0.025$, further brings down this value to $0.4\%$, owing to the accurate capturing of the fluid-solid interface. These values are a significant improvement over those reported in \cite{wang2010interpolation} $(\approx 20\%)$ and of similar order of magnitude after the implementation of a volume correction technique $(\approx 2.5\%)$ in \cite{wang2010interpolation} for a shear modulus of $\mu^s_L=0.05$. The authors in \cite{wang2010interpolation} used an immersed finite element approach to solve the coupled fluid-solid system. The volume conservation errors for \cite{roy2015benchmarking} were in the range of $2-8.5\%$ for varying levels of grid refinement for a shear modulus of $0.1$ in the time interval $t \in [0,8]$. The immersed finite element technique was again the tool of choice in \cite{roy2015benchmarking}. The authors in \cite{griffith2017hybrid} demonstrated the volume conservation errors less than $0.4\%$ for a solid shear modulus of $0.2$ in the time period $t \in [0,10]$. \cite{griffith2017hybrid} using a hybrid finite difference-finite element discretization for the immersed boundary method to solve the coupled system of equations. For the present approach in the time interval $t \in [0,10]$, maximum volume conservation errors are less than $0.5\%$ for all cases except the non-adaptive case of $\mu^s_L=0.3$ with $128 \times 128$ elements, for which it is $0.9\%$.
Figure \ref{solid_cavity_e_mass} illustrates the variation of the mass conservation error for the system with time. The maximum absolute mass error computed via the order parameter varies from $0.4\%$ to $0.024\%$ for the different cases considered.
Hence, we can conclude that the proposed interface-driven adaptive procedure for a diffuse interface-based fully Eulerian approach performs better than most hybrid methods in terms of conservation properties.


\subsection{Bouncing of an Elastic Ball}
\begin{figure}[htbp]
	\centering
%
	\begin{subfigure}[b]{0.47\textwidth}
		\centering		
		\includegraphics[width=\textwidth]{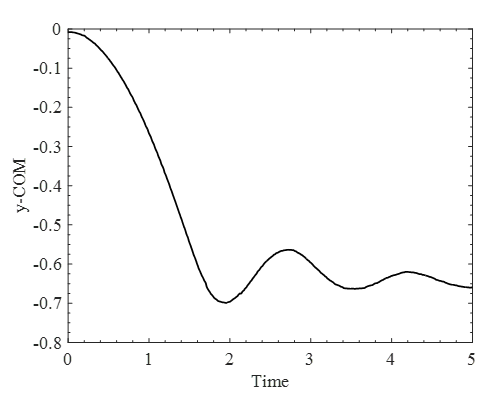}
		\caption{}
		\label{fig_first_case}
	\end{subfigure}
	\hfill
	\begin{subfigure}[b]{0.47\textwidth}
		\centering		
		\includegraphics[width=\textwidth]{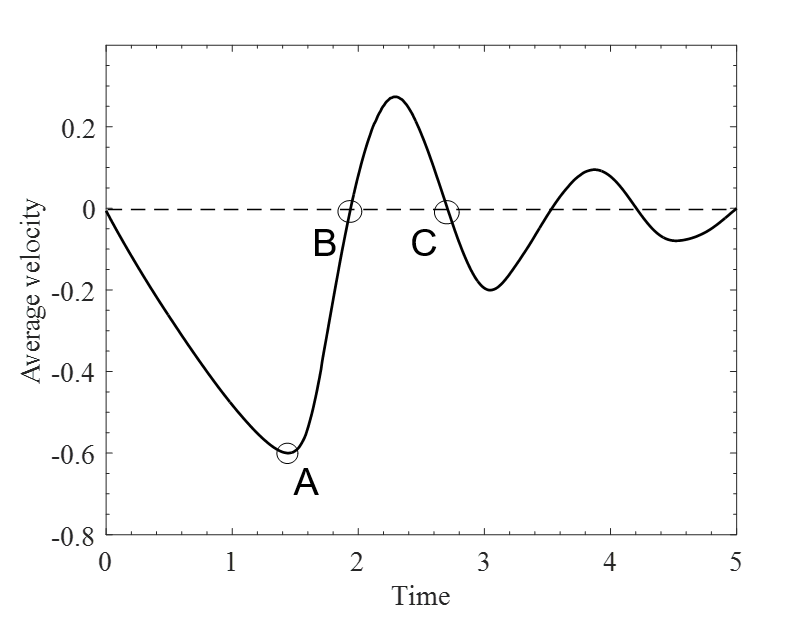}
		\caption{}
		\label{ball_drop_avg_vel}
	\end{subfigure}	
	\caption{Computed quantities for bouncing of an elastic ball: (a) variation of the center of mass of the ball with time and (b) variation of average velocity of the ball with time}
	\label{ball_drop}
\end{figure}

We consider the demonstration of a contact problem between an elastic solid and a rigid wall. For this problem, we place a soft elastic ball at the centre of a closed tank containing a viscous incompressible fluid. The domain is a square of side length 2 on $[-1,1] \times [-1,1]$ and the radius of the ball is 0.4 units initially. The physical parameters for the problem are: fluid viscosity $\mu^f=10$, solid viscosity $\mu^s=0$, shear modulus $\mu^s_L=10^4$, fluid density $\rho^f=1000$, solid density $\rho^s=5000$ and acceleration due to gravity $\boldsymbol{g}=(0,-0.98)$. We have no-slip boundary conditions for velocity and Neumann boundary conditions for the order parameter on all four walls. The initial condition for the order parameter is considered as
\begin{equation}
	\phi(x,y,0) = \tanh\bigg( \frac{0.4-\sqrt{(x-0)^2 + (y-0)^2}}{\sqrt{2}\varepsilon}\bigg) ,
\end{equation}
Since gravity is the only external force acting, the ball experiences ``free-fall" inside the viscous fluid. The ball begins to move down due to the action of gravity, with increasing speed. It comes in contact with the floor at point A (Fig.\ref{ball_drop_avg_vel}) and begins to compress. Due to the smooth, elastic nature of the ball, it stores its energy during compression and re-uses it to rise up during the rebound process. The important instances during the motion of the ball have been marked in Fig. \ref{ball_drop_avg_vel}. The bottom-most part of the ball comes in contact with the floor for the first time at point A and compresses from A to B; the ball begins to rise at point B where we observe a change in direction of the velocity and finally attains its maximum height at point C where the velocity again becomes zero. The loss of the elastic energy of the solid to the surrounding viscous fluid gradually dampens the solid motion and brings it to a stationary state. 


\begin{figure}[htbp]
	\centering
	\begin{tikzpicture}[scale=0.65]
	\begin{pgfonlayer}{nodelayer}
		\node [style=none] (0) at (0, 0) {};
		\node [style=none] (1) at (12, 0) {};
		\node [style=none] (2) at (12, 12) {};
		\node [style=none] (3) at (0, 12) {};
		\node [style=none] (6) at (12.25, 0) {};
		\node [style=none] (7) at (13.25, 0) {};
		\node [style=none] (8) at (12.25, 12) {};
		\node [style=none] (9) at (13.25, 12) {};
		\node [style=none] (10) at (12.75, 12) {};
		\node [style=none] (11) at (12.75, 0) {};
		\node [style=none] (12) at (0, 12.25) {};
		\node [style=none] (13) at (0, 13.25) {};
		\node [style=none] (14) at (12, 12.25) {};
		\node [style=none] (15) at (12, 13.25) {};
		\node [style=none] (16) at (12, 12.75) {};
		\node [style=none] (17) at (0, 12.75) {};
		\node [style=none, label={above:\large $L_x$}] (18) at (6, 13) {};
		\node [style=none, label={right:\large $L_y$}] (19) at (13, 6) {};
		\node [style=none, label={below: $X$}] (31) at (2, 0) {};
		\node [style=none, label={left: $Y$}] (32) at (0, 2) {};
		\node [style=none] (33) at (6, 6) {};
		\node [style=none, label={above: Incompressible fluid}] (34) at (4.5, 9.25) {};
		\node [style=none, label={below: Elastic ball}] (35) at (9.25, 2) {};
		\node [style=none] (36) at (7.5, 4) {};
		\node [style=none, label={below: $\Omega^f$}] (37) at (2, 8.25) {};
		\node [style=none, label={below: $(\rho^f,\mu^f)$}] (38) at (2, 7.25) {};
		\node [style=none, label={below: $\Omega^s$}] (39) at (6, 7.55) {};
		\node [style=none, label={below: $(\rho^s,\mu^s,\mu^s_{L})$}] (40) at (6, 6.5) {};
		\node [style=none, label={right: $g$}] (41) at (16, 6) {};
	\end{pgfonlayer}
	\begin{pgfonlayer}{edgelayer}
		\draw [thick] (0.center) to (1.center);
		\draw [thick] (1.center) to (2.center);
		\draw [thick] (0.center) to (3.center);
		\draw (8.center) to (9.center);
		\draw (6.center) to (7.center);
		\draw (13.center) to (12.center);
		\draw (15.center) to (14.center);
		\draw [thick, style=latex] (10.center) to (11.center);
		\draw [thick, style=latex] (17.center) to (16.center);
		\draw [very thick, style=stealth] (0.center) to (31.center);
		\draw [very thick, style=stealth] (0.center) to (32.center);
		\draw [thick] (3.center) to (2.center);
		\draw [style=stealth] (35.center) to (36.center);
		\draw [thick] (6,6) circle (2.5cm);
		\draw [thick, style=stealth] (16,7) to (16,5);
	\end{pgfonlayer}
\end{tikzpicture}
\caption{Computational domain for the bouncing of an elastic ball.}
\label{fig:ball_drop_domain}
\end{figure}
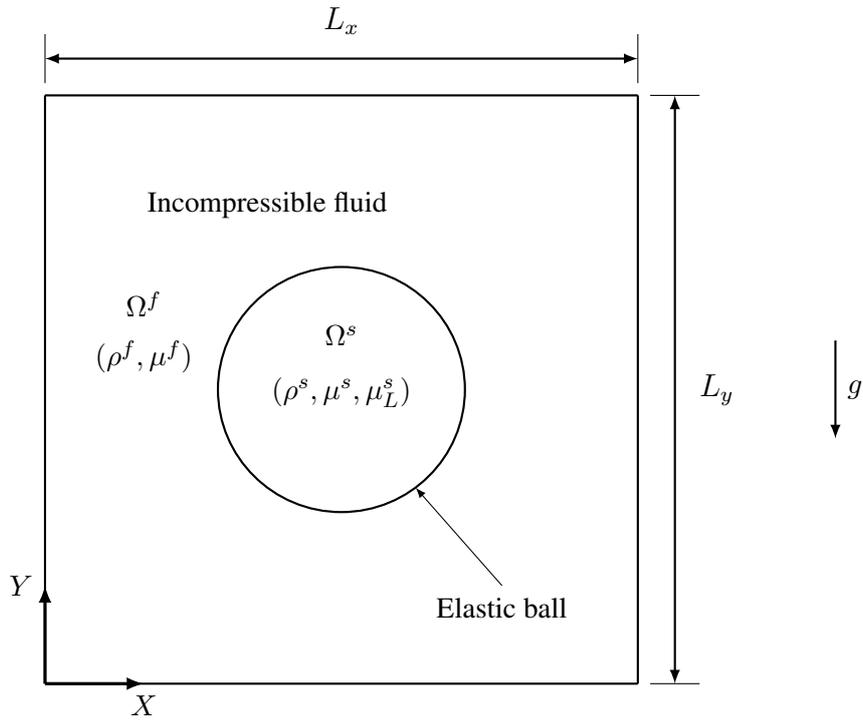

\begin{figure}[htbp]
	\centering		
		\includegraphics[scale=0.35]{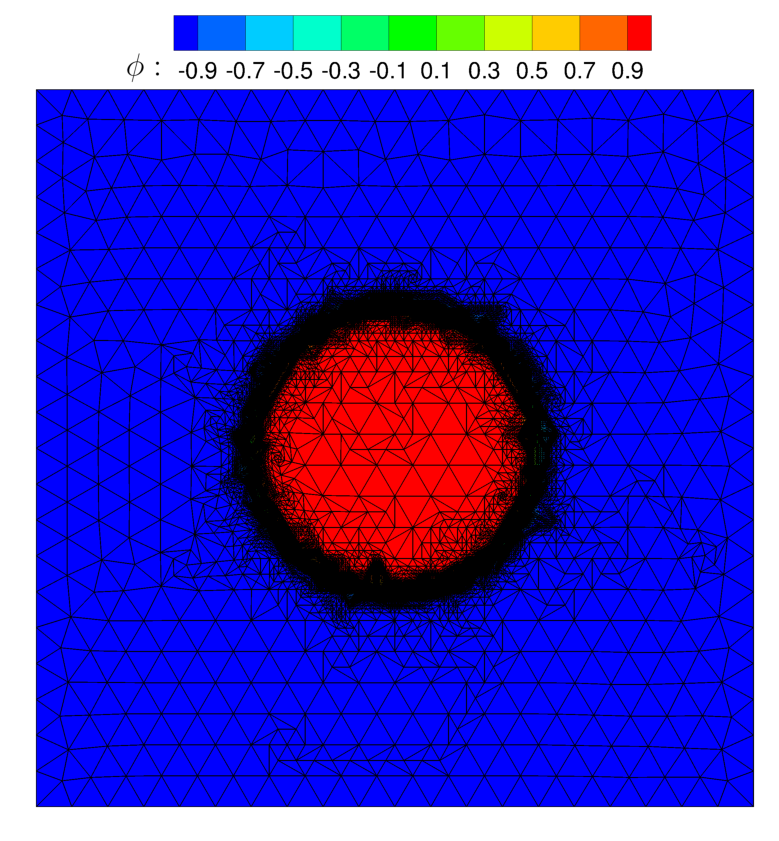}
		\caption{An elastic ball falling problem: representative adaptive grid and order parameter distribution at initial time step}
		\label{adaptive_grid_ball_drop}
\end{figure}

\begin{figure}[htbp]
	\centering
	\begin{subfigure}{0.49\textwidth}
		\centering		
		\includegraphics[width=\textwidth]{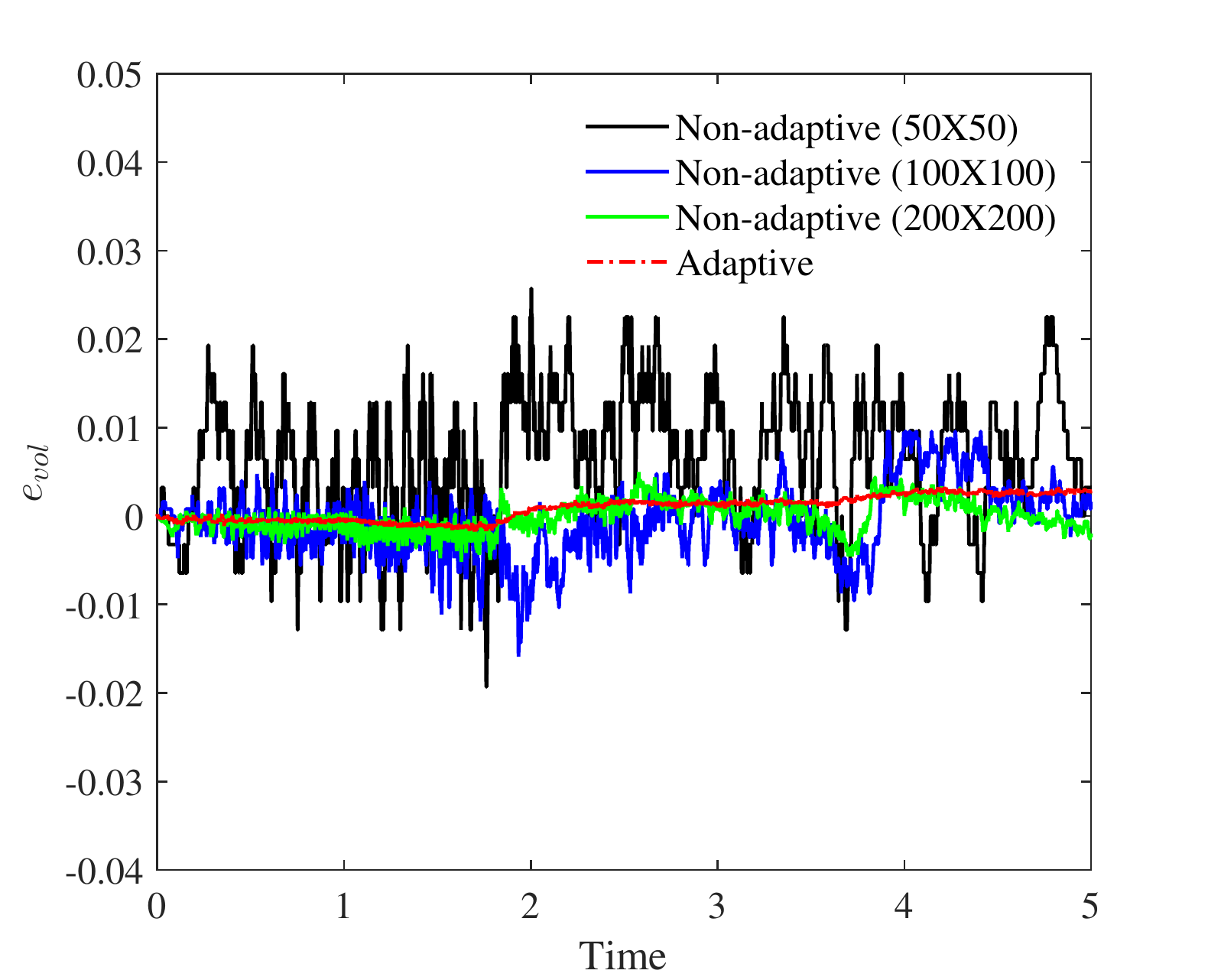}
		\caption{}
		\label{ball_drop_e_vol}
	\end{subfigure}
	\hfill
	\begin{subfigure}{0.49\textwidth}
		\centering		
		\includegraphics[width=\textwidth]{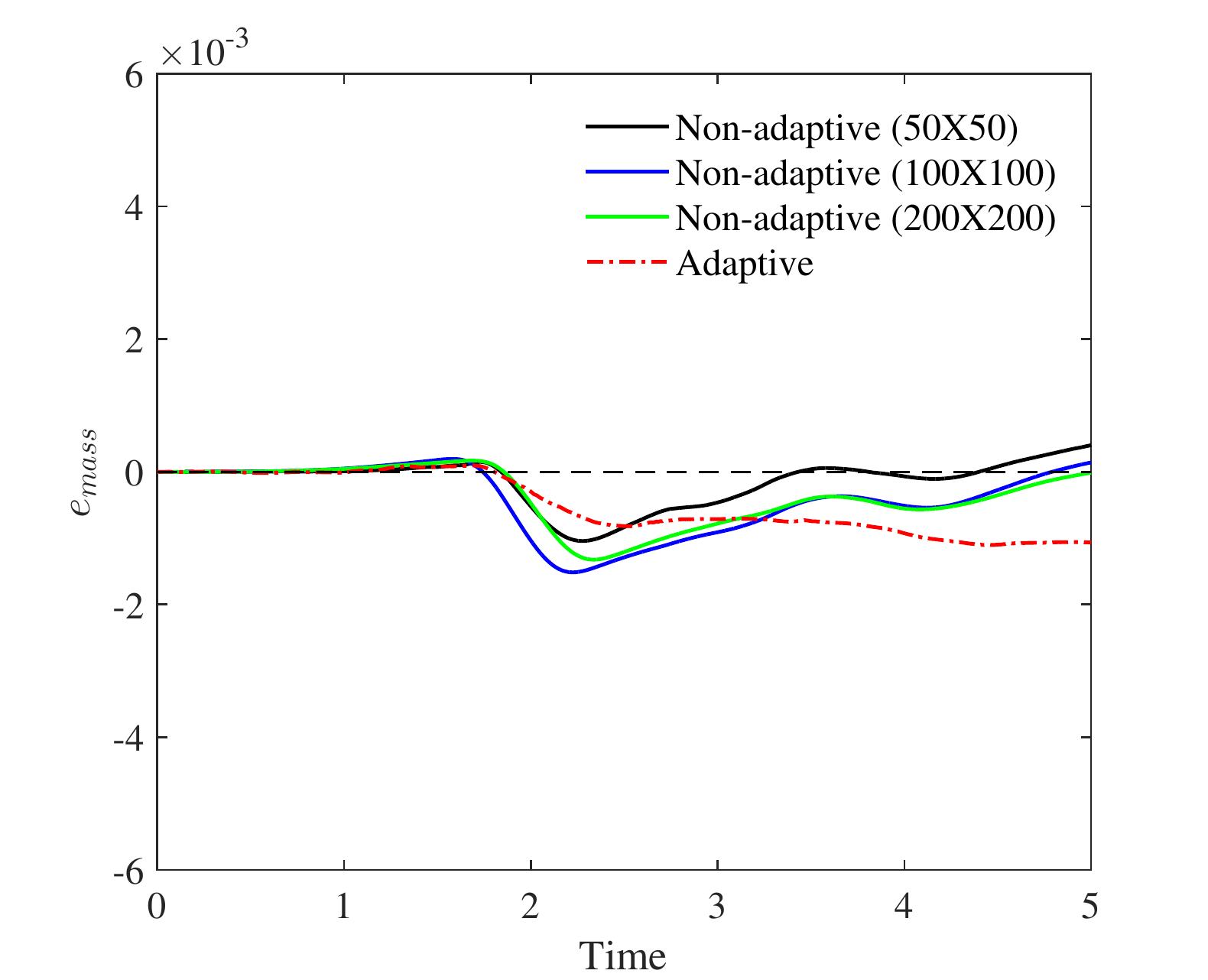}
		\caption{}
		\label{ball_drop_e_mass}
	\end{subfigure}
	\caption{Adaptive grid results for an elastic ball falling problem: (a) comparison of relative error for the ball volume for adaptive and non-adaptive cases with time (with an increase in refinement); (b) comparison of mass error in the entire domain for the adaptive and non-adaptive cases.}
	\label{ball_drop_adapt}
\end{figure}


We use adaptive mesh refinement for this problem as well (Fig. \ref{adaptive_grid_ball_drop}). The interface thickness parameter $\varepsilon$ is taken as $0.01$, for an initial grid size of $h_{init}=0.04$. The user-defined parameters for the D{\"o}rfler criteria are taken as $\theta=0.5$ and $\theta_c=0.05$. A significant improvement in the conservation properties is observed with the introduction of adaptive refinement in the grid as illustrated in Fig. \ref{ball_drop_e_vol}. This can be again attributed to the accurate resolution of the fluid-solid interface with the introduction of the adaptive procedure. The error in the conservation of the ball volume lies in a range of $-0.2\%$ to $+0.3\%$ for the adaptive grid case. It is worth noting that the fluctuations in the volume conservation error reduce greatly with an increase in refinement. The maximum error in the mass conservation in the domain is around $0.1\%$ for the adaptive case.

\subsection{Bouncing of Two Balls in a Liquid Tank}
\begin{figure}[htbp]
	\centering
	\begin{subfigure}[b]{0.49\textwidth}
		\centering
		\begin{tikzpicture}[scale=0.45]
	\begin{pgfonlayer}{nodelayer}
		\node [style=none] (0) at (0, 0) {};
		\node [style=none] (1) at (12, 0) {};
		\node [style=none] (2) at (12, 12) {};
		\node [style=none] (3) at (0, 12) {};
		\node [style=none] (6) at (12.25, 0) {};
		\node [style=none] (7) at (13.25, 0) {};
		\node [style=none] (8) at (12.25, 12) {};
		\node [style=none] (9) at (13.25, 12) {};
		\node [style=none] (10) at (12.75, 12) {};
		\node [style=none] (11) at (12.75, 0) {};
		\node [style=none] (12) at (0, 12.25) {};
		\node [style=none] (13) at (0, 13.25) {};
		\node [style=none] (14) at (12, 12.25) {};
		\node [style=none] (15) at (12, 13.25) {};
		\node [style=none] (16) at (12, 12.75) {};
		\node [style=none] (17) at (0, 12.75) {};
		\node [style=none, label={above:\large $L_x$}] (18) at (6, 13) {};
		\node [style=none, label={right:\large $L_y$}] (19) at (13, 6) {};
		\node [style=none, label={below: $X$}] (31) at (2, 0) {};
		\node [style=none, label={left: $Y$}] (32) at (0, 2) {};
		\node [style=none] (33) at (6, 6) {};
		\node [style=none, label={above: Incompressible fluid}] (34) at (3.5, 4.25) {};
		\node [style=none, label={below: Elastic balls}] (35) at (9.25, 10) {};
		\node [style=none] (36) at (7.5, 4) {};
		\node [style=none, label={below: $\Omega^f$}] (37) at (3, 3.25) {};
		\node [style=none, label={below: $(\rho^f,\mu^f)$}] (38) at (3, 2.25) {};
		\node [style=none, label={below: $\Omega^s$}] (39) at (3, 10.55) {};
		\node [style=none, label={below: $(\rho^s,\mu^s,\mu^s_{L})$}] (40) at (3, 9.5) {};
		\node [style=none, label={below: $\Omega^s$}] (41) at (9, 4.55) {};
		\node [style=none, label={below: $(\rho^s,\mu^s,\mu^s_{L})$}] (42) at (9, 3.5) {};
		\node [style=none, label={right: $g$}] (43) at (15, 6) {};
	\end{pgfonlayer}
	\begin{pgfonlayer}{edgelayer}
		\draw [thick] (0.center) to (1.center);
		\draw [thick] (1.center) to (2.center);
		\draw [thick] (0.center) to (3.center);
		\draw (8.center) to (9.center);
		\draw (6.center) to (7.center);
		\draw (13.center) to (12.center);
		\draw (15.center) to (14.center);
		\draw [thick, style=latex] (10.center) to (11.center);
		\draw [thick, style=latex] (17.center) to (16.center);
		\draw [very thick, style=stealth] (0.center) to (31.center);
		\draw [very thick, style=stealth] (0.center) to (32.center);
		\draw [thick] (3.center) to (2.center);
		\draw [style=stealth] (7.2,9.8) to (5,10.2);
		\draw [style=stealth] (10,8.8) to (8.9,5.2);
		\draw [thick] (9,3) circle (2cm);
		\draw [thick] (3,9) circle (2cm);
		\draw [thick, style=stealth] (15,7) to (15,5);
	\end{pgfonlayer}
	\end{tikzpicture}
		\caption{}
		\label{fig:2_balls_domain}
	\end{subfigure}
	\hfill
	\begin{subfigure}[b]{0.49\textwidth}
		\centering		
		\includegraphics[scale=0.216]{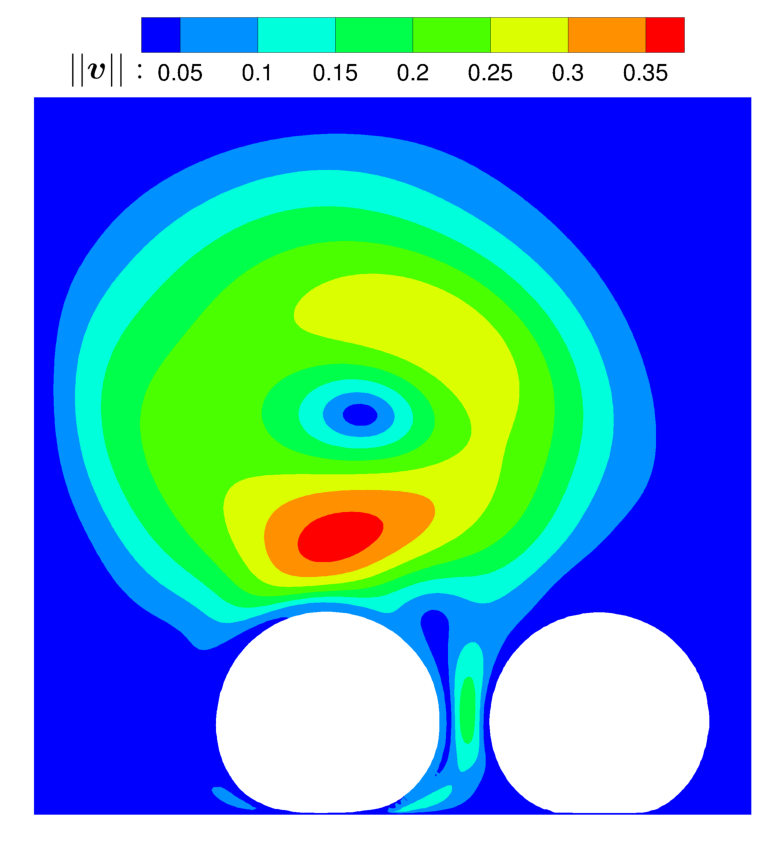}
		\caption{}
		\label{2_balls_mag_vel}
	\end{subfigure}
	\caption{Bouncing of two balls in a liquid tank: (a) schematic of the computational domain; (b) final values of magnitude of velocity in the domain}
	\label{fig_sim}
\end{figure}
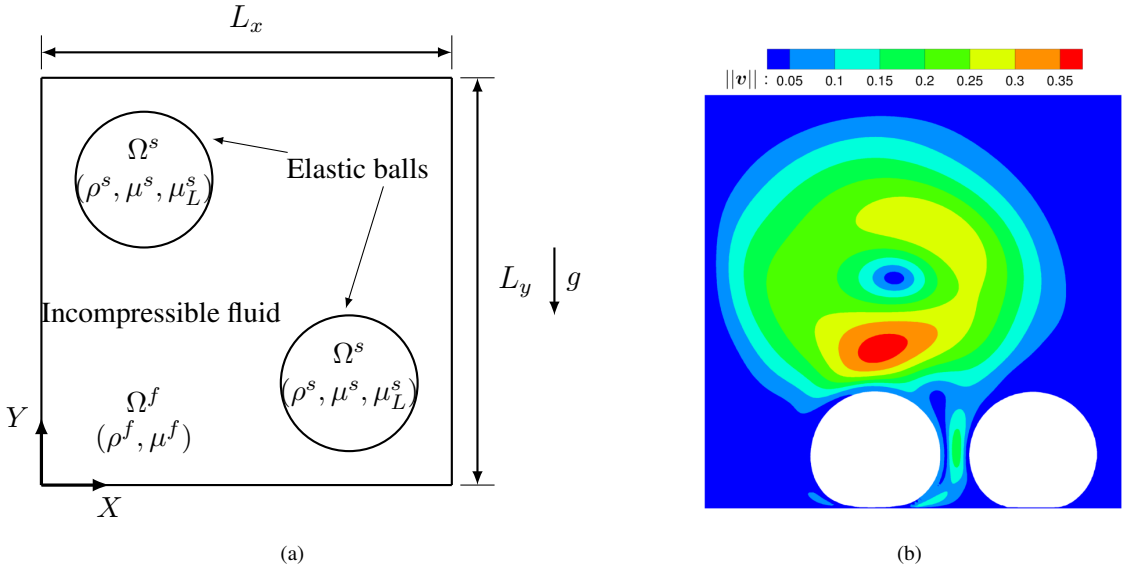

\begin{figure}[htbp]
	\centering
	\begin{subfigure}[b]{0.49\textwidth}
		\centering		
		\includegraphics[width=\textwidth]{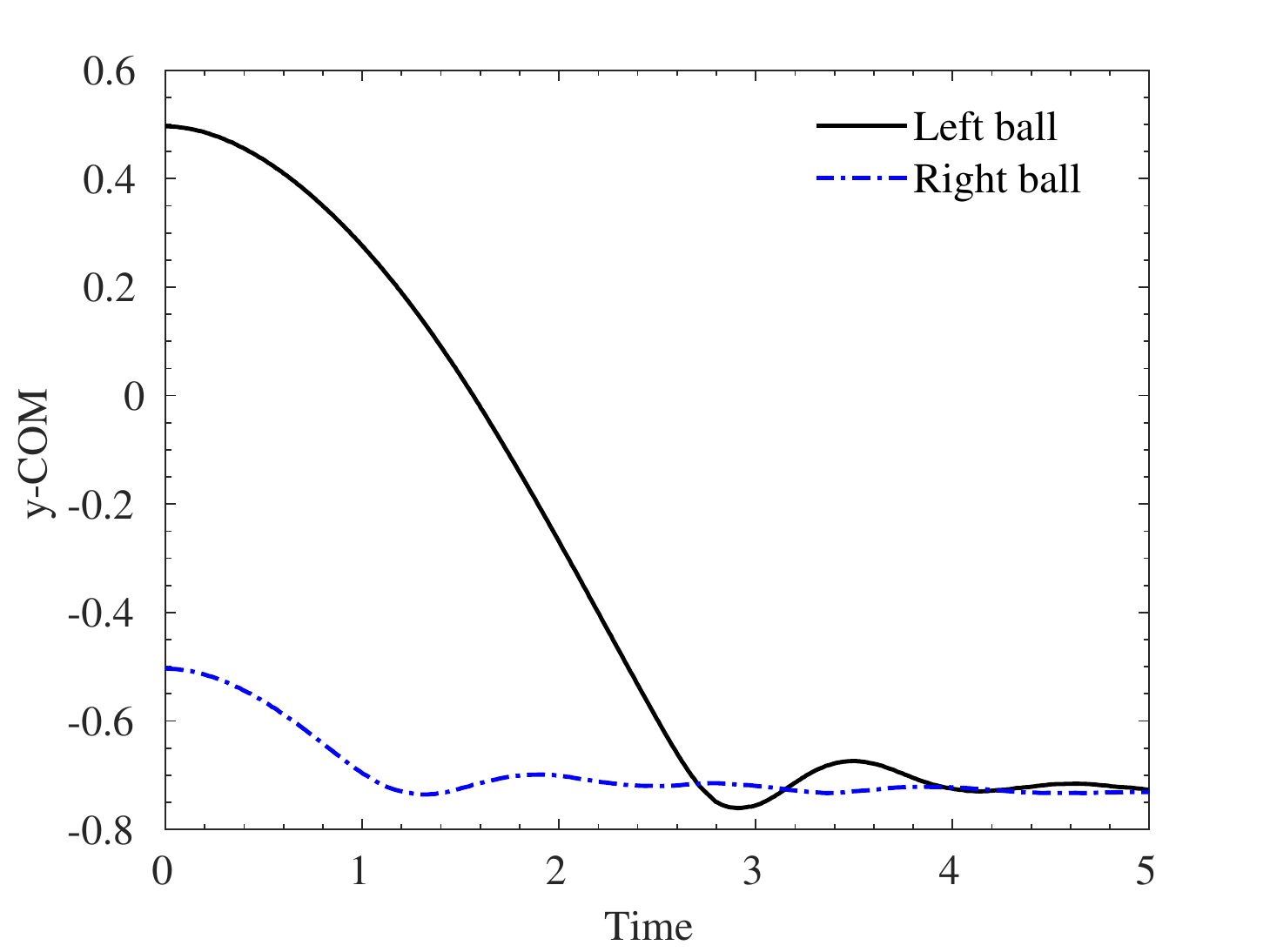}
		\caption{}
		\label{y_com}
	\end{subfigure}
	\hfill
	\begin{subfigure}[b]{0.49\textwidth}
		\centering		
		\includegraphics[width=\textwidth]{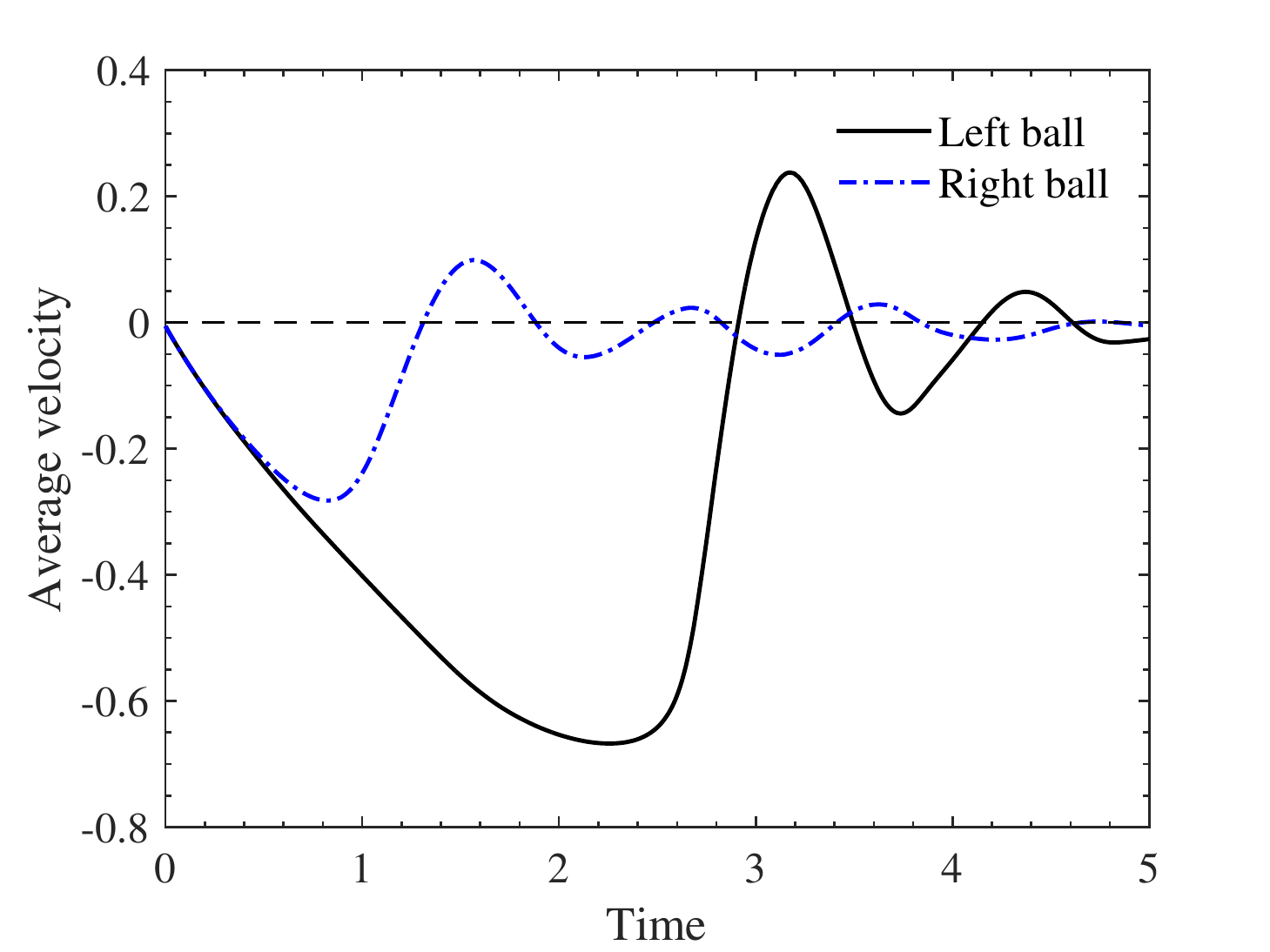}
		\caption{}
		\label{avg_vel}
	\end{subfigure}
\caption{Computed quantities for bouncing of two balls in a liquid tank: (a) variation of the center of mass of the balls with time and (b) variation of the average velocity of the balls with time }
	\label{fig:2_balls_motions_history}
\end{figure}

Next, we present a demonstration problem of multiple elastic bodies with free motion solved using the phase-field fully Eulerian approach. Here we observe the motion of two elastic balls placed inside a liquid tank containing a viscous incompressible fluid. The radius of each ball is 0.3 units initially. The balls are released from different initial heights and are allowed to fall under the action of gravity. The initial condition for the order parameter is assumed as:
\begin{equation}
\phi(x,y,0) = 1 + \tanh\bigg( \frac{0.3-\sqrt{(x+0.5)^2 + (y-0.5)^2}}{\sqrt{2}\varepsilon}\bigg) + \tanh\bigg( \frac{0.3-\sqrt{(x-0.5)^2 + (y+0.5)^2}}{\sqrt{2}\varepsilon}\bigg) .
\end{equation}
Similar to the previous case, the balls experience collision and rebound from the bottom wall by virtue of the stored elastic energy during the deformation process. The higher ball experiences a stronger rebounce due to larger initial energy compared to the ball placed at a lower height. We also observe that the higher ball drifts away from the wall as it moves downwards. Figure \ref{2_balls_mag_vel} illustrates the contour plot for the magnitude of velocity at the final time step. Figure \ref{fig:2_balls_motions_history} demonstrates the motion history of the two individual balls. The right ball undergoes several minor rebounds from the bottom wall, whereas the left ball undergoes one major rebound during the time of observation.

\begin{figure}[htbp]
	\centering		
		\includegraphics[scale=0.35]{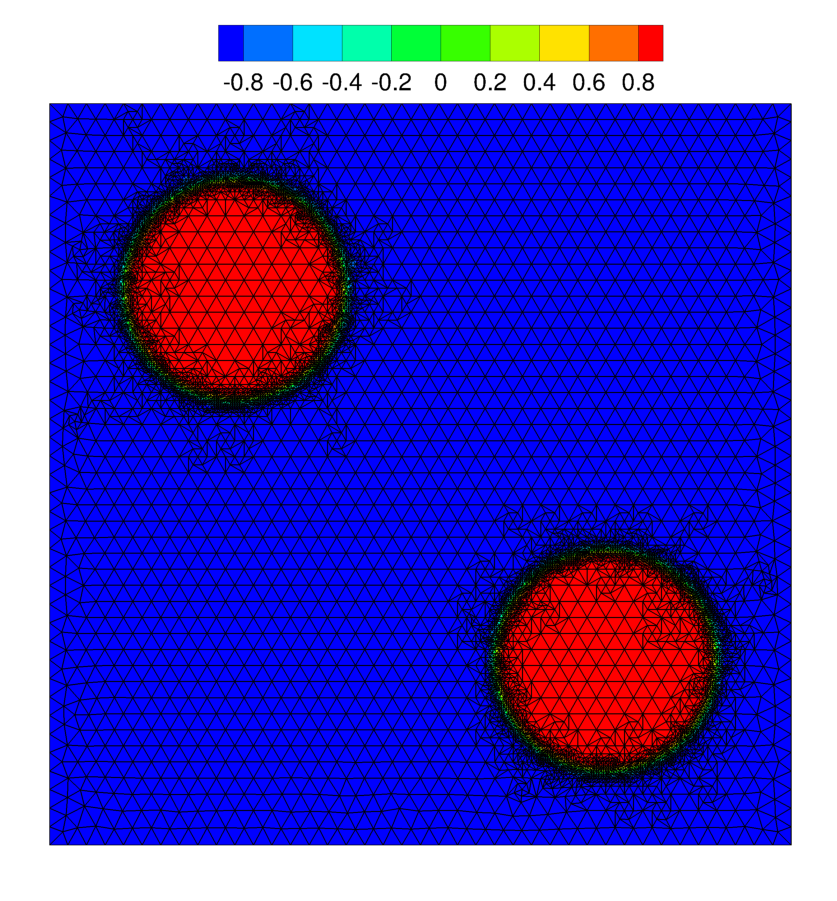}
		\put (-233,285) {$\phi :$}
		\caption{Bouncing of two balls in a liquid tank: representative adaptive grid and order parameter distribution at initial time step}
		\label{2_balls_adapt_grid}
\end{figure}

\begin{figure}[htbp]
	\centering
	\begin{subfigure}[b]{0.495\textwidth}
		\centering		
		\includegraphics[width=\textwidth]{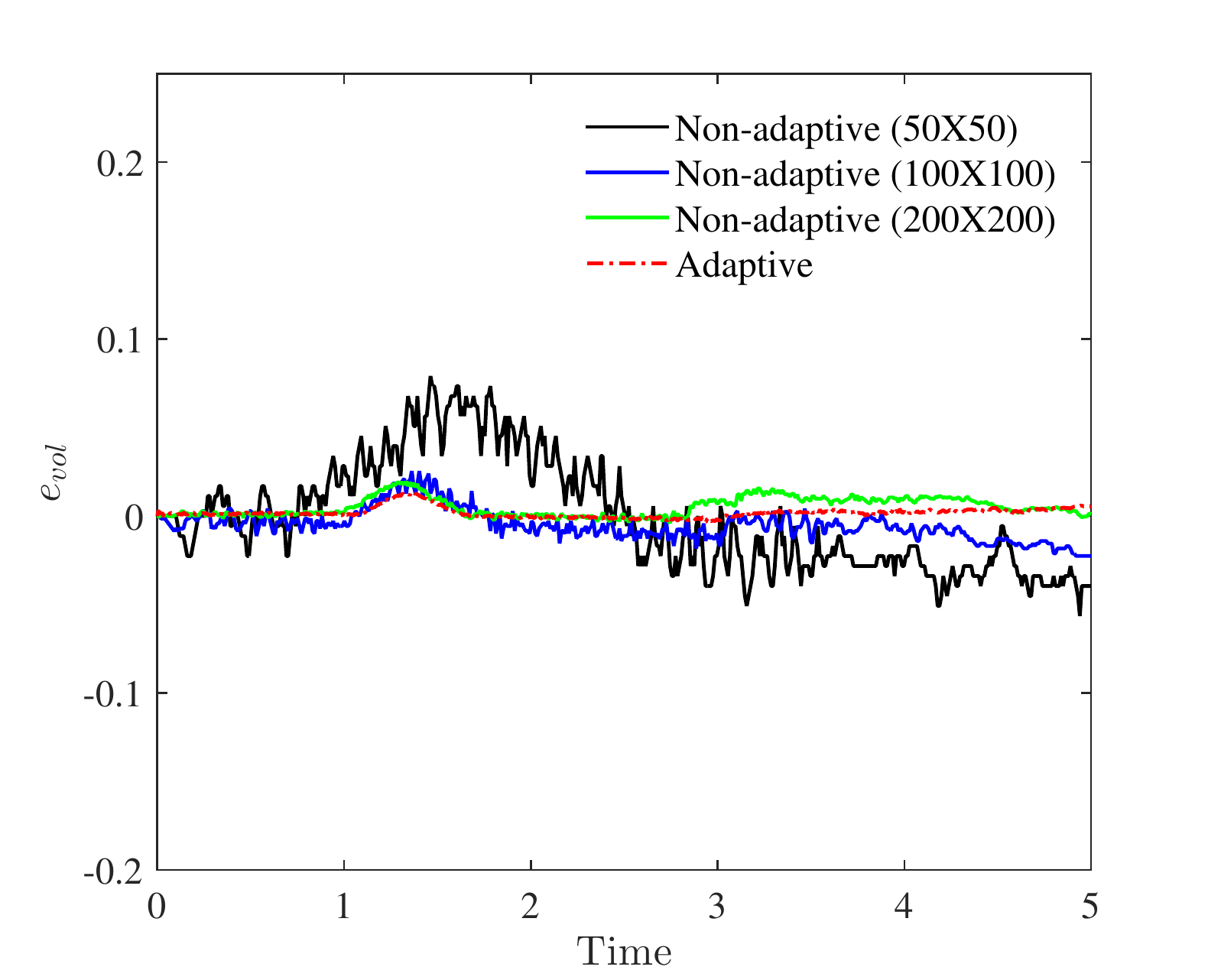}
		\caption{}
		\label{e_vol}
	\end{subfigure}
	\hfill
	\begin{subfigure}[b]{0.495\textwidth}
		\centering		
		\includegraphics[width=\textwidth]{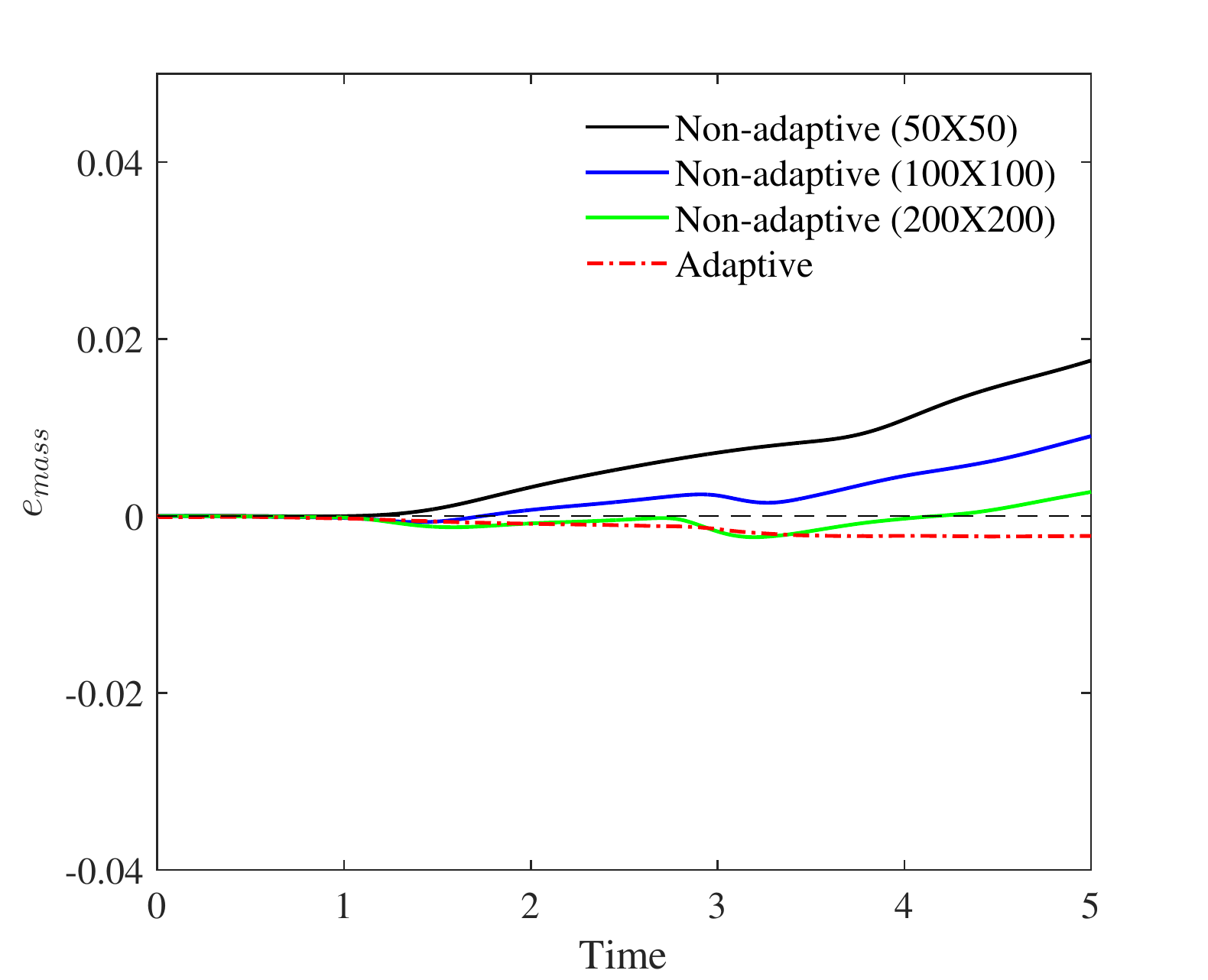}
		\caption{}
		\label{e_mass}
	\end{subfigure}
	\caption{Adaptive grid results for bouncing of two balls in a liquid tank: (a) comparison of relative volume error for the left ball for adaptive and non-adaptive cases with time (with an increase in refinement); and (b) comparison of the mass error in the entire domain for the adaptive and non-adaptive cases.}
	\label{fig:2_balls_adaptive_error}
\end{figure}

Adaptive refinement is carried out for this problem with $\varepsilon=0.01$ for an initial grid size of $h_{init}=0.05$. The user-defined parameters for the D{\"o}rfler criteria are taken as: $\theta=0.4$ and $\theta_c=0.1$. A restriction of 45000 maximum number of elements is enforced in the domain for the adaptive procedure. The average number of degrees of freedom for the adaptive case during the simulation time was around 17530, whereas it was 40401 for the finest non-adaptive mesh. The adaptive mesh at the start of the time loop is shown in Fig. \ref{2_balls_adapt_grid}. Figure \ref{fig:2_balls_adaptive_error} illustrates the improvement in conservation properties with the introduction of adaptive mesh refinement. We observe a significant reduction in volume conservation error of the left ball (Fig. \ref{e_vol}) and the total mass of the order parameter (Fig. \ref{e_mass}) for the adaptive case compared to the non-adaptive cases, owing to the increased density of elements at the fluid-solid interface. The error in volume conservation is within $1.3\%$ for the left ball in the adaptive refinement case.
Similarly, the maximum error in mass conservation for the domain is $-0.22\%$ for the adaptive case in contrast to the same being nearly $1-2\%$ for the non-adaptive cases.


The presented implicit variational fully Eulerian algorithm is general in its formulation and can be extended to three dimensions, without any restriction from the proposed variational formulation. However, the geometric complexities involved in refining and coarsening of the grid in three dimensions pose some challenges during parallelization \cite{chen2008ifem, zhang2009parallel}. These are some of the aspects for future consideration along with the incorporation of constitutive models for complex solid deformations, rupture phenomenon, and flexible multibody contact problems.

\section{Conclusions}
\label{conclusions}
In the current work, we presented an adaptive variational fully Eulerian approach for solving fluid-structure interaction problems on an unstructured grid. Both fluid and solid domains are described on an Eulerian grid and the interface is represented using the phase-field formulation. The solid strains are evolved using the left Cauchy-Green deformation tensor to evaluate the solid stresses and deformations. We introduced an adaptive mesh refinement to improve accuracy and conservation properties originally lost due to the Eulerian representation of the solid. The adaptive error indicator relies on the residual of the Allen-Cahn equation and hence the procedure effectively results in an interface-driven adaptive grid. The refinement and coarsening steps are avoided in intermediate Newton iterations to capture the non-linearities of the governing differential equations. We presented several test cases with increasing order of complexity to demonstrate the effectiveness and robustness of the present approach. The second-order spatial accuracy of the solver was established through a systematic convergence study. The comparison studies yielded a nearly three times reduction in computational effort and about 4.5 times improvement in mass loss error for the adaptive cases compared to the non-adaptive ones. We performed a parametric study to suggest the optimum range for the interface resolution ($\varepsilon$) for different initial grid sizes to minimize the mass loss error. We demonstrated that irrespective of the initial grid size, we have a recommended range of the interface thickness parameter, $\varepsilon = 0.004-0.014$, to perform FSI simulations with adaptive refinement. We solved the problem of a deformable solid in a driven cavity to illustrate the robustness of the solver in handling large solid deformations and showcased the enhanced conservation properties achieved via the adaptive procedure. Finally, we demonstrated the handling of solid-wall contact in the final section.

\section*{Acknowledgement}
The authors would like to acknowledge the Natural Sciences and Engineering Research Council of Canada (NSERC) and Seaspan Shipyards for the funding. This research was supported in part through computational resources and services provided by Advanced Research Computing at the University of British Columbia. 
\appendix
\setcounter{equation}{0}
\setcounter{figure}{0}
\section{Equivalence of the Transport Equations for Evolving Solid Strain}
\label{appendixA}
In this appendix, we show that the evolution of the solid strains can be carried out by either convecting $\boldsymbol{\xi}$, $\boldsymbol{F}$ or $\boldsymbol{B}$ and that the three methods are analytically identical to each other. As mentioned previously, the above transport equation requires that the material derivative of the {inverse map function} $(\boldsymbol{\xi})$ is zero, subject to an initial condition:
\begin{equation} \label{xi_eqn}
	\frac{\partial \boldsymbol{\xi}}{\partial t} + (\boldsymbol{v}^s \cdot \boldsymbol{\nabla} ) \boldsymbol{\xi} = 0,
\end{equation}
\begin{equation}
	\boldsymbol{\xi}(\boldsymbol{x},t=0)=\boldsymbol{x}=\boldsymbol{X} .
\end{equation}
To obtain the transport of the {deformation gradient tensor} $(\boldsymbol{F})$, we take the gradient of Equation \ref{xi_eqn},
\begin{equation}
	\nabla \left(\frac{\partial \boldsymbol{\xi}}{\partial t}\right) + \nabla ((\boldsymbol{v}^s \cdot \nabla) \boldsymbol{\xi}) = 0,
\end{equation}
\begin{equation}
	\frac{\partial (\nabla \boldsymbol{\xi})}{\partial t} + (\boldsymbol{v}^s \cdot \nabla)(\nabla \boldsymbol{\xi}) + \nabla \boldsymbol{\xi} \nabla \boldsymbol{v}^s = 0 .
\end{equation}
Using the fact that $\boldsymbol{F} = (\nabla \boldsymbol{\xi})^{-1}$, we have
\begin{equation}
	\frac{\partial (\boldsymbol{F}^{-1})}{\partial t} + (\boldsymbol{v}^s \cdot \nabla)\left(\boldsymbol{F}^{-1} \right) = - \boldsymbol{F}^{-1} \nabla \boldsymbol{v}^s  .
\end{equation}
The above equation can also be written as:
\begin{equation}
	\dot{\boldsymbol{F}}^{-1} = \frac{D \boldsymbol{F}^{-1}}{D t} = - \boldsymbol{F}^{-1} \nabla \boldsymbol{v}^s .
\end{equation}
From $\boldsymbol{F} \boldsymbol{F}^{-1} = \boldsymbol{I}$ and taking the total derivative of the above equation, we get
\begin{equation}
	\dot{\boldsymbol{F}} \boldsymbol{F}^{-1} + \boldsymbol{F} \dot{\boldsymbol{F}}^{-1} = \boldsymbol{0} .
\end{equation}
Hence, we have
\begin{equation}
\begin{aligned}
	\dot{\boldsymbol{F}} \boldsymbol{F}^{-1} &= - \boldsymbol{F} \dot{\boldsymbol{F}}^{-1} ,\\
	\dot{\boldsymbol{F}} \boldsymbol{F}^{-1} &= - \boldsymbol{F}\left(-\boldsymbol{F}^{-1} \nabla \boldsymbol{v}^s \right) = \nabla \boldsymbol{v}^s .
\end{aligned}
\end{equation}
Post-multiplying both sides by $\boldsymbol{F}$, we obtain the transport equation for $\boldsymbol{F}$:
\begin{equation} \label{F_eqn}
	\dot{\boldsymbol{F}} = \frac{\partial \boldsymbol{F}}{\partial t} + (\boldsymbol{v}^s \cdot \nabla ) \boldsymbol{F} = \nabla \boldsymbol{v}^s \boldsymbol{F} .
\end{equation}
Equation (\ref{F_eqn})  is the transport equation for the deformation gradient tensor.

Now for the transport of the {left Cauchy-Green deformation tensor} $(\boldsymbol{B})$, we first take the transpose of Eq. (\ref{F_eqn}):
\begin{equation}
	\dot{\boldsymbol{F}^T} = \frac{\partial \boldsymbol{F}^T}{\partial t} + (\boldsymbol{v}^s \cdot \nabla ) \boldsymbol{F}^T = \boldsymbol{F}^T (\nabla \boldsymbol{v}^s)^T .
\end{equation}
By taking the material derivative of the left cauchy-Green tensor $\boldsymbol{B} = \boldsymbol{F} \boldsymbol{F}^T$, we can write:
\begin{subequations}
\begin{align}
	\dot{\boldsymbol{B}} = \frac{\partial \boldsymbol{B}}{\partial t} + (\boldsymbol{v}^s \cdot \nabla) \boldsymbol{B} &= \frac{\partial \left(\boldsymbol{F} \boldsymbol{F}^T \right)}{\partial t} + (\boldsymbol{v}^s \cdot \nabla ) \left(\boldsymbol{F} \boldsymbol{F}^T \right) ,\\	
	&= \frac{\partial \boldsymbol{F}}{\partial t} \boldsymbol{F}^T + \boldsymbol{F}\frac{\partial \boldsymbol{F}^T}{\partial t} + ((\boldsymbol{v}^s \cdot \nabla) \boldsymbol{F}) \boldsymbol{F}^T + \boldsymbol{F} \left((\boldsymbol{v}^s \cdot \nabla) \boldsymbol{F}^T \right) ,\\
	&= \bigg(\frac{\partial \boldsymbol{F}}{\partial t} + (\boldsymbol{v}^s \cdot \nabla ) \boldsymbol{F} \bigg) \boldsymbol{F}^T + \boldsymbol{F} \bigg(\frac{\partial \boldsymbol{F}^T}{\partial t} + (\boldsymbol{v}^s \cdot \nabla ) \boldsymbol{F}^T \bigg) ,\\
	&= ((\nabla \boldsymbol{v}^s) \boldsymbol{F}) \boldsymbol{F}^T + \boldsymbol{F} \left(\boldsymbol{F}^T (\nabla \boldsymbol{v}^s)^T \right) ,\\
	&= \nabla \boldsymbol{v}^s \boldsymbol{F} \boldsymbol{F}^T + \boldsymbol{F} \boldsymbol{F}^T (\nabla \boldsymbol{v}^s)^T ,
\end{align}
\end{subequations}
which gives the transport equation for $\boldsymbol{B}$ as follows:
\begin{equation}
	\frac{\partial \boldsymbol{B}}{\partial t} + (\boldsymbol{v}^s \cdot \nabla) \boldsymbol{B} = \nabla \boldsymbol{v}^s \boldsymbol{B}+ \boldsymbol{B} (\nabla \boldsymbol{v}^s)^T .
\end{equation}
Hence, the three equations of $\boldsymbol{\xi}$, $\boldsymbol{F}$ or $\boldsymbol{B}$ for evolving the solid strains are equivalent in a continuum sense.

\section{Discretization of Solid Shear Stresses}
\label{appendixB}
We expand the solid shear stresses in terms of the left-Cauchy Green tensor to be incorporated in the unified continuum equations.
The shear stress in the solid, according to incompressible Neo-Hookean model, is given by:
\begin{equation}
	\boldsymbol{\sigma}^s_{sh} = \mu^s_{L} (\boldsymbol{F}^{\mathrm{n}+\alpha} (\boldsymbol{F}^{\mathrm{n}+\alpha})^T - \boldsymbol{I}) = \mu^s_{L} (\boldsymbol{B}^{\mathrm{n}+\alpha} - \boldsymbol{I}) .
\end{equation}
We need to evaluate $\boldsymbol{B}^{\mathrm{n}+\alpha}$ to substitute in the above equation to calculate the stresses. For this, we make use of the generalized-$\alpha$ time integration and the evolution equation for left Cauchy-Green tensor. From the generalized-$\alpha$ time integration, we have
\begin{subequations}
\begin{align}
	\boldsymbol{B}^{\mathrm{n}+\alpha} &= \boldsymbol{B}^{\mathrm{n}} + \alpha(\boldsymbol{B}^{\mathrm{n}+1} - \boldsymbol{B}^{\mathrm{n}} ) ,\\
	&= \boldsymbol{B}^{\mathrm{n}} + \alpha \Delta t (\partial_t \boldsymbol{B}^{\mathrm{n}} + \varsigma (\partial_t \boldsymbol{B}^{\mathrm{n+1}} - \partial_t \boldsymbol{B}^{\mathrm{n}})) ,\\
	&= \boldsymbol{B}^{\mathrm{n}} + \alpha \Delta t \left(\partial_t \boldsymbol{B}^{\mathrm{n}} + \frac{\varsigma}{\alpha_m} (\partial_t \boldsymbol{B}^{\mathrm{n+\alpha_m}} - \partial_t \boldsymbol{B}^{\mathrm{n}})\right) ,\\
	&= \boldsymbol{B}^{\mathrm{n}} + \alpha \Delta t \left( \left(1-\frac{\varsigma}{\alpha_m} \right)\partial_t \boldsymbol{B}^{\mathrm{n}} + \frac{\varsigma}{\alpha_m}\partial_t \boldsymbol{B}^{\mathrm{n+\alpha_m}}\right) , \label{B_n+alpha}
\end{align}
\label{gen_alpha_B}
\end{subequations}
Using the transport equation for the left Cauchy-Green tensor, we have
\begin{equation}
	\partial_t \boldsymbol{B}^{\mathrm{n+\alpha_m}} = \nabla \boldsymbol{v}^{\mathrm{n+\alpha}}\boldsymbol{B}^{\mathrm{n+\alpha}} + \boldsymbol{B}^{\mathrm{n+\alpha}} (\nabla \boldsymbol{v}^{\mathrm{n+\alpha}})^T -(\boldsymbol{v}^{\mathrm{n+\alpha}} \cdot \nabla)\boldsymbol{B}^{\mathrm{n+\alpha}} .
\label{B_transport}
\end{equation}
For simplification, we assume $\varsigma=\alpha_m$ $(\rho_{\infty}=1)$. Substituting Equation (\ref{B_transport}) in Equation (\ref{B_n+alpha}) with the above simplification, we obtain
\begin{equation}
	\boldsymbol{B}^{\mathrm{n}+\alpha} = \boldsymbol{B}^{\mathrm{n}} + \alpha \Delta t \bigg(\nabla \boldsymbol{v}^{\mathrm{n+\alpha}}\boldsymbol{B}^{\mathrm{n+\alpha}} + \boldsymbol{B}^{\mathrm{n+\alpha}} (\nabla \boldsymbol{v}^{\mathrm{n+\alpha}})^T -(\boldsymbol{v}^{\mathrm{n+\alpha}} \cdot \nabla)\boldsymbol{B}^{\mathrm{n+\alpha}} \bigg) .
\end{equation}
The expression for solid shear stress finally comes out as
\begin{equation}
	\boldsymbol{\sigma}^s_{sh} = \mu^s_{L} \bigg( \alpha \Delta t \bigg(\nabla \boldsymbol{v}^{\mathrm{n+\alpha}}\boldsymbol{B}^{\mathrm{n+\alpha}} + \boldsymbol{B}^{\mathrm{n+\alpha}} (\nabla \boldsymbol{v}^{\mathrm{n+\alpha}})^T -(\boldsymbol{v}^{\mathrm{n+\alpha}} \cdot \nabla)\boldsymbol{B}^{\mathrm{n+\alpha}} \bigg) + \boldsymbol{B}^{\mathrm{n}} - \boldsymbol{I} \bigg) .
\end{equation}


\bibliography{references}

\end{document}